\documentclass[12pt]{amsart}
\usepackage{graphicx}
\usepackage{fourier}
\usepackage{graphicx}
\usepackage{epsfig}

\newtheorem{thm}{Theorem}[section]

\usepackage{amsmath}
\usepackage{amsxtra}
\usepackage{amscd}
\usepackage{amsthm}
\usepackage{amsfonts}
\usepackage{amssymb}
\usepackage{eucal}
\newcommand{\bmb}{\left( \begin{array}{rr}}
\newcommand{\enm}{\end{array}\right)}

\newcommand{\Z}{{\mathbb Z}}

\newcommand{\R}{{\mathbb R}}

\newcommand{\al}{{\alpha}}

\newcommand{\qq}{{\mathfrak{q}}}
\newcommand{\uu}{{u}}

\numberwithin{equation}{section}

\begin{document}

\title[Tangent method for $q$-weighted paths]{A tangent method derivation of the arctic curve \\for $q$-weighted paths with arbitrary starting points}
\author{Philippe Di Francesco} 
\address{
Department of Mathematics, University of Illinois, Urbana, IL 61821, U.S.A. 
and 
Institut de physique th\'eorique, Universit\'e Paris Saclay, 
CEA, CNRS, F-91191 Gif-sur-Yvette, FRANCE\hfill
\break  e-mail: philippe@illinois.edu
}
\author{Emmanuel Guitter}
\address{
Institut de physique th\'eorique, Universit\'e Paris Saclay, 
CEA, CNRS, F-91191 Gif-sur-Yvette, FRANCE.
\break  e-mail: emmanuel.guitter@ipht.fr
}

\begin{abstract}
We use a tangent method approach to obtain the arctic curve in a model of non-intersecting lattice paths within the first quadrant,
including a $q$-dependent weight associated with the area delimited by the paths. Our model is characterized by an arbitrary 
sequence of starting points along the positive horizontal axis, whose distribution involves an arbitrary piecewise differentiable function. 
We give an explicit expression for the arctic curve in terms of this arbitrary function and of the parameter $q$. A particular emphasis 
is put on the deformation of the arctic curve upon varying $q$, and on its limiting shapes when $q$ tends to $0$ or infinity. Our analytic 
results are illustrated by a number of detailed examples.
\end{abstract}

\maketitle
\date{\today}
\tableofcontents

\section{Introduction}
\label{sec:introduction}
The study of two-dimensional non intersecting lattice path (NILP) configurations is a subject of constant investigation, in particular because they provide
alternative descriptions for a number of statistical models, including tiling problems \cite{CEP,JPS} or dimer models on regular lattices. Quite generally, their statistics
exhibits a number of interesting properties, among which is the remarkable \emph{arctic curve phenomenon}
which may be described as follows: for prescribed boundary conditions (obtained for instance by fixing the starting and ending points of the paths),
the paths may by construction visit only a fixed domain $D$ in the lattice. In the thermodynamic limit, i.e.\ for a large number of paths and under the appropriate scaling, 
this accessible domain $D$ is then split into one or several
liquid disordered phases in which paths may fluctuate with a finite entropy, and frozen (crystalline) ordered phases in which paths develop some underlying order 
generally imposed by some nearby boundary. Frozen phases may correspond either to fully filled regions with a compact arrangement of the paths 
characterized by a fixed common orientation or, on the contrary, to regions not visited by paths. In the thermodynamic limit, the transition between frozen and 
liquid phases is sharp and takes place along a well defined \emph{arctic curve} (with possibly several connected components) whose shape depends only on the 
boundary conditions and on some local weights possibly attached to the paths. The arctic curve phenomenon was described in a quite general setting in 
\cite{KOS,KO1,KO2}. Several methods were designed to obtain, for specific NILP problems, the precise location of their arctic curve. These are in general based 
on the identification of the various phases \emph{in the bulk} and their implementation, which requires the evaluation of bulk expectation values, is achieved by use of 
quite involved techniques such as inversion of the Kasteleyn matrix, or more recently by exploiting the underlying cluster integrable system structure of the equivalent dimer problem \cite{DFSG,KP}. 
\medskip

On the other hand, an elegant new technique, referred to as the \emph{tangent method}, was recently invented by Colomo and Sportiello \cite{COSPO}: it produces
the arctic curve via a simple geometric construction, without recourse to any bulk order parameter evaluation. 
The idea is the following: many NILP problems have several 
equivalent formulations involving different families of paths and a given portion of the arctic curve may always in practice be understood, for the appropriate path family,
as the separation between a liquid phase and a region empty of all paths. In particular, the shape of the arctic curve is dictated by the most likely trajectories of \emph{outermost paths} in the NILP 
configuration since these are precisely the paths which delimit the visited region. Based on this remark, the tangent method consists in reconstructing the arctic curve from the location 
of the outermost path trajectories for the various equivalent path families defining the model.
In practice, the trajectory of the outermost path is obtained by perturbing it upon moving one of its endpoints outside of the originally allowed domain $D$, so as
to force it to cross the empty region before it eventually exits $D$. The perturbed and unperturbed trajectories are expected to share a common part before 
they eventually split \emph{tangentially} (hence the name of the method). After splitting, the perturbed trajectory which takes place in some empty region is 
somewhat trivial as it is no longer influenced by the other paths: as a consequence, it follows a geodesic 
and one may thus easily reconstruct the position of the tangency (splitting) point from that of the point 
where the path most likely exits $D$. The latter is determined by a variational principle.
By varying the displaced endpoint, one then reconstructs the entire unperturbed outermost trajectory as the envelope 
of the family of geodesics thus produced, yielding the desired portion of arctic curve.
The tangent method was tested successfully in a number of problems \cite{COSPO,DFLAP,DFGUI} where it was shown to reproduce already known 
results and yielded new explicit predictions.
\medskip

In a recent paper \cite{DFGUI}, we concentrated on a particular NILP problem involving paths traveling up and left along the edges of the first 
quadrant of a regular square lattice and with an \emph{arbitrary sequence of starting points} along the positive horizontal axis, with abscissa $a_0=0,a_1,a_2,...,a_n$, and with the fixed sequence of endpoints along the positive vertical axis
at positions $0,1,2,...,n$. Applying the tangent method,
we were able to obtain the corresponding arctic curve in terms of the asymptotic distribution of starting points in the thermodynamic limit.
In particular, this allowed us to recover via simple geometrical constructions the results of \cite{DM1} and {\it de facto} to validate the tangent method.

In the present paper, we address the same question of the arctic curve, for the same NILP problem, but including a new $q$-dependent weight for 
the NILP configurations, associated with the \emph{area under the paths}. 
More precisely, let $\mathcal{A}_i$ be the area delimited by the coordinate axes and a path $\mathcal{P}_i$ in the first quadrant. A NILP configuration then receives a total statistical weight 
$q^{\sum_i \mathcal{A}_i}$ where the sum runs over all the paths in the configuration.  
A small value of $q$ favors configurations in which the paths are squeezed towards the origin of the first quadrant so as to lower the 
cumulative area $\sum_i \mathcal{A}_i$. On the contrary, a large value of $q$ pushes the paths away from this origin. Such choice of $q$-dependent weight is quite 
natural and was already considered in \cite{MkPe17} in a more specific situation\footnote{This situation corresponds in fact to a particular instance of our general framework 
with a sequence of starting points corresponding to so-called freezing boundaries only.}. There the model is presented in its equivalent tiling formulation, 
which may itself be viewed as a plane partition, or equivalently as a three-dimensional piling of elementary cubic bricks (see \cite{MkPe17,DFGUI}). In this language, the above cumulative area 
$\sum_i \mathcal{A}_i$ has a nice geometrical interpretation as a measure of the \emph{volume} below the surface of the brick piling (and above some appropriate base plane, see \cite{MkPe17} for details).
\medskip

Our main result is an explicit parametric expression for the arctic curve in terms of the 
(arbitrary, piecewise differentiable) distribution of starting points $\al(u)=\lim_{n\to \infty} a_{\lfloor n\,u\rfloor }/n$, $u\in [0,1]$, and of the renormalized parameter $\qq =q^{1/n}$:

\begin{thm}\label{mainthm}
Let $x(t)$ be  the $\qq$-deformed exponential moment-generating function for the distribution $\al(u)$ of starting points,
namely:
\begin{equation}\label{eq:defx} x(t):=\qq^{\textstyle{-t \, \int_0^1 \frac{du}{t-\qq^{\al(u)}}}} \ .
\end{equation}
The arctic curve for the asymptotic configurations of NILP with prescribed endpoints is given in the following parametric form $(X(t),Y(t))$, for admissible ranges of $t\in \R$:
\begin{equation} \qq^{X(t)}=  \frac{t^2 \, x'(t)}{t\, x'(t)+x(t)(1-x(t))}\ , \quad 
\qq^{Y(t)}=\frac{t\, x'(t)+1-x(t)}{t\, x'(t)+x(t)(1-x(t))} \ .
\label{eq:arcticthm}
\end{equation}
\end{thm}
The precise relevant admissible domains for $t$ are discussed in the paper.
Using this result,
we may follow the deformation of the arctic curve for varying $\qq$, and obtain its limiting shape whenever $\qq$ tends to $0$ or to infinity. 

The paper is organized as follows. In Section~\ref{sec:partfunc}, we give a precise definition of the NILP problem under study, which is first presented in 
its ``original" form (Section~\ref{sec:directform}) involving a first family of paths along the edges of the first quadrant of a regular square lattice, and then reformulated 
in terms of a second, dual family of paths (Section~\ref{sec:alternativeform}), with a detailed analysis of the mapping between these two formulations. 
The model is entirely characterized by a its fixed arbitrary sequence of starting points as well as by the weight parameter $q$ and we give in Section~\ref{sec:directform}
an explicit expression for its partition function. Section~\ref{sec:onepoint} is devoted to the computation of the basic quantities required to apply the tangent method to our problem.
These include in particular the so-called \emph{one-point function}, computed in Section~\ref{sec:onepointexact}, which enumerates path configurations in which the outermost path is perturbed so as to exit the allowed domain $D$ at a prescribed exit point. The associated scaling expression in the thermodynamic limit of a large number of paths is discussed in Section~\ref{sec:saddle} where
we also analyze the position of the most likely exit point. Section~\ref{sec:arctic} proves our main result, namely the above parametric equation \eqref{eq:arcticthm} for the arctic curve . Its derivation requires computing the equation for  ``geodesics" (Section~\ref{sec:freetrajectory}),
i.e.\ free trajectories of the (perturbed) outermost path within an unvisited region empty of all the other paths. The arctic curve is then obtained from the tangent method principle
as the envelope of the geodesics passing via the previously identified most likely exit points (Section~\ref{sec:tangentmethod}). 
The above construction, based exclusively on the original path family of Section~\ref{sec:directform}, produces only one portion of the arctic curve, its so-called ``right part".
We show in Section~\ref{sec:otherportion} how to get other portions of the arctic curve, a generic "left part" (Section~\ref{sec:leftpart}) obtained from outermost trajectories in the second path family of 
Section~\ref{sec:alternativeform}, as well as possible additional portions (Section~\ref{sec:freezing}) arising for so-called "freezing boundaries" in the presence 
of either fully filled intervals or gaps in the sequence of starting points. Section~\ref{sec:limitsgen} is devoted to the description of the arctic curve in the limit where $q$ tends to $0$
or to infinity, either via heuristic arguments (Section~\ref{sec:heuristicgen}) based on the identification of the most likely limiting NILP configuration, or via a rigorous 
treatment analyzing the limit of the arctic curve equation \eqref{eq:arcticthm} when $\qq$ becomes large (Section~\ref{sec:analyticgen}) or small (Section~\ref{sec:analyticgenbis}).
Section~\ref{sec:examples} presents a number of explicit examples of this deformation of the arctic curve when $q$ varies for a fairly generic class of starting point distributions 
(Section~\ref{sec:limitslinear}),
including situations with freezing boundaries resulting from a fully filled interval in the starting point sequence (Section~\ref{sec:freez1}) or from a gap (Section~\ref{sec:freez2}).
As a  final example we revisit the path formulation of the classical rhombus tiling problem of a hexagonal domain \cite{CLP} in Section~ \ref{sec:ellipse}. We show how the arctic curve, known to be an ellipse for 
$q=1$ is deformed for large or small $q$ as a result of the invasion of the liquid phase by frozen regions.
We gather a few concluding remarks in Section~\ref{sec:conclusion}.

\medskip

\noindent{\bf Acknowledgments.}  

\noindent  We thank L. Petrov for suggesting the $q$-deformed problem and A. Sportiello 
for useful discussions. PDF is partially supported by the Morris and Gertrude Fine endowment and the NSF grant DMS18-02044. EG acknowledges the support of the grant ANR-14-CE25-0014 (ANR GRAAL).

\section{Partition function for $q$-weighted non-intersecting lattice paths}
\label{sec:partfunc}
\subsection{Direct path formulation}
\label{sec:directform}
As in \cite{DFGUI}, we consider configurations of non-intersecting lattice paths consisting of $(n+1)$ paths $\mathcal{P}_i$,  $i=0,1,\cdots,n$, making west- or 
north-oriented unit steps along the edges of the regular square lattice $\Z^2$, starting at respective position $O_i=(a_i,0)$ along the $x$-axis and ending at position $E_i=(0,i)$
along the $y$-axis. Here $(a_i)_{0\le i\le n}$ denotes an arbitrarily fixed
\emph{strictly increasing sequence of integers} with $a_0=0$. The paths are non-intersecting in the sense that any two distinct paths may not share a common vertex.
Clearly, the domain $D$ accessible to the paths is a rectangle of size $a_n\times n$ in the first quadrant, with its lower left corner at the origin.

\begin{figure}
\begin{center}
\includegraphics[width=10cm]{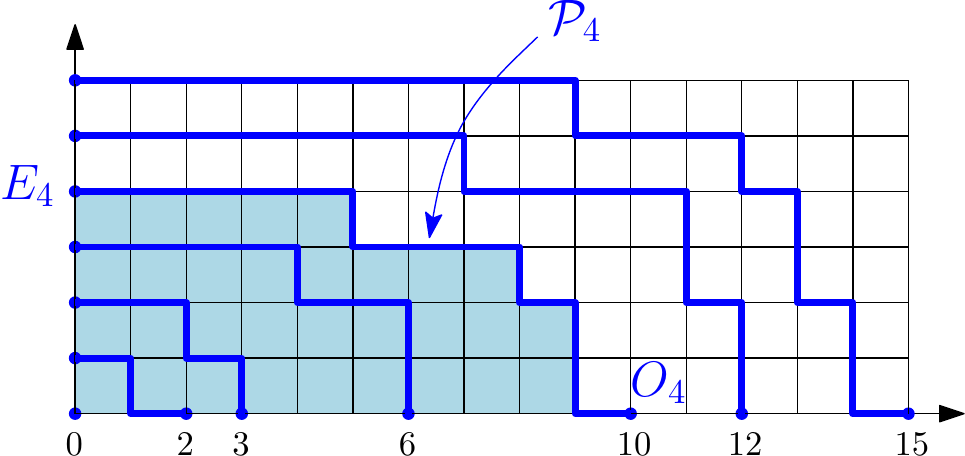}
\end{center}
\caption{\small A sample configuration of $n+1=7$ non-intersecting lattice paths made of west- or north-oriented unit steps. The $i$-th path $\mathcal{P}_i$ starts at position $O_i=(a_i,0)$ and ends 
at position $E_i=(0,i)$ (here for the sequence $(a_i)_{0\leq i\leq n}=(0,2,3,6,10,12,15)$). For illustration, we colored the domain ``to the left of the path" $\mathcal{P}_4$ whose number of unit
squares defines the area $\mathcal{A}_4$ (here $=31$). The weight of the configuration is $q^{\sum_{i=0}^n\mathcal{A}_i}$ (here $q^{0+1+5+16+31+53+75}=q^{181}$). 
}
\label{fig:paths}
\end{figure}

The novelty of the present paper is that each path $\mathcal{P}_i$ now receives a weight $q^{\mathcal{A}_i}$, 
where $q$ is some arbitrary \emph{positive real number} and $\mathcal{A}_i$ 
measures the area ``to the left of the path" $\mathcal{P}_i$, i.e.\ the number of unit squares in the domain delimited by the path $\mathcal{P}_i$ and its projection along the $y$-axis 
(see figure \ref{fig:paths}). 
Note that in the present case, this area may also be viewed as the area ``under the path", i.e. the number of unit squares in the domain delimited by $\mathcal{P}_i$ and its projection 
along the $x$-axis. The total weight of a NILP configuration is then the product of its path weights, namely $q^{\sum_{i=0}^n\mathcal{A}_i}$. 
Alternatively, the weight $q^{\mathcal{A}_i}$ of the path $\mathcal{P}_i$ may be obtained by assigning to each north-oriented step $(x,y)\to (x,y+1)$ of the path a local weight $q^x$.
Since this latter formulation involves only local edge weights, the partition function $Z_n(q):=Z_n(q;(a_i)_{0\leq i\leq n})$ of the model may be obtained via the famous   
Lindstr\"om-Gessel-Viennot (LGV) lemma \cite{LGV1,GV} as 
\begin{equation}
Z_n(q)=\det\left(\left(A_{i,j}\left(q\right)\right)_{0\leq i,j\leq n}\right)
\label{eq:Znq}
\end{equation}
where $A_{i,j}(q)$ denotes the partition function of a \emph{single path} $\mathcal{P}$ (made of west- and north-oriented steps) connecting $O_i$ to $E_j$,
and with weight $q^{\mathcal{A}}$ if $\mathcal{A}$ is the area to the left of the path $\mathcal{P}$.
Since a path from $O_i$ to $E_j$ is made of a total of $a_i+j$ steps among which exactly $j$ are oriented north, we have clearly
\begin{equation*}
A_{i,j}(q)= {a_i+j \brack j}_q
\end{equation*}
in terms of the $q$-binomial
\begin{equation}
{a\brack b}_q:= 
\prod_{s=1}^{b} \frac{q^{s+a-b}-1}{q^{s}-1} \quad \hbox{for}\ a\geq b\geq 0
\label{eq:qbinomial}
\end{equation}
and ${a\brack b}_q:=0$ otherwise\footnote{Note that the product expression for the $q$-binomial is in practice valid for all $a\geq 0$ as it gives $0$ for $0\leq a<b$. Note also that 
${a\brack b}_q={a\brack a-b}_q$.}.
As in \cite{DFGUI}, the value of the determinant \eqref{eq:Znq} is easily obtained by performing the LU decomposition of the matrix $A(q)$ with elements $A_{i,j}(q)$ above, i.e.\ upon
writing $A(q)$ as the product of a uni\footnote{By uni-lower triangular, we mean a lower triangular matrix with all its diagonal elements equal to $1$.}-lower triangular square matrix $L(q)$ by an upper triangular square matrix $U(q)$, so that 
$Z_n(q)=\prod_{i=0}^n U_{i,i}(q)$. 

Let us show that we may take for $L(q)$ the inverse of the uni-lower triangular matrix $L^{-1}(q)$ with matrix elements
\begin{equation}
L^{-1}(q)_{i,j}=\left\{
\begin{matrix}\frac{\displaystyle{\prod\limits_{s=0}^{i-1}(q^{a_i}-q^{a_s})}}{\displaystyle{\prod\limits_{s=0\atop s\neq j}^{i}(q^{a_j}-q^{a_s})}} & \hbox{for}\ i\geq j\ ,\\
0 & \hbox{for}\ i<j\ ,
\end{matrix}
\right.
\label{eq:Lmat}
\end{equation}
i.e.\ that $U(q):=L^{-1}(q)\, A(q)$ is upper triangular. 
We may compute directly
\begin{eqnarray*}
U_{i,j}(q)&=&\left(L^{-1}\left(q\right)A\left(q\right)\right)_{i,j}=\sum_{k=0}^i \frac{\prod\limits_{s=0}^{i-1} (q^{a_i}-q^{a_s})}{\prod\limits_{s=0\atop s\neq k}^i(q^{a_k}-q^{a_s})}\,
{a_k+j \brack j}_q \\
&=&\prod_{s=0}^{i-1} (q^{a_i}-q^{a_s}) \oint_{{\mathcal C}(q^{a_0},q^{a_1},\cdots ,q^{a_i})} \frac{dt}{2{\rm i}\pi } \,\frac{1}{\prod\limits_{s=0}^i (t-q^{a_s})} \prod_{s=1}^j \frac{t\, q^{s}-1}{q^{s}-1}\ ,
\end{eqnarray*}
where the contour ${\mathcal C}(q^{a_0},q^{a_1},\cdots ,q^{a_i})$ encircles all the finite poles $q^{a_0}, q^{a_1}, \cdots, q^{a_i}$ of the integrand. 
The contour integral is then easily obtained as minus the residue of its integrand at $t=\infty$, 
which clearly vanishes if $j<i$ since the integrand is an $O(t^{j-i-1})$ at large $t$: this shows that $U(q)$ is upper triangular as announced. Moreover, picking the residue at $t=\infty$ when
$j=i$, we also have
\begin{equation*} 
U_{i,i}(q)= \prod_{s=0}^{i-1} (q^{a_i}-q^{a_s}) \prod_{s=1}^i \frac{q^{s}}{q^{s}-1}\,=\,  q^{i^2} \prod_{s=0}^{i-1}\frac{q^{a_i}-q^{a_s}}{q^i-q^s} 
\end{equation*}
and the partition function finally reads
\begin{equation}
Z_n(q;(a_i)_{0\leq i\leq n})=\prod_{i=0}^n U_{i,i}(q)=q^{\frac{1}{6}n(n+1)(2n+1)}\, \frac{\Delta(q^{a_0} ,q^{a_1} ,q^{a_2} ,\cdots,q^{a_n})}{\Delta(1 ,q ,q^2 ,\cdots,q^n)}\ ,
\label{eq:partf}
\end{equation}
where $\Delta(x_0 ,x_1 ,x_2 ,\cdots,x_n)=\prod_{i<j} (x_j-x_i)$ is the Vandermonde determinant.

\subsection{Alternative path formulation}
\label{sec:alternativeform}
\begin{figure}
\begin{center}
\includegraphics[width=13cm]{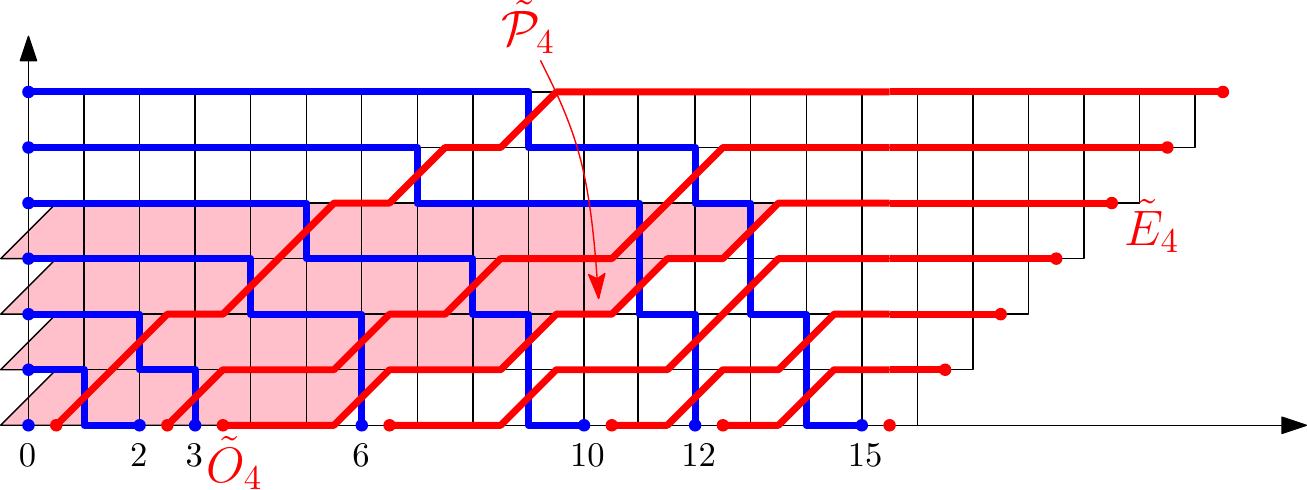}
\end{center}
\caption{\small The second set of paths (in red) associated to the original configuration (in blue) of figure \ref{fig:paths}. For illustration, we colored the domain ``to the left of the path" $\tilde{\mathcal{P}}_4$, with
area $\tilde{\mathcal{A}}_4=39$. The weight of the new configuration is $q^{\sum_{i=0}^6\tilde{\mathcal{A}}_i}=q^{0+14 + 26 + 34 + 39 + 40 + 28}=q^{181}$, equal to that of the original configuration.}
\label{fig:secondpaths}
\end{figure}
As explained in \cite{DFGUI}, the NILP configurations of our model may be bijectively transformed into particular
tiling configurations which in turn may be reformulated into \emph{alternative path configurations}. Here we shall concentrate 
on one particular alternative path description of our model, referred to as the ``second set" of paths in \cite{DFGUI}. Its
configurations consist again of $(n+1)$ NILP $\tilde{\mathcal{P}}_i$,  $i=0,1,\cdots,n$, now made of northeast- and east-oriented unit steps,
 with respective starting points $\tilde{O}_i$ of coordinates $(a_{n-i}+1/2,0)$ along the $x$-axis and endpoints
$\tilde{E}_i$ of coordinates $(a_n+1/2+i,i)$ along the line $y=x-a_n-1/2$ (see figure \ref{fig:secondpaths}). 
The bijection between the original NILP configurations and these second set of non-intersecting paths may be obtained directly as follows:
given the original NILP configuration, the $i$-th path $\tilde{\mathcal{P}}_i$ in the 
associated second set of paths is obtained, starting from $\tilde{O}_i$, by performing east-oriented unit steps as long as these steps do not intersect a path of the first original set and by overpassing any encountered such path via a northeast-oriented step crossing a north-oriented step of the original path (see figure \ref{fig:secondpaths}).  The procedure is continued until 
the final point $\tilde{E}_i$ is reached (after $i$ crossings, so that $\tilde{E}_i$ has the desired $y$-coordinate $i$). Note that, as opposed to the original path numbering from left to right, the paths in the second set 
are now numbered from right to left. It is clear that the mapping from $\{\mathcal{P}_i\}_{0\leq i\leq n}$ to $\{\tilde{\mathcal{P}}_i\}_{0\leq i\leq n}$ is a bijection since, from the data
of any $\{\tilde{\mathcal{P}}_i\}_{0\leq i\leq n}$ in the second set of paths, we may easily reconstruct its unique pre-image $\{\mathcal{P}_i\}_{0\leq i\leq n}$ by a similar construction.

\begin{figure}
\begin{center}
\includegraphics[width=13cm]{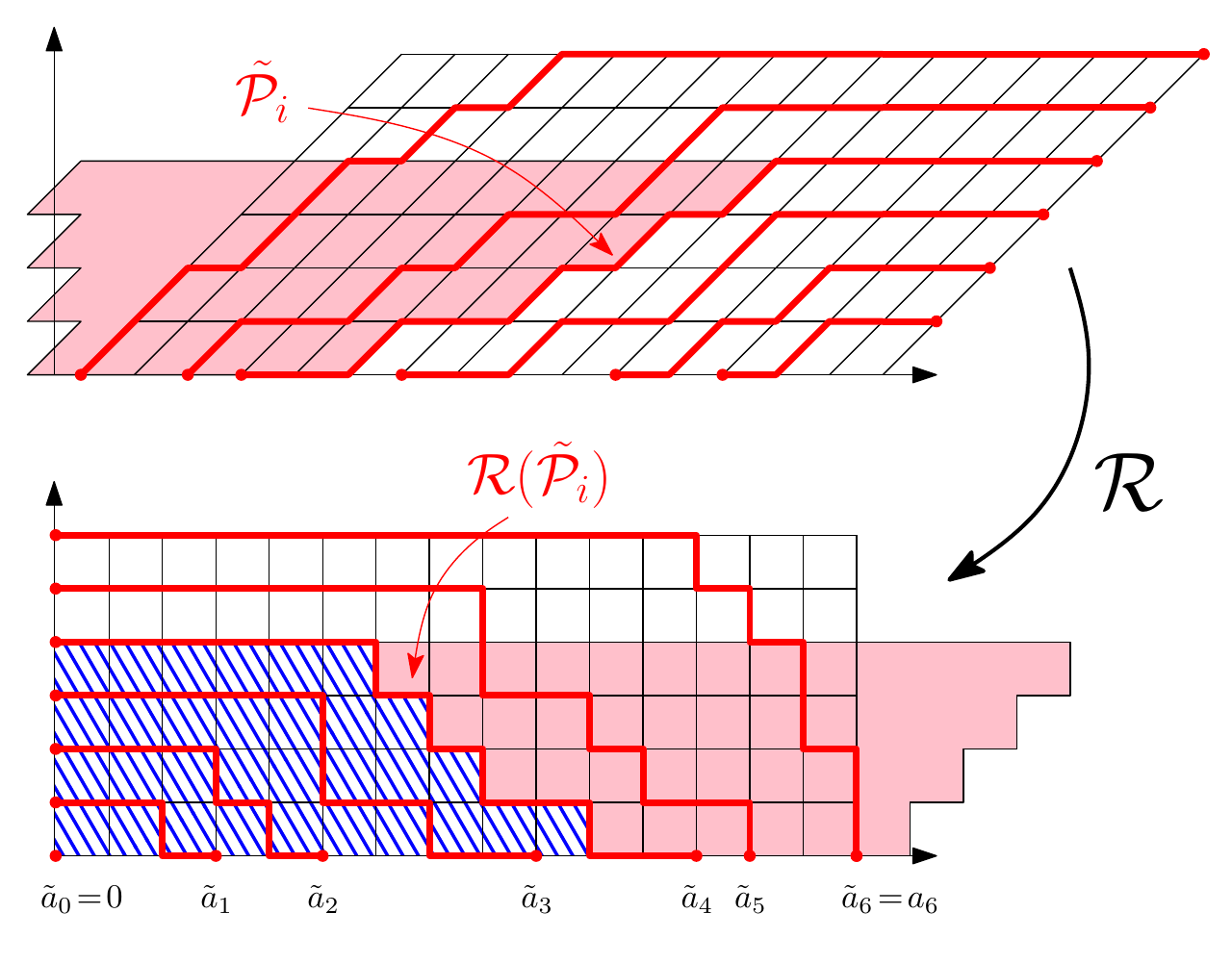}
\end{center}
\caption{\small The mapping $\mathcal{R}$ from the configuration of figure \ref{fig:secondpaths} to a NILP configuration made of north- and west-oriented steps, now
associated to the sequence $(\tilde{a}_i)_{0\leq i\leq 6}=(0,3,5,9,12,13,15)$. The area to the left of the transformed
path $\mathcal{R}(\tilde{\mathcal{P}}_i)$ (shaded domain) is given by $\left(i\, \left(a_6+1\right)+\sum_{y=0}^{i-1}y\right)-\tilde{\mathcal{A}}_i$ where $\tilde{\mathcal{A}}_i$ is the area to the left
of $\tilde{\mathcal{P}}_i$ before mapping (colored domain).}
\label{fig:reflection}
\end{figure}

Let us now discuss how to transfer the weight of the original NILP configuration to its image by the above bijection: this weight is clearly recovered in the second setting by assigning to each northeast-oriented step $(x-1/2,y)\to (x+1/2,y+1)$ a weight $q^x$ as any such step
is ``dual" to a north-oriented step $(x,y)\to (x,y+1)$ in the original configuration.
By performing a simple shear of the original unit squares into elementary rhombi of the same unit area,
this in turn corresponds to assigning a weight $q^{\tilde{\mathcal{A}}_i}$ to each path $\tilde{\mathcal{P}}_i$ of the new configuration, 
where $\tilde{\mathcal{A}}_i$ denotes again the area to the left of $\tilde{\mathcal{P}}_i$, now defined as the total area (number of rhombi) of the domain 
delimited by the path $\tilde{\mathcal{P}}_i$ and its projection along the ``vertical sawtooth line" surrounding the $y$-axis (see figures \ref{fig:secondpaths} and \ref{fig:reflection}). Again the total weight of a NILP configuration is the product of its path weights, namely $q^{\sum_{i=0}^n\tilde{\mathcal{A}}_i}$. 
With this weight, the partition function of the second path configurations is, by construction, identical to that, $Z_n(q;(a_i)_{0\leq i\leq n})$,
of the first path configurations, namely given by \eqref{eq:partf}.

\medskip
As in \cite{DFGUI}, we may, on the other hand, transform a path configuration in the second NILP set back into a configuration made of north- and west-oriented steps in 
a much more straightforward way: this direct transformation is achieved by first performing a shear transformation $(x,y)\mapsto (x-y,y)$ (transforming northeast-oriented steps into north-oriented steps) followed 
by a reflection $(x,y)\to (a_n+1/2-x,y)$ (transforming east-oriented steps into west-oriented steps), as displayed in figure \ref{fig:reflection}. The resulting mapping
\begin{equation*}
\mathcal{R}:\ (x,y)\mapsto (a_n+1/2+y-x,y)
\end{equation*}
sends the endpoints $\tilde{E}_i$ to $\mathcal{R}(\tilde{E}_i)=(0,i)$ and the starting points $\tilde{O}_i$ to $\mathcal{R}(\tilde{O}_i)=(\tilde{a}_i,0)$ involving the strictly increasing sequence of integers (with $\tilde{a}_0=0$)
\begin{equation}
\tilde{a}_i:=a_n-a_{n-i}\ .
\label{eq:atilda}
\end{equation} 
We thus recover path configurations as those of the original setting but \emph{with a new
set of starting points} now characterized by the sequence $(\tilde{a}_i)_{0\le i\le n}$.
 
As for the weight $q^x$ assigned to any northeast-oriented step $(x-1/2,y)\to (x+1/2,y+1)$ of, say, the path $\tilde{\mathcal{P}}_i$, it is attached after the mapping $\mathcal{R}$, 
to a north-oriented step $(a_n+1+y-x,y)\to (a_n+1+y-x,y+1)$ of the path $\mathcal{R}(\tilde{\mathcal{P}}_i)$, In other words\footnote{Note that $\mathcal{R}$ is an involution, hence setting 
$(a_n+1+y-x,y)=(\tilde{x},\tilde{y})$
amounts to setting $(x,y)=(a_n+1+\tilde{y}-\tilde{x},\tilde{y})$.}, any north-oriented step $(\tilde{x},\tilde{y})\to (\tilde{x},
\tilde{y}+1)$ of
the path $\mathcal{R}(\tilde{\mathcal{P}}_i)$ receives a weight $q^{a_n+1+\tilde{y}-\tilde{x}}$. Since the path $\mathcal{R}(\tilde{\mathcal{P}}_i)$ has exactly $i$ north-oriented steps $(\tilde{x},\tilde{y})\to (\tilde{x},
\tilde{y}+1)$ whose ordinates 
$\tilde{y}$ take the respective integer values $j=0,1,\cdots,i-1$, the above weight is recovered by assigning a weight $q^{-\tilde{x}}$ to each north-oriented steps $(\tilde{x},\tilde{y})\to (\tilde{x},
\tilde{y}+1)$  together with a global weight 
\begin{equation*}
\sum_{i=0}^{n}\left(i\, \left(a_n+1\right)+\sum_{j=0}^{i-1}j \right)=\sum_{i=0}^{n}i\, \frac{2a_n+i+1}{2}= \frac{1}{6}n(n+1)(3a_n+n+2)\ .
\end{equation*}
We deduce the identity 
\begin{equation}
Z_n(q;(a_i)_{0\leq i\leq n})=q^{\frac{1}{6}n(n+1)(3a_n+n+2)}\, Z_n(q^{-1};(\tilde{a}_i)_{0\leq i\leq n})
\label{eq:Zident}
\end{equation}
relating the partition functions of NILP configurations in the \emph{same original setting} made of north- and west-oriented steps but associated with \emph{different sequences} 
$(a_i)_{0\leq i\leq n}$ and $(\tilde{a}_i)_{0\leq i\leq n}$ respectively.
This identity may also be verified by a direct calculation from the explicit expression \eqref{eq:partf} and the relation \eqref{eq:atilda} between $a_i$ and $\tilde{a}_i$. 

The above (back and forth) bijective mappings between NILP configurations of the two different settings may appear here as a pure academic exercise but they will prove 
very useful in Section \ref{sec:leftpart} when using the second set of paths to compute the so-called ``left part" of the arctic curve.

\section{One-point function and free trajectory partition function}
\label{sec:onepoint}

\subsection{Exact expressions}
\label{sec:onepointexact}
\begin{figure}
\begin{center}
\includegraphics[width=10cm]{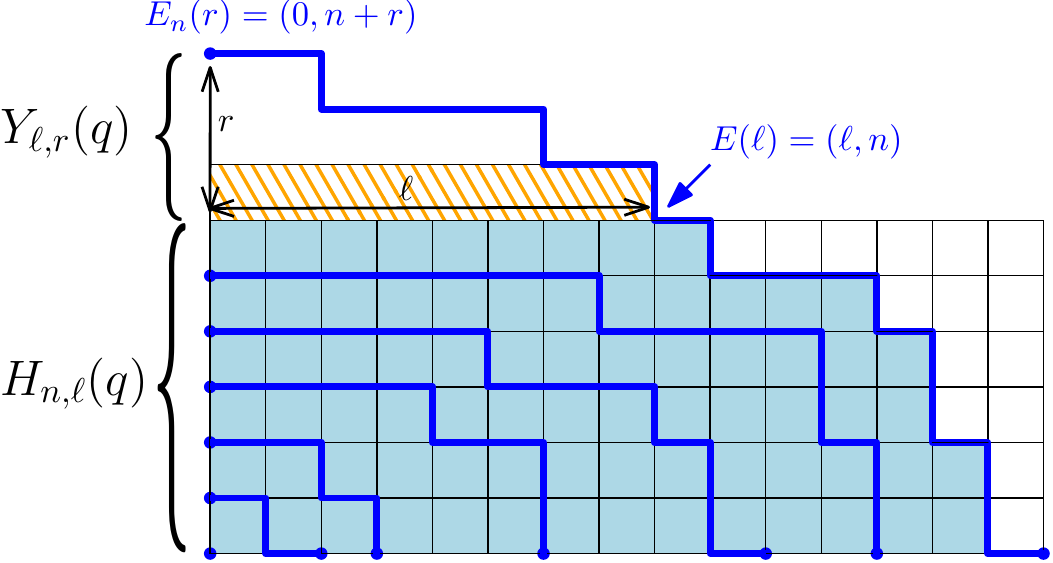}
\end{center}
\caption{\small A modified NILP configuration where the $n$-th path ends at position $E_n(r)=(0,n+r)$. This forces this path
to exit the domain $y\leq n$ by a north-oriented step at $E(\ell)=(\ell,n)$ for some $\ell$ between $0$ and $a_n$. The (normalized) partition function for the part of the configuration below
the $y=n$ line is given by $H_{n,\ell}(q)$, including a weight $q^{\mathcal{A}_n}$ corresponding to the area $\mathcal{A}_n$ to the left of the portion of the $n$-th path below this line (colored domain).
The partition function for the part of the configuration above
the $y=n$ line is given by $Y_{\ell,r}(q)$, including a weight $q^{\ell}$ for the first (shaded) strip.
}
\label{fig:tangent}
\end{figure}
The tangent method consists in slightly modifying the NILP configurations by moving the endpoint $E_n$ of the $n$-th path $r$ steps north to the position $E_n(r)=(0,n+r)$. This forces this path
to exit the domain $y\leq n$ (hence the domain $D$) by a north-oriented step at some $x$-coordinate $\ell$ between $0$ and $a_n$. Let us denote by $E(\ell)=(\ell,n)$ this ``exit point"
(see figure \ref{fig:tangent}).
As in \cite{DFGUI}, the so-called \emph{one-point function} $H_{n,\ell}(q):=H_{n,\ell}(q;(a_i)_{0\leq i\leq n})$ corresponds precisely to the partition function
for configurations where we let the $n$-th path $\mathcal{P}_n$ \emph{stop at a fixed exit point $E(\ell)$}, normalized by the original partition function $Z_n(q)$ (so that $H_{n,0}(q)=1$ since $E(0)=E_n$).
Here, the weight of the truncated path $\mathcal{P}_n$ is chosen to be $q^{\mathcal{A}_n}$, where $\mathcal{A}_n$ denotes the number of unit squares in the region delimited by this truncated path 
$\mathcal{P}_n$ and its projection along the $y$-axis. This corresponds to our notion of area ``to the left of the path", but note that 
it is no longer identical to the area ``under the (truncated) path" whenever $\ell>0$ (the difference between the two areas being ${n\ell}$).

Denoting by $Z_n(q,\ell):=Z_n(q,\ell;(a_i)_{0\leq i\leq n})$ the partition function of these configurations with exit point $E(\ell)$, the one-point function $H_{n,\ell}$
is simply obtained as the ratio
\begin{equation*}
H_{n,\ell}(q) =\frac{Z_n(q,\ell)}{Z_n(q)}=\frac{\det\left(A(q,\ell)\right)}{\det\left(A(q)\right)}\ ,
\end{equation*}
where the new LGV matrix $A(q,\ell)$ differs from $A(q)$ only in its last column:
\begin{equation*}
A_{i,j}(q,\ell)=\left\{ \begin{matrix} A_{i,j}(q) & {\rm for}\  j<n\ , \\
\\
\displaystyle{q^{n\ell}\, {a_i+n-\ell\brack n}_q }& {\rm for}\ j=n \ .\end{matrix}\right. 
\end{equation*}
Since $A(q,\ell)$ and $A(q)$ differ only in their last column, the matrix $U(q,\ell):= L^{-1}(q) A(q,\ell)$ differs also from $U(q)= L^{-1}(q) A(q)$
in its last column only, hence it is upper triangular, leading immediately to $H_{n,\ell}(q)=U_{n,n}(q,\ell)/U_{n,n}(q)$,
where  
\begin{eqnarray*}
U_{n,n}(q,\ell)&=&\sum_{k=0}^n L^{-1}_{n,k}(q) A_{k,n}(q,\ell)=q^{n\ell}\sum_{k=0}^n
\frac{\prod\limits_{s=0}^{n-1} (q^{a_n}-q^{a_s})}{\displaystyle\prod\limits_{s=0\atop s\neq k}^{n} (q^{a_k}-q^{a_s})}\, 
{a_k+n-\ell\brack n}_q\ \\
&=& q^{n\ell}\prod_{s=0}^{n-1} (q^{a_n}-q^{a_s})\, \oint_{{\mathcal C}(q^{a_k} | a_k\geq \ell)}\frac{dt}{2{\rm i}\pi} \, \prod_{s=0}^n \frac{1}{t-q^{a_s}}
\prod_{s=1}^n \frac{t\, q^{s-\ell}-1}{q^s-1} \ .
\end{eqnarray*}
Here the contour ${\mathcal C}(q^{a_k} | a_k\geq \ell)$ encircles the finite poles $q^{a_k}$ of the integrand \emph{only for values of $k$ such that $a_k\geq \ell$}.
Other values of $k$ (with $a_k<\ell$) are indeed absent de facto from the sum in the first line due to the vanishing of the $q$-binomial ${a_k+n-\ell\brack n}_q$ whenever $a_k<\ell$.

This yields the desired expression
\begin{equation}
H_{n,\ell}(q)=\frac{U_{n,n}(q,\ell)}{U_{n,n}(q)}=q^{n\ell-n(n+1)/2}\oint_{{\mathcal C}(q^{a_k} | a_k\geq \ell)}\frac{dt}{2{\rm i}\pi} \, \prod_{s=0}^n \frac{1}{t-q^{a_s}}
\prod_{s=1}^n (t\, q^{s-\ell}-1) \ .
\label{eq:Hnl}
\end{equation}
Note finally that the last product in the integrand vanishes for $t=q^a$ when $a=\ell-n,\cdots,\ell-1$ so that the contour ${\mathcal C}(q^{a_k} | a_k\geq \ell)$ may be extended
to ${\mathcal C}(q^{a_k} | a_k\geq \ell-n)$ by also encircling poles $q^{a_k}$ with $\ell-n\leq a_k<\ell$ since these poles contribute $0$ to the integral.

\medskip
To obtain the full partition function for NILP configurations where the $n$-th path ends at the shifted position $E_n(r)=(0,n+r)$, we also need the partition function $Y_{\ell,r}(q)$ of
 the remaining part of the $n$-th path,
leading from $E(\ell)=(\ell,n)$ to $E_n(r)$, hereafter referred to as the ``free trajectory" of the $n$-th path as it is not affected by the other paths.
It is simply given by
\begin{equation*}
 Y_{\ell,r}(q)= q^{\ell}\, {\ell+r-1\brack \ell}_q
 \end{equation*}
since the first step must be north-oriented (with weight $q^\ell$) and the $q$-binomial precisely incorporates the desired weight $q^{\mathcal{A}}$
for the area $\mathcal{A}$ to the left of the new portion of path lying above the $y=n+1$ line (see figure \ref{fig:tangent}).

The modified (normalized) partition function for configurations with a fixed shifted endpoint $E_n(r)$ for the $n$-th path is simply obtained by summing
over all possible intermediate positions $E(\ell)$ of the exit point, namely given by
\begin{equation}
\sum_{\ell=0}^{a_n} H_{n,\ell}(q)\, Y_{\ell,r}(q)\ .
\label{eq:modpartf}
\end{equation}

\subsection{Scaling limit}
\label{sec:saddle}
The tangent method uses the most likely value $\ell$ for the exit point $E(\ell)$, i.e.\ that which maximizes the modified partition function \eqref{eq:modpartf} 
for fixed $r$. The relation between the optimal $\ell$ and $r$ is easily obtained in the limit of large $n$ by analyzing the asymptotics 
of the various functions at hand under the appropriate scaling, namely
\begin{equation*}
\ell=\xi\, n,\ r=z\, n, \ a_i=n\, \al(i/n)
\end{equation*}
with $\xi$ and $z$ remaining finite, and where $\al(u)$ is an \emph{increasing piecewise differentiable function} for $u\in[0,1]$ such that its derivative, when defined, satisfies $\al'(u)\geq 1$
since the sequence $(a_i)_{0\leq i\leq n}$ is strictly increasing.
To get a non-trivial large $n$ limit, it is also necessary to adjust the weight $q$ by setting:
\begin{equation*}
q=\qq^{1/n}
\end{equation*}
with a finite $\qq$.

From the product expression \eqref{eq:qbinomial} for the $q$-binomial, we immediately deduce the asymptotic equivalent:
\begin{eqnarray*}
&Y_{\ell,r}(q)&\sim \ e^{n S_1(\xi,z)}\ ,\\  & S_1(\xi,z)&= \int_0^\xi d\uu \, {\rm Log}\left( \frac{\qq^{\uu+z}-1}{\qq^\uu -1}\right) 
\end{eqnarray*} 
while, from the expression \eqref{eq:Hnl}, we deduce
\begin{equation}
\begin{split}
&H_{n,\ell}(q) \sim \oint \frac{dt}{2{\rm i}\pi}\, e^{n S_0(t,\xi)}\ ,\\
&S_0(t,\xi)=\left(\xi-\frac{1}{2}\right){\rm Log}(\qq)+ \int_{0}^1 d\uu \, {\rm Log}\left(\frac{t\, \qq^{\uu-\xi}-1}{t-\qq^{\al(\uu)}}\right)\ .
\end{split}
\label{eq:Hnlasymp}
\end{equation}
Here the contour must encircle only those $\qq^{\al(u)}$ such that $\al(u)\geq \xi$. For $\qq>1$ (i.e.\ $q>1$), it must therefore surround the segment 
$[\qq^\xi,\qq^{\al(1)}]$, hence cross the real axis anywhere in the 
interval $]\qq^{\xi-1},\qq^{\xi}[$ (recall indeed that the poles $q^{a_k}$ for $\ell-n \leq a_k <\ell$ do not contribute to the integral) and in the interval $]\qq^{\al(1)},+\infty[$
(there are no poles larger than $q^{a_n}$).
Similarly, for $\qq<1$ (i.e.\ $q<1$), it must  surround the segment 
$[\qq^{\al(1)},\qq^\xi]$, hence cross the real axis in the interval $]-\infty,\qq^{\al(1)}[$ and in the 
interval $]\qq^{\xi},\qq^{\xi-1}[$. At large $n$, the integral is estimated by a saddle-point method, namely
\begin{eqnarray*}
&H_{n,\ell}(q)& \!\!\! \sim\ e^{n S_0(t^*,\xi)}\ ,\\
&\frac{\displaystyle{\partial S_0(t,\xi)}}{\displaystyle{\partial t}}\Big\vert_{t=t^*}& \!\!\! =0\ .
\end{eqnarray*}
The optimal value of $\xi$ for fixed $z$ is then obtained by extremizing $S_0(t^*,\xi)+S_1(\xi,z)$ with respect to $\xi$ at fixed $z$. The two (saddle-point and extremization)
operations may be performed \emph{simultaneously} by solving the two extremization conditions:
\begin{eqnarray*}\frac{\partial S_0(t,\xi)}{\partial t}&=&0=\int_0^1 d\uu \left\{ \frac{\qq^{\uu-\xi}}{t\, \qq^{\uu-\xi}-1}-\frac{1}{t-\qq^{\al(\uu)}}\right\}\\
&&\ \ \  =\frac{1}{t\, {\rm Log}(\qq)}\ {\rm Log}\left(\frac{t\, \qq-\qq^{\xi}}{t-\qq^\xi }\right)- \int_0^1 d\uu \frac{1}{t-\qq^{\al(\uu)}}\ ,\\
\frac{\partial (S_0(t,\xi)+S_1(\xi,z))}{\partial \xi}&=&0= {\rm Log}\left(\qq\,  \frac{\qq^{\xi+z}-1}{\qq^\xi -1}\right)-t\, {\rm Log}(\qq)\, \int_0^1 d\uu \frac{\qq^{\uu-\xi}}{t\, \qq^{\uu-\xi}-1}\ .
\end{eqnarray*}
Using the definition \eqref{eq:defx} for the $\qq$-defomed moment-generating function of the distribution $\al$, namely
\begin{equation*}
x(t)=\qq^{\textstyle{-t \, \int_0^1 d\uu \frac{1}{t-\qq^{\al(\uu)}}}}\ ,
\end{equation*}
the above equations reduce to
\begin{equation*}
\frac{t\, \qq-\qq^\xi}{t-\qq^\xi}\, x(t)=1\ ,\qquad  \qq\, \frac{\qq^{\xi+z}-1}{\qq^{\xi}-1}\, x(t)=1\ ,
\end{equation*}
which yield the parametric solution $(\xi(t), z(t))$ for the optimal $\xi$ at fixed $z$:
\begin{equation}\label{sol} \qq^{\xi(t)}=t\, \frac{\qq\, x(t)-1}{x(t)-1}\ ,\qquad \qq^{z(t)}=\frac{t+x(t)-1}{t\, \qq\,  x(t)} \ .
\end{equation}
Since $\qq^{\xi(t)}$ and $\qq^{z(t)}$ must be real, $t$ must be real and therefore lie in the specific intervals mentioned above when discussing the intersection of the $t$-contour with the real axis. 
It is easily checked that $(\qq^{\xi(t)}-t)/(t-\qq^{\xi(t)-1})=-\qq\, x(t) <0 $ (since $x(t)>0$), hence $t$ cannot lie in the interval $]\qq^{\xi-1},\qq^{\xi}[$ for $\qq>1$ (respectively
$]\qq^{\xi},\qq^{\xi-1}[$ for $\qq<1$). The solution above is thus \emph{valid only for a parameter $t$ in the range $]\qq^{\al(1)},+\infty[$ if $\qq>1$ and for a parameter 
$t$ in the range $]-\infty,\qq^{\al(1)}[$ whenever $\qq<1$}.

\section{Arctic curve: first portion}
\label{sec:arctic}
\subsection{Geodesic equation for the free trajectory}
\label{sec:freetrajectory}
So far we obtained in \eqref{sol} the most likely exit point $E(\ell=n\, \xi)$ for a fixed shifted endpoint $E_n(r=n\, z)$ in the scaling limit. The tangent method relies on the assumption that 
the ``geodesic path" connecting $E(\ell)$ to $E_n(r)$, i.e.\ the most likely free trajectory passing through these two points,
is \emph{tangent to the arctic curve} at their meeting point. In other words, the $n$-th path (travelled backwards from $E_n(r)$) continues to follow a geodesic trajectory below the $y=n$ line until it
meets the other paths of the NILP configuration tangentially along the arctic curve. Here it is important to note that, as opposed to the case $q=1$ considered in \cite{DFGUI}, the geodesic path
is no longer a straight line but follows a certain curve depending on $n$, $\ell$, $r$ and \emph{on the parameter $q$}.

To compute the equation of this most likely free trajectory, let us consider the intersection point between the path from $(\ell,n+1)$ (recall that the first step after $E(\ell)$ is a north-oriented step)
 to $E_n(r)=(0,n+r)$ and, say, a vertical line $x=m$ for $m$ between $0$ and $\ell$. If $(m,n+p)$ denotes this intersection point (with $p$ between $1$ and $r$), the free trajectory
 partition function
reads
\begin{equation*}
Y_{\ell,r}(q)= \sum_{p=1}^r q^{m(p-1)}\, {r-p+m\brack r-p}_q\, {\ell-m+p-1\brack p-1}_q \ .
\end{equation*}

At large $n$, we use again scaling variables $\ell=\xi \, n$, $m=\mu\, n$, $r=z\, n$, $p=\phi\, n$ and $q=\qq^{\frac{1}{n}}$ to write
\begin{eqnarray*}
Y_{\ell,r}(q) &\sim& \int_0^z d\phi\,  e^{n\, S(\phi,\mu;\xi,z)}\ ,\\
 S(\phi,\mu;\xi,z)&=&\mu\,\phi\, {\rm Log}(\qq)+ \int_{0}^{z-\phi}d\uu\, {\rm Log}\left(\frac{\qq^{\uu+\mu}-1}{\qq^\uu-1} \right)
+\int_{0}^{\phi}d\uu\, {\rm Log}\left(\frac{\qq^{\uu+\xi-\mu}-1}{\qq^\uu-1} \right)\ .
\end{eqnarray*}
For fixed $\xi$ and $z$, the most likely free trajectory $\phi=\phi(\mu)$ is obtained as the saddle-point of the integrand via
\begin{equation*}
\frac{\partial S(\phi,\mu;\xi,z)}{\partial \phi}={\rm Log}\left(\qq^\mu\, \frac{\qq^{z-\phi}-1}{\qq^{z-\phi+\mu}-1}\,
\frac{\qq^{\phi+\xi-\mu}-1}{\qq^\phi-1} \right) =0\ ,
\end{equation*}
namely
\begin{equation*}
(1-\qq^\xi)\qq^\phi +(1-\qq^z)\qq^\mu =1-\qq^{z+\xi}\ .
\end{equation*}
Using rescaled cartesian coordinates $X=x/n$, $Y=y/n$, this gives, for fixed $\xi$ and $z$, the most likely free (rescaled) trajectory $(X,Y)=(\mu,1+\phi)$ 
by letting $\mu$ vary between $0$ and $\xi$ (or equivalently  letting $\phi$ vary between $0$ and $z$). 
The above trajectory is equivalently rewritten as
\begin{equation}
\frac{1-\qq^{X}}{1-\qq^\xi}+\frac{1-\qq^{Y-1}}{1-\qq^z}=1
\label{eq:tgcurve}
\end{equation}
with $0\leq X \leq \xi$ (or equivalently $1\leq Y\leq 1+z$). The above expression for the geodesic path emphasizes the fact that the rescaled endpoints $(X,Y)=(0,1+z)$ 
(corresponding to $E_n(r)$) and $(X,Y)=(\xi,1)$ (corresponding to $E(\ell)$) lie on the curve, as wanted. The geodesic trajectory is straightforwardly extended to 
values of $X>\xi$ ($Y<1$) and describes the most likely rescaled position of the $n$-th path until it reaches the other paths. 

\subsection{Tangent method and arctic curve}
\label{sec:tangentmethod}
We are now ready to apply the tangent method principles: the arctic curve is obtained as the envelope of the above geodesic trajectories \eqref{eq:tgcurve} for varying endpoints
(characterized by $z$ in the scaling limit) and their associated most likely exit point (characterized by $\xi$), i.e.\ for varying values of $\xi$ and $z$ 
\emph{related via the parametric equation \eqref{sol}}. 
Letting $t$ vary in \eqref{sol} yields a family of ``tangent curves" with equation
\begin{equation*}
\frac{1-\qq^{X}}{1-\qq^{\xi(t)}}+\frac{1-\qq^{Y-1}}{1-\qq^{z(t)}}=1
\end{equation*}
parametrized by $t$. Substituting the solution \eqref{sol} for $\xi(t)$ and $z(t)$,
we end up with the particularly simple equation for the tangent curves:
\begin{equation}
x(t)\, \qq^Y +\frac{1-x(t)}{t}\, 
\qq^X -1 =0
\label{eq:family}
\end{equation}
with $x(t)$ as in \eqref{eq:defx}.
The envelope of these curves is the solution of the linear (in $\qq^X$ and $\qq^Y$) system:
\begin{eqnarray*}
t\, x(t)\, \qq^Y +(1-x(t))\, \qq^X -t  &=&0\ , \\
(t\, x'(t)+x(t))\, 
\qq^Y -x'(t)\, \qq^X -1&=&0 \ ,  
 \end{eqnarray*}
leading to the following  explicit parametric equation for the arctic curve $(X(t),Y(t))$ in terms of the quantity $x(t)$ defined in 
\eqref{eq:defx}:
\begin{equation}
 \qq^{X(t)}=  \frac{t^2 \, x'(t)}{t\, x'(t)+x(t)(1-x(t))}\ , \quad 
\qq^{Y(t)}=\frac{t\, x'(t)+1-x(t)}{t\, x'(t)+x(t)(1-x(t))}\ ,
\label{eq:arctic}
\end{equation}
with, as already discussed, $t\in ]\qq^{\al(1)},+\infty[$ whenever $\qq>1$ and $t\in ]-\infty,\qq^{\al(1)}[$ whenever $\qq<1$.
This proves a first instance of Theorem \ref{mainthm}, for the indicated ranges of $t$.

\begin{figure}
\begin{center}
\includegraphics[width=10cm]{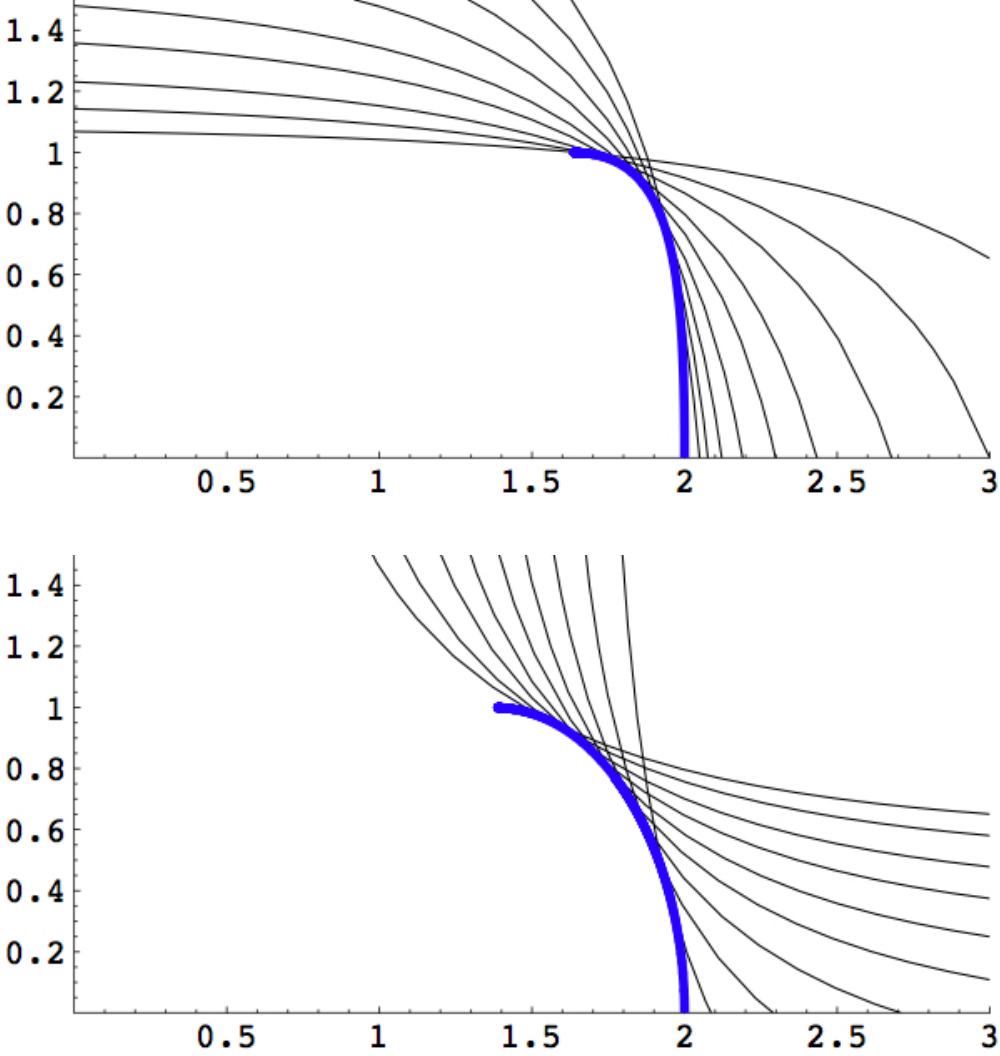}
\end{center}
\caption{\small The ``right part" of the arctic (thick solid blue line) as given by \eqref{eq:arctic} for the appropriate 
domain of $t$ (see text) for $\qq=3$ (top) and $\qq=1/3$ (bottom) in the particular case $\al(u)=2 u$. The extremities
of this portion of curve are at $({\rm Log}\left(\qq(\qq+1)/2\right)/{\rm Log}(\qq),1)$ and $(2,0)$. We also indicated members of the family
of tangent curves (thin lines) whose envelope defines the portion of arctic curve at hand.}
\label{fig:tangentcurves}
\end{figure}

\medskip
For illustration, let us discuss the simple case where the sequence of starting points is taken as $a_i=2\ i$,
$i=1,\cdots,n$. This results in a linear function $\al(u)=2 u$ and the function $x(t)$ is easily computed 
from its general expression \eqref{eq:defx} as
\begin{equation*}
x(t)=\frac{1}{\qq}\, \sqrt{\frac{t-\qq^2}{t-1}}\ .
\end{equation*}  
The corresponding arctic curve \eqref{eq:arctic} is displayed in figure \ref{fig:tangentcurves} together with the
associated family of tangent curves (as given by \eqref{eq:family}) for $\qq=3$ and $\qq=1/3$ respectively. 
Note that these tangent curves are concave for $\qq=3$ and convex for $\qq=1/3$, which is consistent with a tendency 
for a free trajectory with fixed endpoints to increase the area to its left when $\qq>1$ and, on the contrary, to decrease it whenever $\qq<1$.
Note also that the parameter $t$ (in both \eqref{eq:arctic} and  \eqref{eq:family}) varies in 
$]\qq^2,+\infty[=]9,+\infty[$ for $\qq=3$  and $]-\infty,\qq^2[=]-\infty,1/9[$ for $\qq=1/3$.
As apparent in figure \ref{fig:tangentcurves}, restricting $t$ to the above ranges builds only one portion
of the arctic curve, its so-called ``right part". This is due to the particular geometry that we used to apply the tangent method, namely
by shifting north the
endpoint $E_n$ of the outermost path in the original NILP formulation of the model. 
As explained in \cite{DFGUI}, other geometries may be used and lead to other portions of the arctic curve.
Let us now discuss how to obtain these other portions in practice. 

\section{Other portions of the arctic curve}
\label{sec:otherportion}
\subsection{Left part of the arctic curve}
\label{sec:leftpart}
\begin{figure}
\begin{center}
\includegraphics[width=16cm]{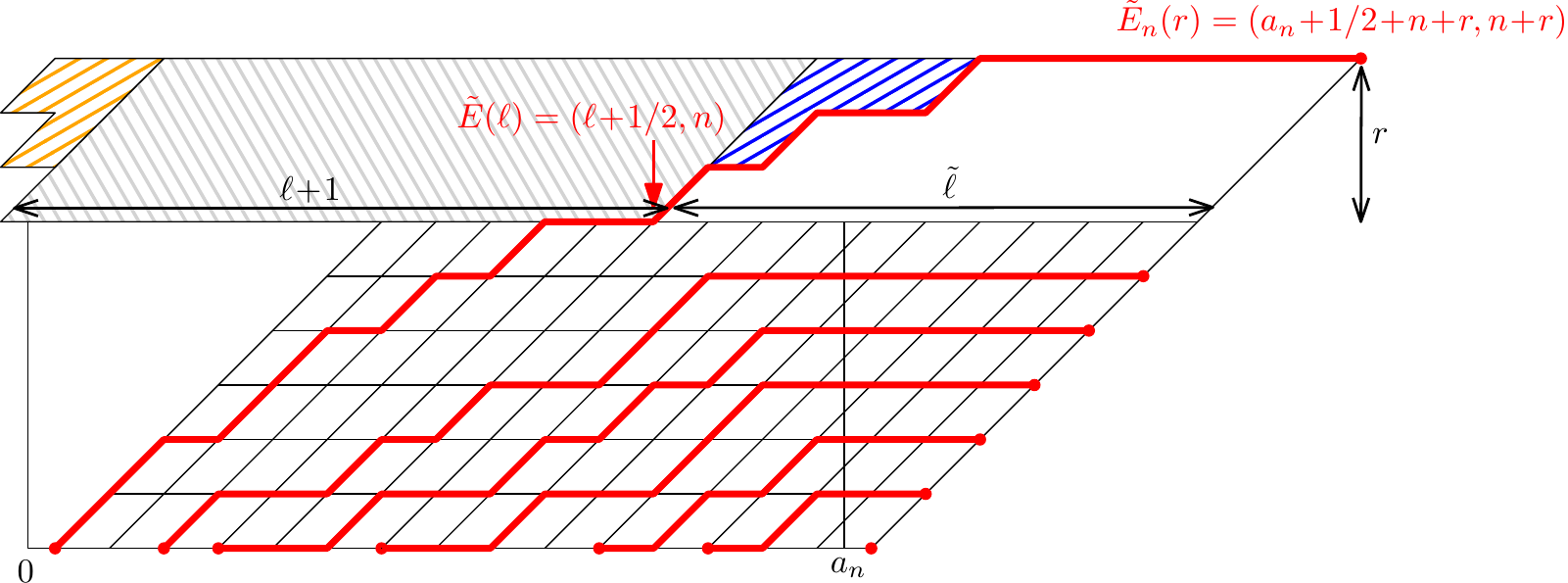}
\end{center}
\caption{\small A modified NILP configuration where the endpoint of the $n$-th path is moved to position $\tilde{E}_n(r)=(a_n+1/2+n+r,n+r)$. 
The partition function $\tilde{H}_{n,\ell}(q)$ for the lower part of the configuration with exit point $\tilde{E}(\ell)$ is obtained via some general symmetry principle (see text).
The partition function $\tilde{Y}_{\ell,r}(q)$ of the upper part involves the area of the shaded domain, divided for convenience into three regions. The leftmost shaded
region is responsible for a weight $q^{r(r-1)/2}$ and the central shaded region for a weight $q^{r(\ell+1)}$. As for the rightmost part, which involves a summation over path
configurations from
$(\ell+3/2,n+1)$ to $(a_n+1/2+n+r,n+r)$ with area equal to the (indicated in blue) rightmost shaded region, it yields, by a simple up-down reflection of the path, 
to a weight ${\tilde{\ell}+r-1\brack r-1}_{q}={\tilde{\ell}+r-1\brack \tilde{\ell}}_{q}$.}
\label{fig:Hnltilde}
\end{figure}
Another portion of the arctic curve, hereafter called its ``left part" for obvious reasons, is obtained by considering the alternative formulation of Section~\ref{sec:alternativeform}
through NILP configurations with northeast- and east-oriented steps. Moving the endpoint $\tilde{E}_n=(a_n+1/2+n,n)$ of the $n$-th path $r$ steps in the northeast direction to the position 
$\tilde{E}_n(r)=(a_n+1/2+n+r,n+r)$ forces this path
to exit the domain $y\leq n$ by a northeast-oriented step at some $x$-coordinate $\ell+1/2$ for some $\ell$ between $n$ and $a_n+n$. Let us denote by $\tilde{E}(\ell)=(\ell+1/2,n)$ 
this exit point (see figure \ref{fig:Hnltilde}).
We denote by $\tilde{H}_{n,\ell}(q):=\tilde{H}_{n,\ell}(q;(a_i)_{0\leq i\leq n})$ the one-point function corresponding, as before, to the (normalized) partition function
for configurations where we let the $n$-th path \emph{stop at a fixed exit point $\tilde{E}(\ell)$} and where the weight of this truncated $n$-th 
path is $q^{\tilde{\mathcal{A}}_n}$ with $\tilde{\mathcal{A}}_n$ the area to the left of the path as before. Note that the normalization condition now implies that
$\tilde{H}_{n,a_n+n}(q)=1$ since $\tilde{E}(a_n+n)=\tilde{E}_n$.

By a straightforward generalization of the argument leading to \eqref{eq:Zident} based on the mapping $\mathcal{R}$, we immediately deduce the relation,
valid for $n\leq \ell\leq a_n+n$:
\begin{equation*}
\tilde{H}_{n,\ell}(q;(a_i)_{0\leq i\leq n})=H_{n,\tilde{\ell}}(q^{-1},(\tilde{a}_i)_{0\leq i\leq n})\ , \quad \tilde{\ell}=a_n+n-\ell
\end{equation*}
with \emph{no $q$-dependent prefactor} since the proportionality factor appearing in \eqref{eq:Zident} eventually drops out in the ratio 
defining the one-point functions (which are normalized partition functions by definition, in particular $\tilde{H}_{n,a_n+n}=H_{n,0}=1$ for any value of the parameter $q$
and of the sequence $(a_i)_{0\leq i\leq n}$). Here $\tilde{\ell}$ is nothing but the $x$-coordinate of $\mathcal{R}(\tilde{E}(\ell))$.
This leads directly from the expression \eqref{eq:Hnl} to
\begin{eqnarray*}
\tilde{H}_{n,\ell}(q)&=&q^{-n\tilde{\ell}+n(n+1)/2}\oint_{{\mathcal C}(q^{-\tilde{a}_k} | \tilde{a}_k\geq \tilde{\ell})}\frac{dt}{2{\rm i}\pi} \, \prod_{s=0}^n \frac{1}{t-q^{-\tilde{a}_s}}
\prod_{s=1}^n (t\, q^{-s+\tilde{\ell}}-1) \\
&=&q^{n\ell -n(n-1)/2}\, q^{-na_n} \oint_{{\mathcal C}(q^{a_k-a_n} | a_k\leq \ell-n)}\frac{dt}{2{\rm i}\pi} \, \prod_{s=0}^n \frac{1}{t-q^{a_{n-s}-a_{n}}}
\prod_{s=1}^n (t\, q^{-s+a_n+n-\ell}-1) \\
&=&q^{n\ell -n(n-1)/2} \oint_{{\mathcal C}(q^{a_k} | a_k\leq \ell-n)}\frac{dt'}{2{\rm i}\pi} \, \prod_{s=0}^n \frac{1}{t'-q^{a_{n-s}}}
\prod_{s=1}^n (t'\, q^{n-s-\ell}-1) \\
&=&q^{n\ell -n(n-1)/2} \oint_{{\mathcal C}(q^{a_k} | a_k\leq \ell-n)}\frac{dt}{2{\rm i}\pi} \, \prod_{s=0}^n \frac{1}{t-q^{a_{s}}}
\prod_{s=0}^{n-1} (t\, q^{s-\ell}-1) \ ,\\
\end{eqnarray*}
where we performed the change of variable $t'=t\, q^{a_n}$ (then called $t$ again in the fourth line). Note that this expression is very similar to that 
\eqref{eq:Hnl} for $H_{n,\ell}(q)$. Apart from minor shifts in the indices, the main difference comes from the contour of integration which now encircles those $q^{a_k}$ with $a_k\leq \ell-n$.
As before, this contour may be extended\footnote{Using this extended domain, it is easily verified by a simple contour deformation that $H_{n,\ell}(q)+\tilde{H}_{n,\ell-1}(q)=1$ for all $\ell$.
This remarkable identity has in fact a simple combinatorial explanation discussed in \cite{DFGUI}.} to the $q^{a_k}$ with $a_k\leq \ell$ since the last product in the integral vanishes for $t=q^{\ell},q^{\ell-1},\cdots, q^{\ell-n+1}$.

We finally need the partition function of the free trajectory, easily computed as (see figure \ref{fig:Hnltilde})
\begin{equation*}
\tilde{Y}_{\ell,r}(q)=q^{r(\ell+1)+r(r-1)/2}{\tilde{\ell}+r-1\brack \tilde{\ell}}_{q}\ .
\end{equation*}

We deduce the asymptotic equivalent
\begin{eqnarray*}
&\tilde{Y}_{\ell,r}(q)&\sim \ e^{n \tilde{S}_1(\xi,z)}\ ,\\  & \tilde{S}_1(\xi,z)&= z(\xi+z/2){\rm Log}(\qq)+\int_0^{\al(1)+1-\xi} d\uu\, {\rm Log}\left( \frac{\qq^{\uu+z}-1}{\qq^\uu -1}\right) \ ,
\end{eqnarray*} 
while
\begin{equation*}
\tilde{H}_{n,\ell}(q) \sim \oint \frac{dt}{2{\rm i}\pi}\, e^{n S_0(t,\xi)}\ ,
\end{equation*}
with \emph{the same} function $S_0(t,\xi)$ as in \eqref{eq:Hnlasymp} for $H_{n,\ell}(q)$. Here however, the contour must encircle only those 
$\qq^{\al(u)}$ such that $\al(u)\leq \xi-1$. For $\qq>1$, it must therefore surround the segment 
$[1,\qq^{\xi-1}]$, hence cross the real axis in the 
interval $]-\infty,1[$ (there are no poles less than $q^{a_0}=1$) and in the interval $]q^{\xi-1},q^{\xi}[$ (the poles $q^{a_k}$ for $\ell-n < a_k \leq \ell$ do not contribute to the integral).
For $\qq<1$, it must  surround the segment 
$[\qq^{\xi-1},1]$, hence cross the real axis in the interval $]\qq^{\xi},\qq^{\xi-1}[$ and in the 
interval $]1,+\infty[$. As before, at large $n$, the integral is estimated by a saddle-point method
and the optimal value of $\xi$ for fixed $z$ is obtained from the two extremization conditions:
\begin{eqnarray*}\frac{\partial S_0(t,\xi)}{\partial t}&=&0=\int_0^1 d\uu \left\{ \frac{\qq^{\uu-\xi}}{t\, \qq^{\uu-\xi}-1}-\frac{1}{t-\qq^{\al(\uu)}}\right\}\ ,\\
\frac{\partial (S_0(t,\xi)+\tilde{S}_1(\xi,z))}{\partial \xi}&=&0= {\rm Log}\left(\qq^{z+1}\,  \frac{\qq^{\al(1)+1-\xi}-1}{\qq^{\al(1)+1-\xi+z} -1}\right)-t\, {\rm Log}(\qq)\, \int_0^1 d\uu \frac{\qq^{\uu-\xi}}{t\, \qq^{\uu-\xi}-1}\ .
\end{eqnarray*}
These equations reduce to:
\begin{equation*}
\frac{t\, \qq-\qq^\xi}{t-\qq^\xi}\, x(t)=1\ ,\qquad  \qq^{z+1}\, \frac{\qq^{\al(1)+1-\xi}-1}{\qq^{\al(1)+1-\xi+z}-1}\, x(t)=1
\end{equation*}
with $x(t)$ as in \eqref{eq:defx}, hence the parametric solution $(\xi(t), z(t))$:
\begin{equation}
\label{solbis} \qq^{\xi(t)}=t\, \frac{\qq\, x(t)-1}{x(t)-1}\ ,\qquad \qq^{z(t)}=\frac{t}{\qq (t\,  x(t)+\qq^{\al(1)}(1-x(t)))} \ .
\end{equation}
As before, the range $t\in ]\qq^{\xi-1},\qq^{\xi}[$ for $\qq>1$ (respectively $t\in ]\qq^{\xi},\qq^{\xi-1}[$ for $\qq<1$) is ruled out since $(\qq^{\xi(t)}-t)/(t-\qq^{\xi(t)-1})=-\qq\, x(t) <0$. The parameter
$t$ is therefore now restricted to the range $t\in ]-\infty,1[$ whenever $\qq>1$ (respectively $t\in ]1,+\infty[$ whenever $\qq<1$).

In order to obtain a new family of tangent curves, we must compute the equivalent of equation \eqref{eq:tgcurve} for the present geometry , i.e.\ find in the present setting
the most likely free (rescaled) trajectory $(X,Y)$ from $(\xi,1)$ (point $\tilde{E}(\ell)$) to $(\al(1)+z,1+z)$ (point $\tilde{E}_n(r)$). Fortunately, a simple symmetry argument allows us 
to get the new equation for geodesics directly from \eqref{eq:tgcurve} by (i) applying to this latter equation the (rescaled) transformation $\mathcal{R}$, i.e.\ the change $(X,Y)\to (\al(1)+Y-X,Y)$ 
and (ii) changing $\qq\to 1/\qq$. Indeed, in configurations enumerated by $\tilde{Y}_{\ell,r}(q)$, the varying part, for fixed $\ell$, of the weight of a free trajectory may be written as $q^{\mathcal{A}}$ 
if ${\mathcal{A}}$ denotes the area \emph{on top of the path} (the rightmost blue shaded domain in figure \ref{fig:Hnltilde}). After the mapping $\mathcal{R}$, this area is still 
on top of the path rather than under it as in the computation leading to \eqref{eq:tgcurve}. This difference simply amounts to changing $q\to 1/q$ 
up to global factor (which is fixed for fixed $\ell$ and $r$). To summarize, we deduce, by applying (i) and (ii) to \eqref{eq:tgcurve}, the new equation for geodesics in the present geometry:
\begin{equation*}
\frac{1-\qq^{-(\al(1)+Y-X)}}{1-\qq^{-(\al(1)+1-\xi)}}+\frac{1-\qq^{-(Y-1)}}{1-\qq^{-z}}=1\ .
\end{equation*}
Picking for $\xi$ and $z$ the values $\xi(t)$ and $z(t)$ of \eqref{solbis}, this yields a parametric equation for a new family of tangent curves, namely after substitution:
\begin{equation*}
x(t)\, \qq^Y +\frac{1-x(t)}{t}\, 
\qq^X -1 =0\ .
\end{equation*}
Remarkably, we obtain for our new family \emph{the same expression} as that obtained in \eqref{eq:family} for the family of tangent curves associated with the first portion (right part) of arctic curve.
The result for the second portion of arctic curve boils down again to equation \eqref{eq:arcticthm} of Theorem \ref{mainthm}, but with now a different
domain of variation for the parameter $t$, namely 
$t\in ]-\infty,1[$ whenever $\qq>1$ and $t\in ]1,+\infty[$ whenever $\qq<1$. 

\begin{figure}
\begin{center}
\includegraphics[width=10cm]{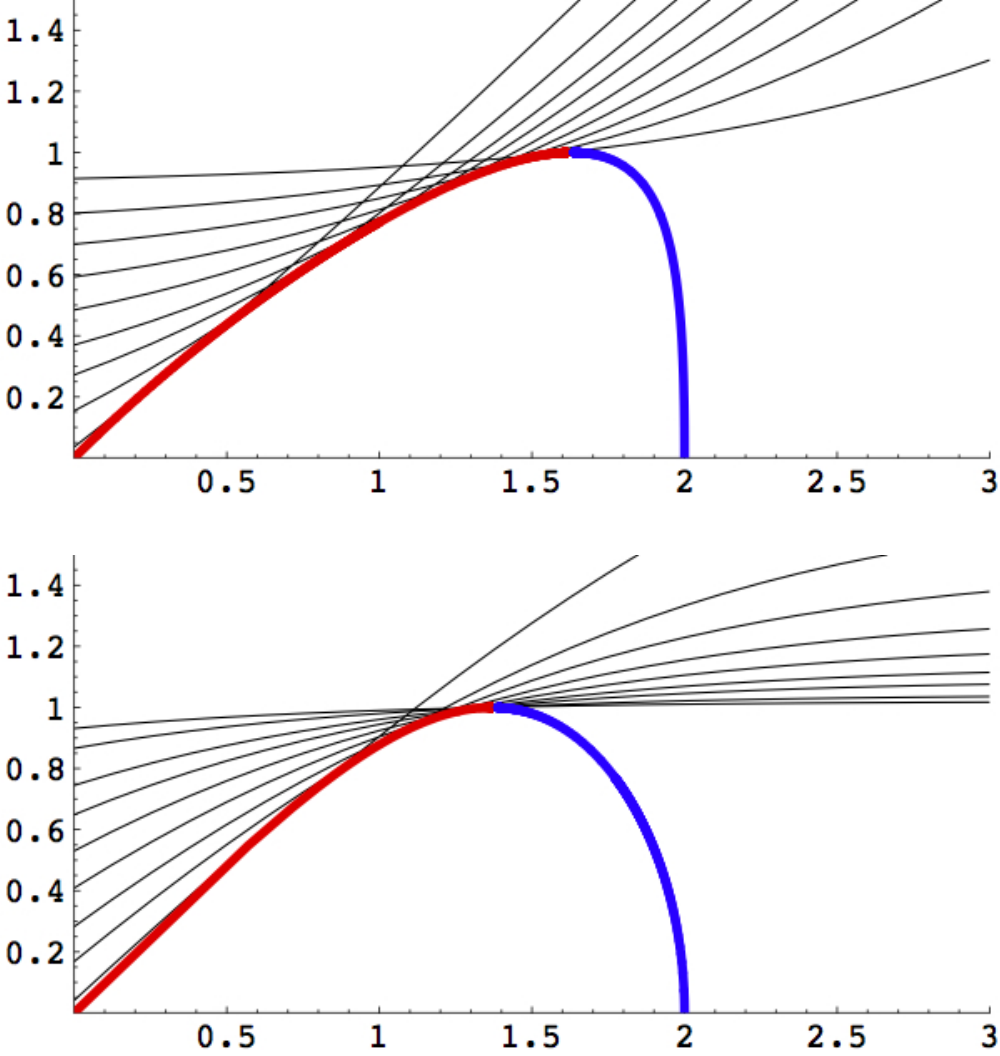}
\end{center}
\caption{\small The complete arctic curve including its right part (thick solid blue) and its left part (thick red line) as given by \eqref{eq:arcticthm} for the appropriate 
respective domains of $t$, here for $\qq=3$ (top) and $\qq=1/3$ (bottom) and in the particular case $\al(u)=2 u$. We also indicated members of the family
of tangent curves (thin lines) whose envelope defines the left part of the arctic curve.}
\label{fig:tangentcurvesall}
\end{figure}
The complete arctic curve, incorporating both the right and left parts, is displayed in figure \ref{fig:tangentcurvesall} in the particular case $\al(u)=2u$.

\subsection{Portions induced by freezing boundaries}
\label{sec:freezing}
Recall that, by construction, the scaling function $\al(u)$ is an increasing piecewise differentiable function for $u\in[0,1]$, such that $\al'(u)\geq 1$ when $\al'(u)$ is defined. 
For a generic such function,
the quantity $x(t)$ given by \eqref{eq:defx} is well-defined and real only for $t$ in the already encountered allowed domains, namely  
$t\in ]-\infty,1[ \cup \allowbreak ]\qq^{\al(1)},+\infty[$ for $\qq>1$ and $t\in ]-\infty,\qq^{\al(1)}[\cup]1,+\infty[$ for $\qq<1$.
This is due heuristically to the fact that $\int_0^1 d\uu\, 1/(t-\qq^{\al(\uu)})$ is generically not defined for $t$ in the interval $[\qq^{\al(0)},\qq^{\al(1)}]=[1,\qq^{\al(1)}]$ for $\qq>1$
(respectively $[\qq^{\al(1)},\qq^{\al(0)}]=[\qq^{\al(1)},1]$ for $\qq<1$) since $\qq^{\al(\uu)}$ spans this interval when $\uu$ varies between $0$ and $1$.
As a consequence, the arctic curve for a generic $\al(u)$ consists only of the two portions computed so far, namely its left and right part above.
\medskip 

As explained in \cite{DFGUI}, there exists however some particular realizations of $\al(u)$ giving rise to extra domains of $t$
for which $x(t)$, as given by \eqref{eq:defx} (possibly through some analytic continuation), remains well defined and real. This in turns 
leads through \eqref{eq:arcticthm} to extra portions of arctic curve by letting $t$ span these new domains.
This phenomenon appears in the particular case of so-called ``freezing boundaries", corresponding to
a situation where the sequence $(a_i)_{0\leq i\leq n}$ contains either macroscopic ``gaps", i.e.\ has no element in one or several intervals of the form 
$\llbracket A_m,A_m+\Delta_m\rrbracket$  with $\Delta_m \propto n$ for large $n$, or, on the contrary, to a situation where 
the sequence has ``fully filled intervals", i.e.\ includes all the successive integer values of one or several intervals $\llbracket A'_m,A'_m+\Delta'_m\rrbracket$.
Both situations correspond to freezing boundaries in the sense that they induce domains just above the $x$-axis where the paths configurations are fully frozen,
which serve as germs for larger frozen domains in the limit of large $n$, hence to new portions of arctic curve (see \cite{DFGUI} for details).    

In terms of the function $\al(u)$, the first situation corresponds to a discontinuity $\delta_m=\Delta_m/n$ 
at the value $u_m=A_m/n$, namely:
\begin{equation*}
\al(u_m^+)-\al(u_m^-)=\delta_m\ .
\end{equation*}
In this case, the quantity $\int_0^1 d\uu\, 1/(t-\qq^{\al(\uu)})$ is now well-defined for $t\in [\qq^{\al(u_m^-)},\qq^{\al(u_m^+)}]$ for $\qq>1$ (respectively  
$t\in [\qq^{\al(u_m^+)},\qq^{\al(u_m^-)}]$ for $\qq<1$) since this interval is no longer spanned by $\qq^{\al(\uu)}$ when $\uu$ varies between $0$ and $1$. 
This in turns creates an extra domain of $t$ on which $x(t)$ remains well-defined and real positive.

The second situation corresponds instead to a function $\al(u)$ with derivative equal to $1$ on some segment (recall that, by definition, $\al'(u)\geq 1$ when defined), 
namely:
\begin{equation*}
\al'(u)=1\ \ \hbox{for}\ u\in ]u'_m,u'_m+\delta'_m[\ .
\end{equation*}
In this case, the quantity $\int_0^1 d\uu 1/(1-\qq^{\al(\uu)})$ has a logarithmic cut for $t$ along $[\qq^{\al(u'_m)},\qq^{\al(u'_m)+\delta'_m}]$ for $\qq> 1$ 
(respectively $ [\qq^{\al(u'_m)+\delta'_m},\qq^{\al(u'_m)}]$ for $\qq < 1$) but, since $\al(\uu)=\al(u'_m)+\uu-u'_m$ for $\uu\in [u'_m,u'_m+\delta'_m]$,
we have along this interval a discontinuity
\begin{equation*}
\int_{u'_m}^{u'_m+\delta'_m}d\uu\frac{1}{t\pm {\rm i}\epsilon-\qq^{\al(\uu)}}=\frac{\delta'_m}{t}-\frac{1}{t\, {\rm Log}(\qq)}\left({\rm Log}\left(\frac{\qq^{\al(u'_m)+\delta'_m}-t}{t-\qq^{\al(u'_m)}}\right)\pm {\rm i}\, \pi \right)
\end{equation*}
which, when exponentiated in \eqref{eq:defx}, contributes to $x(t)$ via a (multiplicative) factor
\begin{equation*}
- \qq^{-\delta'_m}\,\frac{\qq^{\al(u'_m)+\delta'_m}-t}{t-\qq^{\al(u'_m)}}\ ,
\end{equation*}
with a global sign $e^{\pm {\rm i}\pi}=-1$, but \emph{with no cut in $x(t)$} along $[\qq^{\al(u'_m)},\qq^{\al(u'_m)+\delta'_m}]$. 
The quantity $x(t)$ remains thus well-defined and real for $t$ in this interval, but it now takes a negative value.

In both cases of gaps or fully filled intervals, the extra domains of $t$ leading to real values for $x(t)$, once inserted in \eqref{eq:arcticthm}, create new pieces of curve and it was conjectured\footnote{A particular instance of this conjecture was actually proved in \cite{DFGUI} in the case of a fully filled interval placed \emph{at the end} of the sequence of starting points.} in \cite{DFGUI} that these pieces are indeed actual additional portions of the arctic              
curve, separating the liquid phase from new frozen domains directly induced by the boundary conditions (hence the denomination ``freezing boundaries").       
Quite recently, this conjecture was proved in all generality by Debin and Ruelle in \cite{DR} for the $q=1$ version of the model. There it was shown how 
to extend the tangent method to arbitrary freezing boundaries and get these new portions of arctic curve by performing some clever shift below the $x$-axis 
of the starting points for those paths originally originating from one of the extremities of the freezing boundary.
This nice proof clearly extends to the case of arbitrary $q$.
Many examples of freezing boundaries are discussed in \cite{DFGUI} when $q=1$ and we will now
revisit some of them in the present design incorporating a $q$-dependent weight.

\section{The $q\to 0$ and $q\to \infty$ limits}
\label{sec:limitsgen}

\subsection{Heuristic argument}
\label{sec:heuristicgen}
\begin{figure}
\begin{center}
\includegraphics[width=14cm]{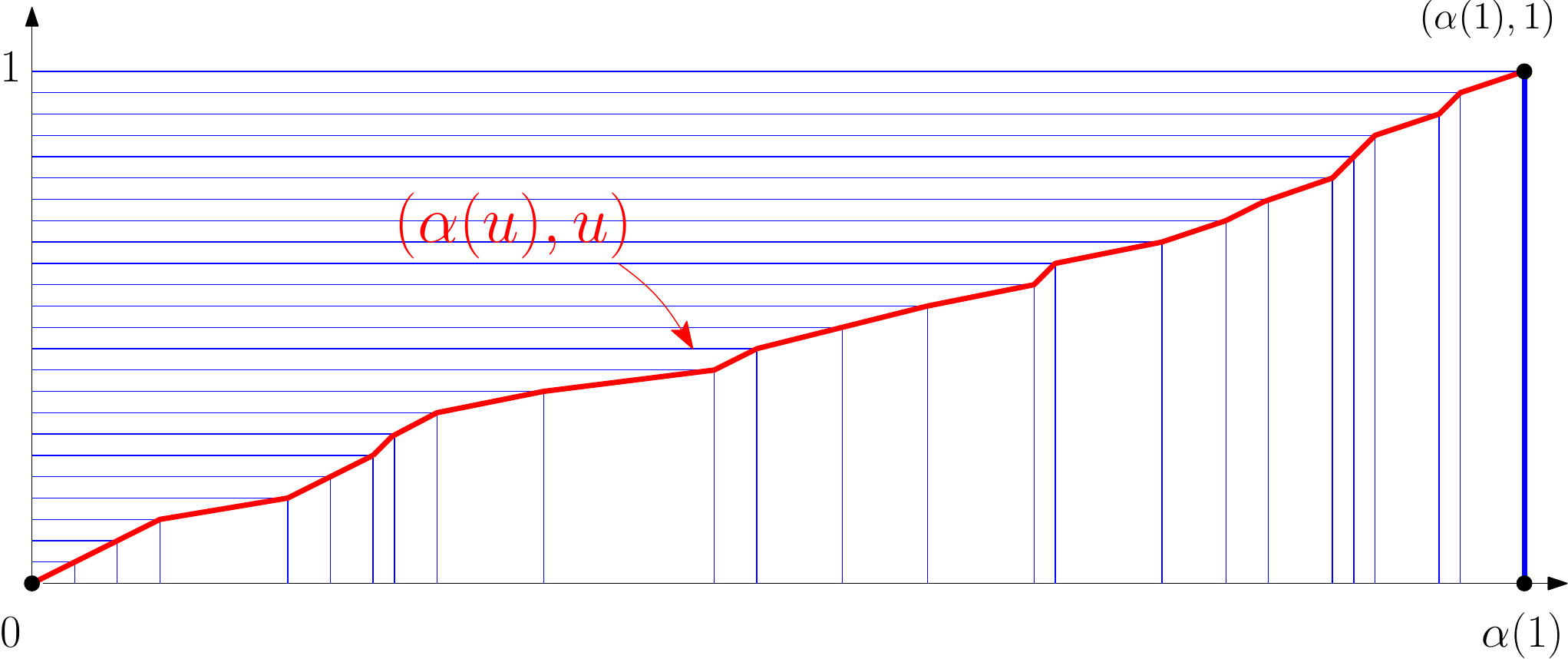}
\end{center}
\caption{\small The NILP configuration with highest weight when $q\to \infty$ for an arbitrary strictly increasing sequence $(a_i)_{0\leq i\leq n}$ whose large $n$ limit
is characterized by the function $\al(u)$. Each path is made of a single vertical north-oriented segment followed by a single horizontal west-oriented segment. In rescaled 
coordinates, the change from vertical to horizontal occurs at position $(\al(u),u)$ with $u\in [0,1]$. The corresponding curve 
connects the point $(0,0)$ to the point $(\al(1),1)$. The thick red curve and the thick blue vertical segment are natural candidates for the $\qq\to \infty$ limit of the 
left and right parts of the arctic curve respectively.}
\label{fig:qinfdiscretgen}
\end{figure}
It is interesting to look at the limit of the arctic curve when $q\to 0$ (i.e.\ $\qq\to 0$) or $q\to \infty$ (i.e.\ $\qq\to \infty$). To address this question,
a first heuristic approach consists in identifying, in each case, the most probable limiting path configuration, i.e. that \emph{with the highest weight}. 
Indeed, let us recall the precise meaning of the left and right parts of the arctic curve for finite $q$ in terms of the original NILP configurations. 
The right part of the arctic curve is the frontier between a liquid phase (below the curve) and a frozen region which is \emph{not visited by any of the 
paths}\footnote{For the second set of paths, this region corresponds instead to paths frozen along horizontal segments.}.
As for the left part, it separates the liquid phase from a frozen region in which the paths all follow \emph{horizontal segments}
towards their respective endpoints\footnote{For the second set of paths, this corresponds indeed to a region not visited by any of the 
paths.}. Finding the arctic curve 
when $q\to 0$ or $q\to \infty$  therefore boils down identifying the location where these separations take place in the most probable limiting path configuration.
\medskip

Let us start with the simplest $q\to \infty$ limit. Letting $q$ tend to infinity selects, in the original NILP setting, a configuration such that each path has the largest possible area compatible with 
the sequence of origins $O_i$ and endpoints $E_i$, i.e.\ is pushed as much as possible towards the upper-right corner $(a_n,n)$. Clearly, 
as displayed in figure \ref{fig:qinfdiscretgen}, this configuration is such that the path $\mathcal{P}_i$ is made of a vertical segment of length $i$ from $O_i$, followed by
a horizontal segment of length $a_i$ to $E_i$. The transition from vertical to horizontal takes place at position $(a_i,i)$ and the curve joining these transition points for increasing $i$
is the limit of the region in which path are frozen horizontally, hence a natural candidate for the $q \to \infty$ limit of the left part of the arctic curve.  In rescaled coordinates, 
this curve is parametrized by $(\al(u),u)$ for $u\in [0,1]$ and goes from $(0,0)$ to $(\al(1),1)$ with slope $1/\al'(u)$ (between $0$ and $1$) at $x$-coordinate $\al(u)$.

On the other hand, the vertical segment joining $(\al(1),1)$ to $(\al(1),0)$ defines the limit of the region visited by the paths
and is therefore a natural candidate for the $q \to \infty$ limit of the right part of the arctic curve. 

To summarize, we expect that the left and right parts of the arctic curve tend for $q \to \infty$ to the above described limiting curve and segment, see figure \ref{fig:qinfdiscretgen}. 
From this analysis, we also expect that the liquid phase, which remains liquid as long as $q$ remains finite, eventually crystallizes right at $q=\infty$ into a sequence of frozen vertical paths 
whose relative spacing is directly measured by the function $\al(u)$.  
\medskip

\begin{figure}
\begin{center}
\includegraphics[width=14cm]{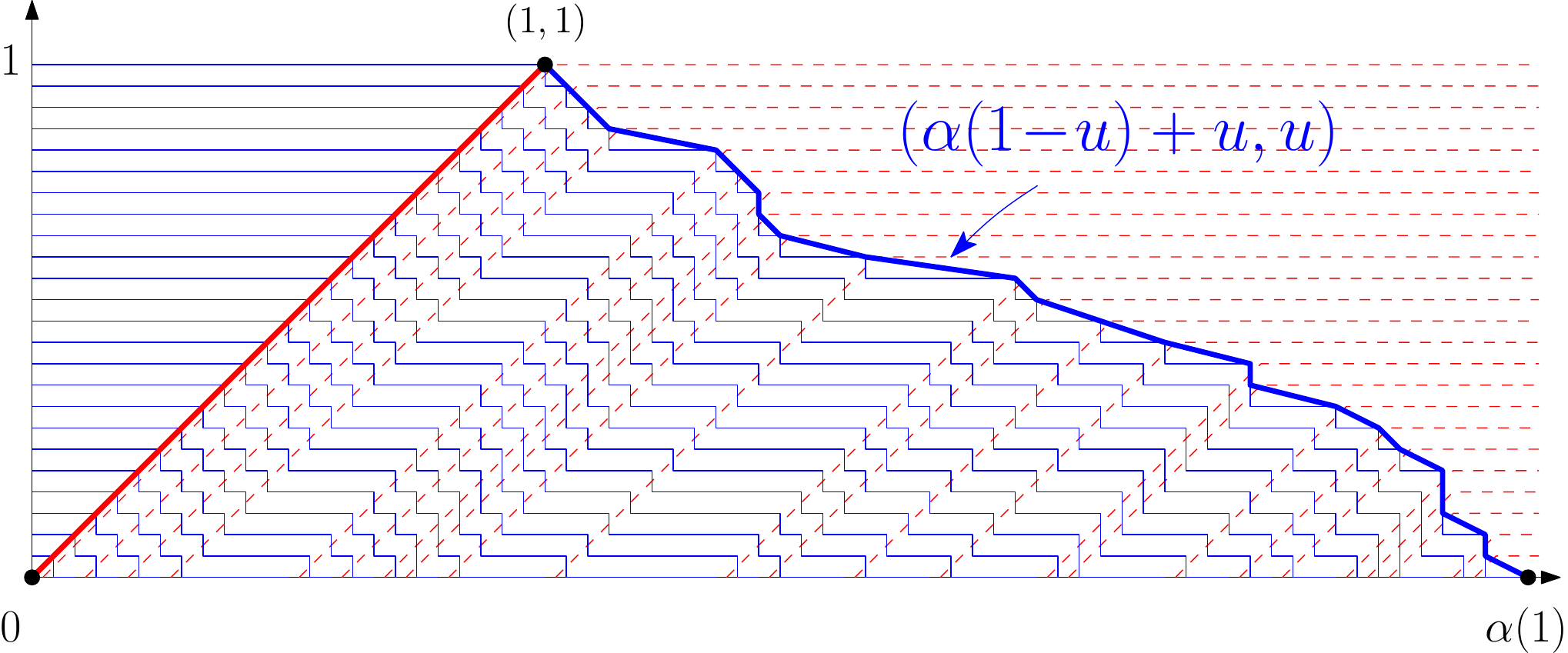}
\end{center}
\caption{\small The NILP configuration with highest weight when $q\to 0$ for an arbitrary strictly increasing sequence $(a_i)_{0\leq i\leq n}$ whose large $n$ limit
is characterized by the function $\al(u)$. The path configuration (solid thin blue) is the pre-image by the bijection of Section \ref{sec:alternativeform} of a configuration of paths (dashed red)
made of a single northeast-oriented segment followed by a single horizontal east-oriented segment (we did not represent here the rightmost parts of these horizontal segments as they carry no 
relevent information). In rescaled 
coordinates, the location of the limit of the region not visited by paths is given by $(\al(1-u)+u,u)$ with $u\in [0,1]$. This 
connects the point $(1,1)$ to the point $(\al(1),0)$. The thick red segment and the thick blue curve are natural candidates for the $\qq \to 0$ limit of the 
left and right parts of the arctic curve respectively.}
\label{fig:q0discretgen}
\end{figure}
Let us now come to the $q\to 0$ limit. This now selects a configuration such that each path has the smallest possible area compatible with the non-intersection constraint, i.e.\
is pushed as much as possible towards the lower-left corner. As displayed in figure \ref{fig:q0discretgen}, this configuration is best described if we now use the second set of paths
made of east- and northeast-oriented steps, as these paths must now be pushed as much as possible towards the upper left corner to reduce the area on their left. Clearly, the path $\tilde{\mathcal{P}}_i$
is then made a northeast-oriented segment from $\tilde O_i=(a_{n-i}+1/2,0)$ to the point $(a_{n-i}+1/2+i,i)$, followed by a horizontal segment towards $\tilde{E}_i$. The curve joining the
transition points $(a_{n-i}+1/2+i,i)$ for increasing $i$ delimits the region where the paths become horizontal, a criterion which, for the original NILP configuration, corresponds instead 
to a region not visited by any of the paths. In other words, this curve is a natural candidate for $q\to 0$ limit of the right part of the arctic curve. In rescaled coordinates, 
it is parametrized by $(\al(1-u)+u,u)$ for $u\in [0,1]$ and connects $(1,1)$ (for $u=1$) to $(\al(1),0)$ (for $u=0$). In particular, it has a slope $-1/(\al'(1-u)-1)$ (between $0$ and $-\infty$) 
at $x$-coordinate $\al(1-u)+u$.

On the other hand, the outermost path $\tilde{\mathcal{P}}_n$ starts, in the most probable configuration, by a northeast-oriented segment from $(1/2,0)$ to $(n+1/2,n)$ which defines the limit
of the region where the original paths are frozen into horizontal lines and this segment is a natural candidate for the $q \to 0$ limit of the left part of the arctic curve.
In rescaled coordinates, it is nothing but the segment joining $(0,0)$ to $(1,1)$.

To summarize, we expect that the right and left parts of the arctic curve tend for $q\to 0$ to the above described curve and segment, see figure \ref{fig:q0discretgen}. We also expect that, below the arctic curve, the liquid phase 
which remains liquid as long as $q >0$, crystallizes right at $q=0$ into a sequence of frozen paths whose shape is the same\footnote{In other
words, the paths are parametrized by $(\al(1-u)+u-v,u-v)$ for $u\in [v,1]$, with $v\in [0,1]$.} as that of the right part of the arctic curve travelled downwards from the point $(1,1)$,  
but are shifted southwest so as to start instead from any point $(1-v,1-v)$ ($v\in [0,1]$)
along the left part of the arctic curve, until they eventually reach the $x$-axis at $(\al(1-v),0)$. In particular, the (negative) slope of the paths is the same along $45^\circ$ oriented lines (see figure \ref{fig:q0discretgen}).   
Let us now validate the above heuristic arguments by a more precise study of the limiting shape of the arctic curve, as given by
\eqref{eq:arcticthm}, when $\qq \to \infty$ or $\qq\to 0$.

\subsection{Analytic treatment for $q\to \infty$}
\label{sec:analyticgen}
For $\qq>1$, the left part of the arctic curve is obtained by letting $t$ vary in $]-\infty,1[$. Let us for convenience decompose this interval into
\begin{equation}
]-\infty,1[\ =\  ]-\infty,-\qq^{\al(1)}]\cup]-\qq^{\al(1)},-1[ \cup [-1,1[
\label{eq:qinfintervals}
\end{equation}
and study the respective portions of arctic curve coming from each of the three subintervals when $\qq\to \infty$. We start with the middle subinterval, which is
best studied by setting $t=-\qq^{\al(\tau)}$ with $\tau \in ]0,1[$. From \eqref{eq:defx}, we may then write
\begin{equation*} 
\begin{split}
{\rm Log}(x(t))&= -{\rm Log}(\qq)\, \int_0^1 d\uu \frac{1}{1+\qq^{\al(\uu)-\al(\tau)}}\\
&= -{\rm Log}(\qq)\, \left(\int_0^\tau d\uu \frac{1}{1+\qq^{\al(\uu)-\al(\tau)}}+\int_\tau^1 d\uu \frac{1}{1+\qq^{\al(\uu)-\al(\tau)}}\right)\\
&\!\!\!\!\!\underset{\qq\to \infty}{\sim} -{\rm Log}(\qq) \ \ \tau\\
\end{split}
\end{equation*}
since, for $\uu \in ]0,\tau[$, we have $\al(\uu)-\al(\tau)<0$ hence the integrand in the first integral tends to $1$, while
for $\uu \in ]\tau,1[$, $\al(\uu)-\al(\tau)>0$ and the integrand in the second integral tends to $0$.
This yields
\begin{equation*}
\begin{split}
x(t) \sim \qq^{-\tau} \ , \qquad & t\, x'(t) \sim -\qq^{-\tau}\frac{1}{\al'(\tau)}\\
\qq^{X(t)} \sim \frac{t^2 x'(t)}{t x'(t)+x(t)} \sim \qq^{\al(\tau)} \frac{1}{\al'(\tau)-1}\ , \qquad & \qq^{Y(t)} \sim \frac{1}{t x'(t)+x(t)} \sim  \qq^{\tau} \frac{\al'(\tau)}{\al'(\tau)-1} \ , \\
\end{split}
\end{equation*}
which implies at leading order
\begin{equation*}
X(\tau)= \al(\tau)\ , \qquad Y(\tau)=\tau\ .
\end{equation*}
When $\tau$ varies between $0$ and $1$, this gives precisely the curve announced in Section \ref{sec:heuristicgen} (with the identification $\tau=u$).  
Here we assumed implicitly that $\al'(\tau)> 1$ but having $\al'(\tau)=1$ at isolated points would not cause any problem. On the other
hand, having $\al'(\tau)=1$ along some interval, which corresponds to a freezing boundary with a fully filled interval, would require a more involved analysis.
We will discuss such a case in Section \ref{sec:freez1} below.  A interesting outcome of our analysis is that, when $\qq\to \infty$, the
left part of the arctic curve seems to be entirely produced by the middle subinterval in the decomposition \eqref{eq:qinfintervals} above. This is indeed the case since, 
as we will now show,
the contribution of the subinterval $]-\infty,-\qq^{\al(1)}[$ reduces to a single point $(\al(1),1)$ at the right extremity of the left part of the arctic curve while that of the subinterval $]-1,1[$
reduces to the point $(0,0)$ at its left extremity. For $t\in ]-\infty,-\qq^{\al(1)}[$, we set $t=-\qq^{\al(1)+\tau}$ with $\tau>0$ and get 
\begin{equation*} 
{\rm Log}(x(t))= -{\rm Log}(\qq)\, \int_0^1 d\uu \frac{1}{1+\qq^{\al(\uu)-\al(1)-\tau}}
\underset{\qq\to \infty}{\sim} -{\rm Log}(\qq)
\end{equation*}
since for $\uu \in [0,1]$, we have $\al(\uu)-\al(1)\leq0$ hence the integrand tends to $1$. We deduce $x(t)\sim \qq^{-1}$. By differentiation, we also have
\begin{equation*} 
\frac{t\, x'(t)}{x(t)}= -{\rm Log}(\qq)\, \int_0^1 d\uu \frac{\qq^{\al(\uu)-\al(1)-\tau}}{\left(1+\qq^{\al(\uu)-\al(1)-\tau}\right)^2}
\underset{\qq\to \infty}{\sim} -{\rm Log}(\qq)\ \qq^{-\tau}  \, \int_0^1 d\uu \qq^{\al(\uu)-\al(1)} \underset{\qq\to \infty}{\sim}  -\frac{1}{\al'(1)}  \qq^{-\tau}
\end{equation*}
since the last integral vanishes\footnote{This may be shown by a saddle point method upon setting $\uu=1-\eta/{\rm Log}(\qq)$ so that the integral 
has asymptotic value
$(1/{\rm Log}(\qq))\int_0^\infty d\eta \, e^{-\al'(1)\eta}=1/(\al'(1)\,  {\rm Log}(\qq))$.} as $1/(\al'(1)\, {\rm Log}(\qq))$.
This now yields
\begin{equation*}
\qq^{X(t)} \sim \frac{t^2 x'(t)}{x(t)} \sim \qq^{\al(1)} \frac{1}{\al'(1)}\ , \qquad  \qq^{Y(t)} \sim \frac{1}{x(t)} \sim  \qq \ , 
\end{equation*}
hence $(X(t),Y(t))$ tends to $(\al(1),1)$ for all $t\in ]-\infty,-\qq^{\al(1)}[$. 
For the last subinterval  $t\in ]-1,1[$, we set $t=\pm \qq^\tau$ with $\tau<0$ and obtain
\begin{equation*} 
{\rm Log}(x(t)) = -{\rm Log}(\qq)\, \int_0^1 d\uu \frac{1}{1\mp \qq^{\al(\uu)-\tau}}\underset{\qq\to \infty}{\sim} \pm {\rm Log}(\qq)\, \qq^{\tau}\, \int_0^1 d\uu \qq^{-\al(\uu)}
\underset{\qq\to \infty}{\sim}  \pm \frac{1}{\al'(0)}  \qq^{\tau}
\end{equation*}
hence $x(t)\sim 1\pm \frac{1}{\al'(0)}  \qq^{\tau}$ and
\begin{equation*}
\begin{split} 
\frac{t\, x'(t)}{x(t)}&= -{\rm Log}(\qq)\, \int_0^1 d\uu \frac{\mp \qq^{\al(\uu)-\tau}}{\left(1\mp \qq^{\al(\uu)-\tau}\right)^2}\\
&={\rm Log}(x(t)) +{\rm Log}(\qq)\, \int_0^1 d\uu \frac{1}{\left(1\mp \qq^{\al(\uu)-\tau}\right)^2}\\
\end{split}
\end{equation*}
so that 
\begin{equation*}
\frac{t\, x'(t)}{x(t)}-{\rm Log}(x(t)) \underset{\qq\to \infty}{\sim}  {\rm Log}(\qq)\, \qq^{2\tau}\, \int_0^1 d\uu \qq^{-2\al(\uu)}
\underset{\qq\to \infty}{\sim}   \frac{1}{2\al'(0)}  \qq^{2\tau}\ .
\end{equation*}
Using ${\rm Log}(x(t))=(x(t)-1)+O\left((x(t)-1)^2\right)$ with $x(t)=1+O(\qq^\tau)$, we now get $t\, x'(t)+x(t)(1-x(t))=O\left(\qq^{2\tau}\right)$, $t\, x'(t)+(1-x(t))=O\left(\qq^{2\tau}\right)$ and
$t^2\, x'(t)=O\left(\qq^{2\tau}\right)$, which implies that 
$\qq^{X(t)}$ and $\qq^{Y(t)}$ tend to finite constants, hence $(X(t),Y(t))$ tends to $(0,0)$ for all $t\in ]-1,1[$. To summarize, the two extremal 
subintervals in \eqref{eq:qinfintervals} contribute only to the two points at the extremities of the left part of the arctic curve.
 
Let us now discuss the limiting shape of the right part of the arctic curve, coming from values of $t$ in the range $]\qq^{\al(1)},+\infty[$.
Writing $t=\qq^{\al(1)}w$ with $w>1$, we may write
\begin{equation*} 
{\rm Log}(x(t))= -{\rm Log}(\qq)\, \int_0^1 d\uu \frac{1}{1-w^{-1}\, \qq^{\al(\uu)-\al(1)}}
\end{equation*}
so the calculation seems at first very similar to that for the interval $]-\infty,-\qq^{\al(1)}[$ and we could be tempted to conclude that this again leads 
to a unique limiting point $(\al(1),1)$. This reasoning however ignores the fact that the denominator in the integrand may remain small for values of $w$ close enough to $1$.
As we shall now see, there exists indeed an appropriate domain of $w$ close to $1$ for which 
the asymptotic value of the integral (otherwise equal to $1$ if $w-1$ does not scale properly with $\qq$) is modified and depends on $w$. More precisely, writing
\begin{equation*} 
{\rm Log}(x(t))= -{\rm Log}(\qq)\, \left(1+\int_0^1 d\uu \frac{\qq^{\al(\uu)-\al(1)}}{w-\, \qq^{\al(\uu)-\al(1)}}\right)\ ,
\end{equation*}
the last integral may be evaluated by a saddle point method upon setting $\uu=1-\eta/{\rm Log}(\qq)$. The asymptotic value of this additional correction reads
\begin{equation*} 
\frac{1}{{\rm Log}(\qq)}\int_0^\infty d\eta \frac{e^{-\al'(1)\eta}}{w-\, e^{-\al'(1)\eta}}=-\frac{1}{\al'(1)} \frac{{\rm Log}\left(1-\frac{1}{w}\right)}{{\rm Log}(\qq)}\ ,
\end{equation*}
which is finite when $w$ is chosen so that $(1-1/w)= \qq^{-\rho}$, i.e.\ $w=1/(1-\qq^{-\rho})$ for some positive $\rho$. Otherwise stated, we have asymptotically
\begin{equation*} 
x(t)\underset{\qq\to \infty}{\sim} \qq^{-1} \left(1-\frac{1}{t\, \qq^{-\al(1)}}\right)^{\frac{1}{\al'(1)}}\ .
\end{equation*}
with a non trivial limiting value when we take $t=\qq^{\al(1)}/(1-\qq^{-\rho})$.  In this case, we obtain directly\footnote{It is indeed easily verified that
$t^2x'(t)\sim \qq^{\al(1)-1+\rho\left(1-\frac{1}{\al'(1)}\right)}$, $t\, x'(t)+(1-x(t))\sim \qq^{\max\left(0,-1+\rho\left(1-\frac{1}{\al'(1)}\right)\right)}$ and
$t\, x'(t)+x(t)(1-x(t))\sim \qq^{-1+\rho\left(1-\frac{1}{\al'(1)}\right)}$.} from \eqref{eq:arcticthm}:
\begin{equation*}
\qq^{X(t)} \sim \qq^{\al(1)}\ , \qquad  \qq^{Y(t)} \sim  \qq^{\max\left(1-\rho\left(1-\frac{1}{\al'(1)}\right),0\right)}
\end{equation*}
which leads to
\begin{equation*}
X(t)=\al(1)\ , \qquad  Y(t)= 1-\rho\left(1-\frac{1}{\al'(1)}\right)\ , \qquad 0< \rho \leq \frac{\al'(1)}{\al'(1)-1}\ .
\end{equation*}
This parametric curve is nothing but the vertical segment from $(\al(1),1)$ to $(\al(1),0)$, which confirms our heuristic result for the $q\to \infty$ limit of
the right part of the arctic curve. Note that the above result requires $\al'(1)>1$. For $\al'(1)=1$, the right part of the arctic curve reduces instead to 
the single point $(\al(1),1)$. We will see such an example in Section \ref{sec:ellipse} below.

\subsection{Analytic treatment for $q\to 0$}
\label{sec:analyticgenbis}
For $\qq<1$, the right part of the arctic curve is now obtained by letting $t$ vary in $]-\infty,\qq^{\al(1)}[$ and we decompose this interval into
\begin{equation}
]-\infty,\qq^{\al(1)}[\ =\  ]-\infty,-1]\cup]-1,-\qq^{\al(1)}[ \cup [-\qq^{\al(1)},\qq^{\al(1)}[
\label{eq:q0intervals}
\end{equation}
to better study the respective portions coming from each of the three subintervals when $\qq\to 0$. Again the non-trivial contribution is that
of the middle subinterval, best expressed by setting $t=-\qq^{\al(\tau)}$ with $\tau \in ]0,1[$. We have indeed
\begin{equation*} 
\begin{split}
{\rm Log}(x(t))&= -{\rm Log}(\qq)\, \int_0^1 d\uu \frac{1}{1+\qq^{\al(\uu)-\al(\tau)}}\\
&= -{\rm Log}(\qq)\, \left(\int_0^\tau d\uu \frac{1}{1+\qq^{\al(\uu)-\al(\tau)}}+\int_\tau^1 d\uu \frac{1}{1+\qq^{\al(\uu)-\al(\tau)}}\right)\\
&\!\!\!\!\!\underset{\qq\to 0}{\sim} -{\rm Log}(\qq) \ \ (1-\tau)\\
\end{split}
\end{equation*}
since, for $\uu \in ]0,\tau[$, we have $\al(\uu)-\al(\tau)<0$ hence the integrand in the first integral tends to $0$, while
for $\uu \in ]\tau,1[$, $\al(\uu)-\al(\tau)>0$ and the integrand in the second integral tends to $1$.
This yields
\begin{equation*}
\begin{split}
x(t) \sim \qq^{\tau-1} \ , \qquad & t\, x'(t) \sim \qq^{\tau-1}\frac{1}{\al'(\tau)}\\
\qq^{X(t)} \sim \frac{t^2 x'(t)}{-(x(t))^2} \sim \qq^{\al(\tau)+1-\tau} \frac{1}{\al'(\tau)}\ , \qquad & \qq^{Y(t)} \sim \frac{t x'(t)-x(t)}{-(x(t))^2} \sim  \qq^{1-\tau} \frac{\al'(\tau)-1}{\al'(\tau)} \ , \\
\end{split}
\end{equation*}
which implies at leading order
\begin{equation*}
X(\tau)= \al(\tau)+1-\tau\ , \qquad Y(\tau)=1-\tau\ .
\end{equation*}
When $\tau$ varies between $0$ and $1$, this gives precisely the curve announced in Section \ref{sec:heuristicgen} (with the identification $\tau=1-u$).  
Let us now discuss
the contribution of the subintervals $]-\infty,-1[$ and $]-\qq^{\al(1)},\qq^{\al(1)}[$.
For $t\in ]-\infty,-1[$, we set $t=-\qq^{\tau}$ with $\tau<0$ and get 
\begin{equation*} 
{\rm Log}(x(t))= -{\rm Log}(\qq)\, \int_0^1 d\uu \frac{1}{1+\qq^{\al(\uu)-\tau}}
\underset{\qq\to 0}{\sim} -{\rm Log}(\qq)
\end{equation*}
since for $\uu \in [0,1]$, we have $\al(\uu)\geq 0$ hence the integrand tends to $1$. We deduce $x(t)\sim \qq^{-1}$. By differentiation, we also have
\begin{equation*} 
\frac{t\, x'(t)}{x(t)}= -{\rm Log}(\qq)\, \int_0^1 d\uu \frac{\qq^{\al(\uu)-\tau}}{\left(1+\qq^{\al(\uu)-\tau}\right)^2}
\underset{\qq\to 0}{\sim} -{\rm Log}(\qq)\ \qq^{-\tau}  \, \int_0^1 d\uu \qq^{\al(\uu)} \underset{\qq\to 0}{\sim}  \frac{1}{\al'(0)}  \qq^{-\tau}
\end{equation*}
since the last integral vanishes\footnote{This again is shown by a saddle point method upon setting $\uu=-\eta/{\rm Log}(\qq)$ so that the integral 
has asymptotic value
$-(1/{\rm Log}(\qq))\int_0^\infty d\eta \, e^{-\al'(0)\eta}=-1/(\al'(0)\,  {\rm Log}(\qq))$.} as $-1/(\al'(0)\, {\rm Log}(\qq))$.
This now yields
\begin{equation*}
\qq^{X(t)} \sim \frac{t^2 x'(t)}{-(x(t))^2} \sim \qq \frac{1}{\al'(0)}\ , \qquad  \qq^{Y(t)} \sim \frac{1}{x(t)} \sim  \qq \ , 
\end{equation*}
hence $(X(t),Y(t))$ tends to $(1,1)$ for all $t\in ]-\infty,-1[$. 
For the other subinterval, i.e.\ for $t\in]-\qq^{\al(1)},\qq^{\al(1)}[$, we set $t=\pm \qq^{\al(1)+\tau}$ with $\tau>0$ and obtain
\begin{equation*} 
{\rm Log}(x(t)) = -{\rm Log}(\qq)\, \int_0^1 d\uu \frac{1}{1\mp \qq^{\al(\uu)-\al(1)-\tau}}\underset{\qq\to 0}{\sim}  \pm {\rm Log}(\qq)\, \qq^{\tau}\, 
\int_0^1 d\uu \qq^{\al(1)-\al(\uu)}
\underset{\qq\to 0}{\sim}  \mp \frac{1}{\al'(1)}  \qq^{\tau}
\end{equation*}
hence $x(t)\sim 1\mp \frac{1}{\al'(1)}  \qq^{\tau}$ and
\begin{equation*}
\begin{split} 
\frac{t\, x'(t)}{x(t)}&= -{\rm Log}(\qq)\, \int_0^1 d\uu \frac{\mp \qq^{\al(\uu)-\al(1)-\tau}}{\left(1\mp \qq^{\al(\uu)-\al(1)-\tau}\right)^2}\\
&={\rm Log}(x(t)) +{\rm Log}(\qq)\, \int_0^1 d\uu \frac{1}{\left(1\mp \qq^{\al(\uu)-\al(1)-\tau}\right)^2}\\
\end{split}
\end{equation*}
so that 
\begin{equation*}
\frac{t\, x'(t)}{x(t)}-{\rm Log}(x(t)) 
\underset{\qq\to 0}{\sim} {\rm Log}(\qq)\, \qq^{2\tau}\, \int_0^1 d\uu\, \qq^{2(\al(1)-\al(\uu))}
\underset{\qq\to 0}{\sim}   -\frac{1}{2\al'(1)}  \qq^{2\tau}\ .
\end{equation*}
Using ${\rm Log}(x(t))=(x(t)-1)+O\left((x(t)-1)^2\right)$ with $x(t)=1+O(\qq^\tau)$, we now get $t\, x'(t)+x(t)(1-x(t))=O\left(\qq^{2\tau}\right)$, $t\, x'(t)+(1-x(t))=O\left(\qq^{2\tau}\right)$ and
$t^2\, x'(t)=O\left(\qq^{\al(1)+2\tau}\right)$, which implies that 
$\qq^{X(t)}\sim \qq^{\al(1)}$ while $\qq^{Y(t)}$ tends to a finite constant, hence $(X(t),Y(t))$ tends to $(\al(1),0)$ for all $t\in ]-1,1[$. We end up with the expected result that the two extremal 
subintervals in \eqref{eq:q0intervals} contribute only to the two extremities of the right part of the arctic curve.
 
Let us conclude our discussion with the limiting shape of the left part of the arctic curve, corresponding to values of $t$ in the range $]1+\infty[$.
Writing directly
\begin{equation*} 
{\rm Log}(x(t))= -{\rm Log}(\qq)\, \int_0^1 d\uu \frac{1}{1-t^{-1}\, \qq^{\al(\uu)}}
\end{equation*}
with $t>1$,  we again have to deal with values of $t$ close enough to $1$ so that the denominator in the integral remains small.
As in the previous section, there exists an appropriate domain of $t$ close to $1$ for which 
the asymptotic value of the integral (otherwise equal to $1$ if $t-1$ does not scale properly with $\qq$) is modified. Writing
\begin{equation*} 
{\rm Log}(x(t))= -{\rm Log}(\qq)\, \left(1+\int_0^1 d\uu \frac{\qq^{\al(\uu)}}{t-\, \qq^{\al(\uu)}}\right)\ ,
\end{equation*}
the last integral may be evaluated by a saddle point method upon setting $\uu=-\eta/{\rm Log}(\qq)$ and its asymptotic value reads
\begin{equation*} 
-\frac{1}{{\rm Log}(\qq)}\int_0^\infty d\eta \frac{e^{-\al'(0)\eta}}{t-\, e^{-\al'(0)\eta}}=\frac{1}{\al'(0)} \frac{{\rm Log}\left(1-\frac{1}{t}\right)}{{\rm Log}(\qq)}\ ,
\end{equation*}
which leads eventually to 
\begin{equation*} 
x(t)\underset{\qq\to 0}{\sim} \qq^{-1} \left(1-\frac{1}{t}\right)^{-\frac{1}{\al'(0)}}\ .
\end{equation*}
Setting $t=1/(1-\qq^{\rho})$ with $\rho > 0$, we obtain directly\footnote{It is easily verified that
$t^2x'(t)\sim \qq^{-1-\rho\left(1+\frac{1}{\al'(0)}\right)}$, $t\, x'(t)+(1-x(t))\sim \qq^{-1-\rho\left(1+\frac{1}{\al'(0)}\right)}$ and
$t\, x'(t)+x(t)(1-x(t))\sim  \qq^{\min\left(-1-\rho\left(1+\frac{1}{\al'(0)}\right),-2-\rho\frac{2}{\al'(0)}\right)}$.} from \eqref{eq:arcticthm}:
\begin{equation*}
\qq^{X(t)} \sim \qq^{\max\left(1-\rho\left(1-\frac{1}{\al'(0)}\right),0\right)}\ , \qquad  \qq^{Y(t)} \sim  \qq^{\max\left(1-\rho\left(1-\frac{1}{\al'(0)}\right),0\right)}
\end{equation*}
which leads to
\begin{equation*}
X(t)= 1-\rho\left(1-\frac{1}{\al'(0)}\right)\ , \qquad  Y(t)= 1-\rho\left(1-\frac{1}{\al'(0)}\right)\ , \qquad 0< \rho \leq \frac{\al'(0)}{\al'(0)-1}\ .
\end{equation*}
This parametric curve is nothing but the segment joining $(0,0)$ to $(1,1)$, which confirms our heuristic result for the $q\to 0$ limit of
the left part of the arctic curve. Note that the above result requires $\al'(0)>1$. For $\al'(0)=1$, the left part of the arctic curve reduces instead to 
the single point $(1,1)$. We will see such an example in Section \ref{sec:ellipse} below.

\section{Examples}
\label{sec:examples}

A quite general situation, which displays most of the interesting phenomena for the arctic curve, corresponds to the case when $\al(u)$ is \emph{piecewise linear}.
More precisely, we demand that $\al(u)$ is made of $k$ linear pieces, i.e.\ satisfies $\al(0)=0$, has slope $p_1$ on $[0,\gamma_1]$, $p_2$ on $[\gamma_1,\gamma_1+\gamma_2]$,
$\dots$,  $p_i$ on $[\gamma_1+\cdots +\gamma_{i-1},\gamma_{1}+\cdots +\gamma_i]$ for $i$ up to $k$. Here the slopes $p_i$ of the various pieces satisfy $p_i\geq 1$, $i=1,\cdots, k$
(to ensure $\al'(u)\geq 1$ when defined), and the widths $\gamma_i$ of these pieces add up to $\sum_{i=1}^k\gamma_i=1$.
In short, we take:
\begin{equation*}
\al(u)=p_i\, u+\sum_{j=1}^{i-1}(p_j-p_i)\gamma_j\ , \qquad \hbox{for}\ u\in [\gamma_1+\cdots +\gamma_{i-1},\gamma_{1}+\cdots +\gamma_i]
\end{equation*}
for $i=1,\cdots, k$.

Note that the case of frozen boundaries of Section \ref{sec:freezing} may be realized in the present setting: the case of a gap $\delta_m$ in $\al(u)$ for 
$u=u_m=\gamma_1+\cdots+\gamma_{m-1}$ is obtained by sending simultaneously $p_m\to \infty$ and $\gamma_m\to 0$, keeping the product $p_m\gamma_m=\delta_m$ finite. 
As for the case of a fully filled interval between $u'_m=\gamma_1+\cdots+\gamma_{m-1}$ and $u'_m+\delta'_m=\gamma_1+\cdots+\gamma_{m}$, it is obtained by simply taking $p_m=1$
and $\gamma_m=\delta'_m$.
Such cases will be discussed in Sections \ref{sec:freez1} and \ref{sec:freez2} below.
\medskip

\begin{figure}
\begin{center}
\includegraphics[width=10cm]{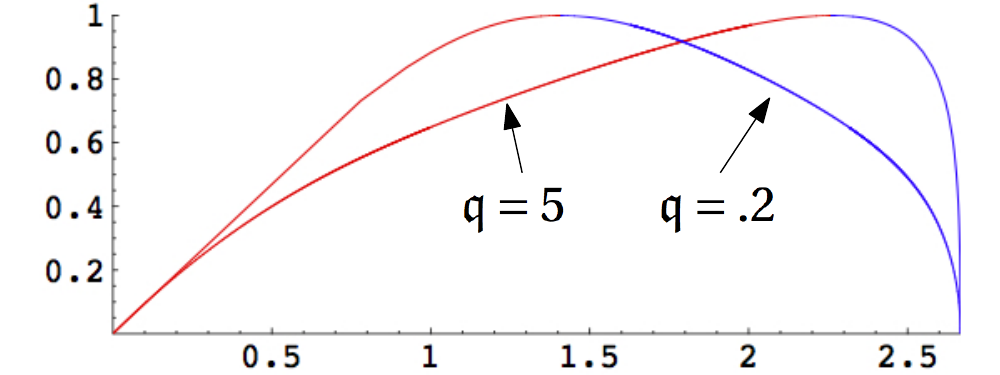}
\end{center}
\caption{\small The left (red) and right (blue) portions of the arctic curve for a piecewise linear $\al(u)$ with $k=3$, $\gamma_1=\gamma_2=\gamma_3=1/3$, 
$p_1=p_3=2$ and $p_2=4$, here for $\qq=5$ and $\qq=.2$.}
\label{fig:pieces}
\end{figure}
Returning to the case of arbitrary $p_i$'s, we introduce for convenience the notation
\begin{equation*}
\theta_i:=\al\left(\sum_{j=1}^i\gamma_j\right)=\sum_{j=1}^i p_j\, \gamma_j, \ , \quad i=1,\cdots, k 
\end{equation*}
 together with $\theta_0:=0$ by convention. We immediately obtain from \eqref{eq:defx} the expression
\begin{equation}
\begin{split}
x(t)&= \qq^{-1} \prod_{i=1}^k \left(\frac{t-\qq^{\theta_i}}{t-\qq^{\theta_{i-1}}}\right)^{\frac{1}{p_i}}\\ & =
\left\{\begin{matrix}
&\qq^{-1} \prod\limits_{i=0}^k \left(1-t^{-1}\qq^{\theta_i}\right)^{\frac{1}{p_i}-\frac{1}{p_{i+1}}}
&\ \hbox{for}\ t<0\ \ \hbox{or}\ \ \left\{\begin{matrix} t>1 & (\qq<1)\\ t>\qq^{\theta_k} & (\qq>1)\\\end{matrix}\right. \\
&\qq^{-1} \prod\limits_{i=0}^k \left(t^{-1}\qq^{\theta_i}-1\right)^{\frac{1}{p_i}-\frac{1}{p_{i+1}}}
&\ \hbox{for}\ \ \left\{\begin{matrix} 0<t<\qq^{\theta_k} 
& (\qq<1)\\ 0<t<1 & (\qq>1)\\\end{matrix}\right. \\
\end{matrix}\right. \\
\end{split}
\label{eq:xforlinear}
\end{equation}
with the convention that $p_0=p_{k+1}=\infty$. The alternative expressions of the second line emphasize that $x(t)$
is well defined and real positive for the indicated domain of $t$.   
Knowing $x(t)$, the two generic, left and right, portions of arctic curve are obtained from the general parametric expression 
\eqref{eq:arcticthm} with $t\in ]-\infty,1[\cup]\qq^{\theta_k},+\infty[$ for $\qq>1$ and $t\in ]-\infty,\qq^{\theta_k}[\cup]1,+\infty[$ for $\qq<1$ since $\al(1)=\theta_k$. 
Figure \ref{fig:pieces} gives an example of such arctic curves in some particular case with $k=3$ linear pieces, for two different values of 
$\qq$ (one larger and one smaller than $1$).
\medskip

\begin{figure}
\begin{center}
\includegraphics[width=14cm]{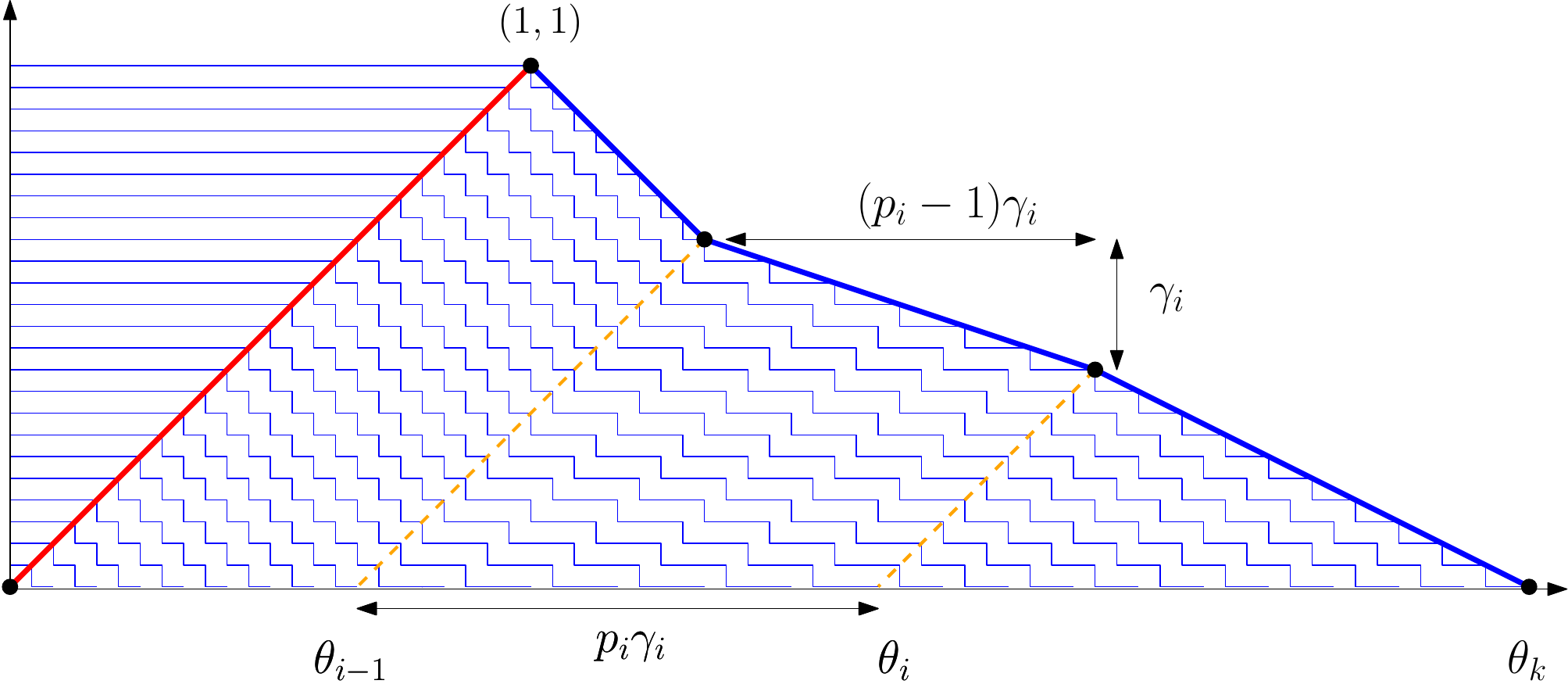}
\end{center}
\caption{\small The configuration with highest weight when $q\to 0$ for a sequence $(a_i)_{0\leq i\leq n}$ whose large $n$ limit
is a piecewise linear function $\al(u)$ as defined in the text. In rescaled coordinates, the outermost path follows a piecewise linear
curve from $(1,1)$ to $(\theta_k,0)$ made of a succession of segments of slope $-1/(p_i-1)$. Each segment is the top side of a $45^\circ$ strip
in which all the paths have the same slope as the segment. The thick red segment and the thick blue curve are the $\qq\to 0$ limit of the 
left and right parts of the arctic curve respectively.}
\label{fig:q0discret}
\end{figure}

\subsection{A look at the $q\to 0$ and $q\to \infty$ limits}
\label{sec:limitslinear}
Here again, it is interesting to have a look at the degenerate limit of the arctic curve when $q\to 0$ or $q\to \infty$. Figure 
\ref{fig:q0discret} displays the configuration selected for $q\to 0$, where each path has the smallest possible area to its left.  
This configuration is clearly made of paths which remain ``parallel" with slope $-1/(p_i-1)$ (i.e.\ are made of
a sequence of blocks consisting in $p_i-1$ west-oriented steps followed by a north-oriented step) within $45^\circ$ strips whose base are, after rescaling,
the segments $[\theta_{i-1},\theta_i]$ for $i=1,\cdots, k$.  In particular, in rescaled coordinates, the outermost path, travelled backwards, is horizontal from $(0,1)$ to $(1,1)$ and then 
follows a piecewise linear curve from $(1,1)$ to $(\theta_k,0)$ 
made of a succession of segments of slope $-1/(p_i-1)$ for $i=1,\cdots, k$. From the discussion of Section \ref{sec:limitsgen}, 
this latter curve corresponds to the $q \to 0$ limit of the right part of the arctic curve while the segment joining $(0,0)$ to $(1,1)$,
constitutes its left part. Below the arctic curve, the liquid phase 
which remains liquid as long as $q>0$, crystallizes right at $q=0$ into a sequence of $45^\circ$ macroscopic strips with a prescribed frozen path 
orientation within each strip, as displayed in figure \ref{fig:q0discret} .   

\begin{figure}
\begin{center}
\includegraphics[width=14cm]{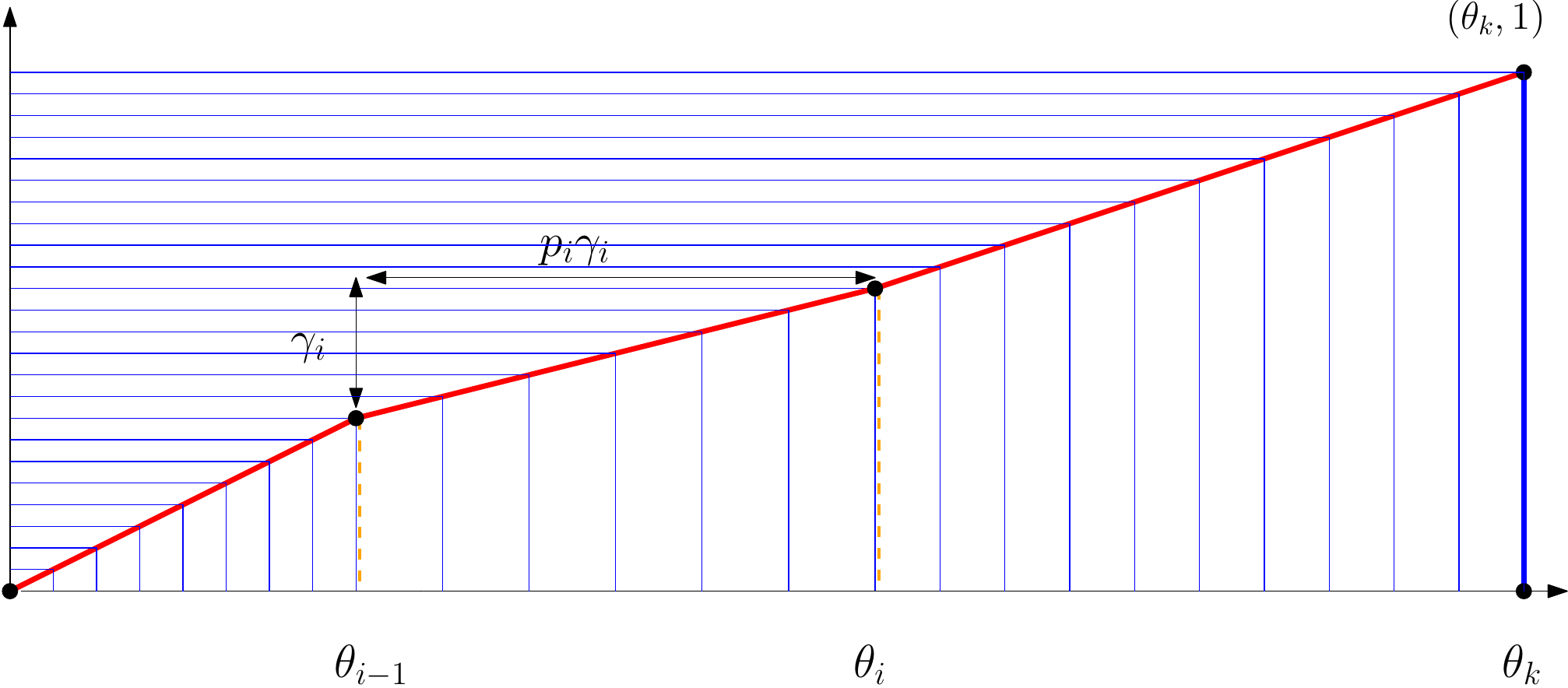}
\end{center}
\caption{\small The configuration with highest weight when $q\to \infty$ for a sequence $(a_i)_{0\leq i\leq n}$ whose large $n$ limit
is a piecewise linear function $\al(u)$ as defined in the text. Each path is made of a single vertical north-oriented segment followed by a single horizontal west-oriented segment. In rescaled 
coordinates, the location of the change from vertical to horizontal follows a piecewise linear
curve from $(0,0)$ to $(\theta_k,1)$ made of a succession of segments of slope $1/p_i$. Each segment is the top side of a vertical strip
in which all the paths are separated by the same spacing. The thick red curve and the thick blue vertical segment are the $\qq\to \infty$ limit of the 
left and right parts of the arctic curve respectively.}
\label{fig:qinfdiscret}
\end{figure}
\medskip

The $q \to \infty$ limit now selects a configuration displayed in  \ref{fig:qinfdiscret}, such that each path has the largest possible area to its left.  
This configuration is made of a vertical segments of length $i$ from $O_i$, followed by
a horizontal segments of length $a_i$ to $E_i$. In rescaled coordinates, the passage from vertical to horizontal follows a piecewise linear curve from $(0,0)$ 
to $(\theta_k,1)$ made of a succession of segments of slope $1/p_i$. This path defines the $q \to \infty$ limit of the left part of the arctic curve while 
the segment joining $(\theta_k,1)$ to $(\theta_k,0)$ now defines its right part. Here again, the liquid phase, which remains liquid as long as $q$
remains finite, is expected to crystallize right at $q=\infty$ into a sequence of macroscopic vertical strips filled with frozen vertical paths, with a prescribed path spacing 
within each strip (see figure \ref{fig:qinfdiscret}).  
\medskip 

We may also obtain the limiting shape of the arctic curve from its analytic expression, as given by \eqref{eq:arcticthm} for the particular 
$x(t)$ of equation \eqref{eq:xforlinear}, taken in the limit $\qq\to 0$ or $\qq\to \infty$. We will not present the details of this analysis since we already performed it in all
generality in Section \ref{sec:limitsgen} but we will still describe its outcome for illustration.

\begin{figure}
\begin{center}
\includegraphics[width=11cm]{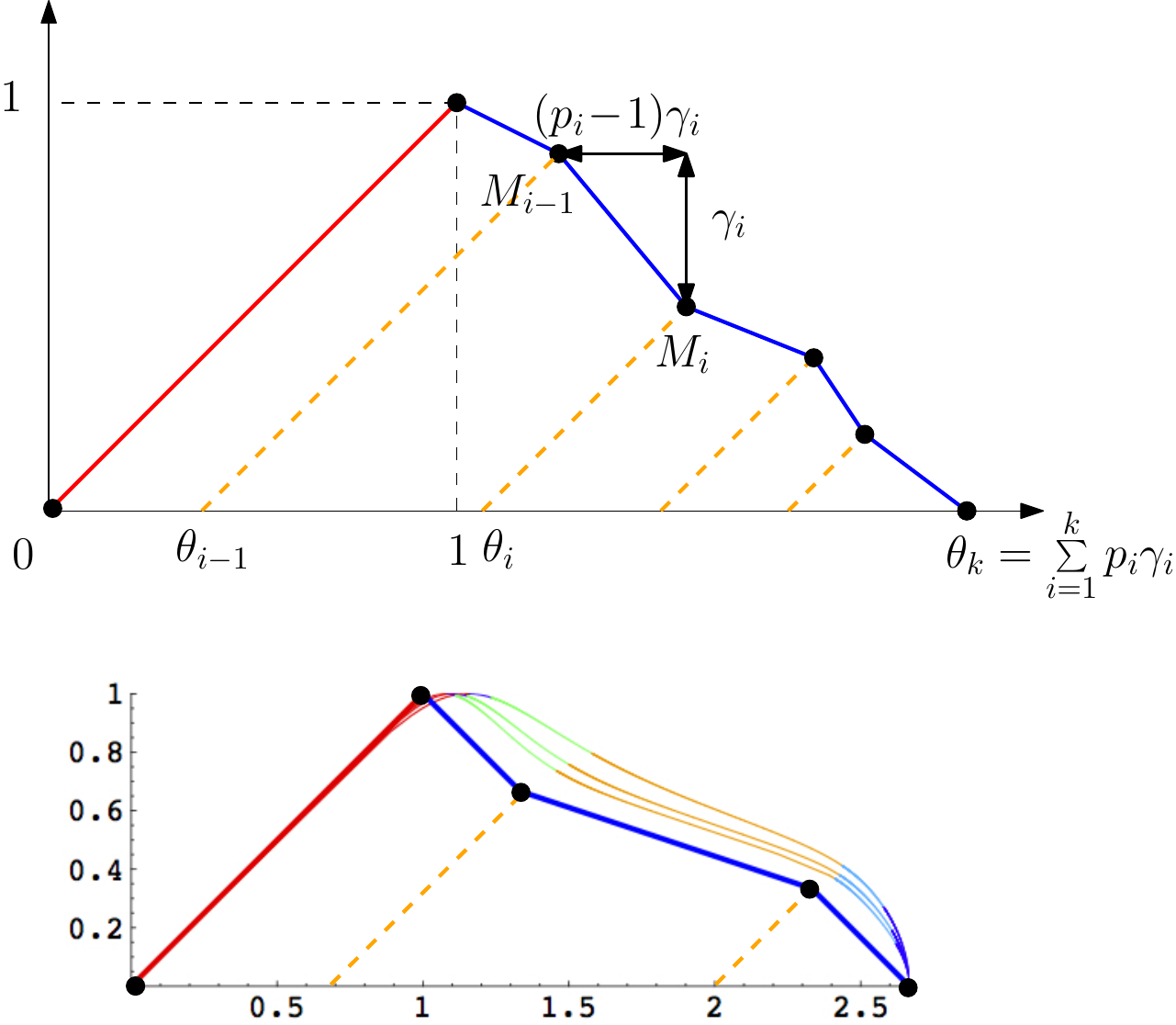}
\end{center}
\caption{\small Top: a schematic picture of the $\qq\to 0$ limiting shape of the left (red) and right (blue) parts of the arctic curve for a piecewise linear $\al(u)$. The liquid phase below the curve eventually 
crystallizes at $\qq=0$ in a configuration as in figure \ref{fig:q0discret}. Bottom: an example of approach of this limit by letting $\qq$ take smaller and smaller values (here $\qq=10^{-2}$,
$10^{-3}$ and $10^{-4}$) for
$k=3$, $\gamma_1=\gamma_2=\gamma_3=1/3$, $p_1=p_3=2$ and $p_2=4$.
The different colors of the right part correspond to the contribution of the various intervals of the parameter $t$ in the decomposition \eqref{eq:decompinter}.}
\label{fig:q0limit}
\end{figure}
 
For $\qq\to 0$, the right part of the arctic curve is obtained by letting $t$ vary in $]-\infty,\qq^{\theta_k}[$, naturally decomposed into
\begin{equation}
]-\infty,\qq^{\theta_k}[\,=\, ]-\infty,-1]\cup\left(\prod_{i=1}^k [-\qq^{\theta_{i-1}},-\qq^{\theta_{i}}] \right) \cup [-\qq^{\theta_k},\qq^{\theta_k}[\ .
\label{eq:decompinter}
\end{equation}
As we know, the two extremal subintervals $]-\infty,-1]$ and $[-\qq^{\theta_k},\qq^{\theta_k}[$ contribute only to the extremal points $(1,1)$ and
$(\theta_k,0)$ of the right part of the arctic curve, whose core is entirely created by the $k$ intermediate subintervals $[-\qq^{\theta_{i-1}},-\qq^{\theta_{i}}]$, $i=1,\cdots, k$.
From the result of Section \ref{sec:limitsgen}, we also know that each such subinterval $[-\qq^{\theta_{i-1}},-\qq^{\theta_{i}}]$ is responsible for a 
portion of arctic curve parametrized by $(\al(\tau)+1-\tau,1-\tau)$ for $\tau$ such that in $\al(\tau)\in [\theta_{i-1},\theta_i]$, i.e.\ $\tau\in [\sum_{j=1}^{i-1} \gamma_j,\sum_{j=1}^{i} \gamma_j]$. 
This now corresponds to a linear portion of arctic curve which is a segment of slope $-1/(p_i-1)$ 
joining the points $M_{i-1}$ and $M_i$ with coordinates
\begin{equation*}
M_i:=\left(1+\sum\limits_{j=1}^{i}(p_j-1) \gamma_j,1-\sum\limits_{j=1}^{i} \gamma_j\right)=\left(\theta_k-\sum\limits_{j=i+1}^{k}(p_j-1) \gamma_j,\sum\limits_{j=i+1}^{k} \gamma_j\right)\ .
\end{equation*}
The concatenation of these segments for $i=1,\cdots, k$ produces the desired piecewise linear curve from $(1,1)$ to $(\theta_k,0)$ displayed in figure \ref{fig:q0limit}. 
As for the left part of the arctic curve, it tends as we know to the segment joining $(0,0)$ to $(1,1)$.

The way the arctic curve approaches its limit is illustrated in figure
\ref{fig:q0limit} which displays in some particular case the actual arctic curves for decreasing values of $\qq$. A particular emphasis was put on the contribution
of the various subintervals so as to follow their deformation toward the associated limiting portion of arctic curve.   

\begin{figure}
\begin{center}
\includegraphics[width=11cm]{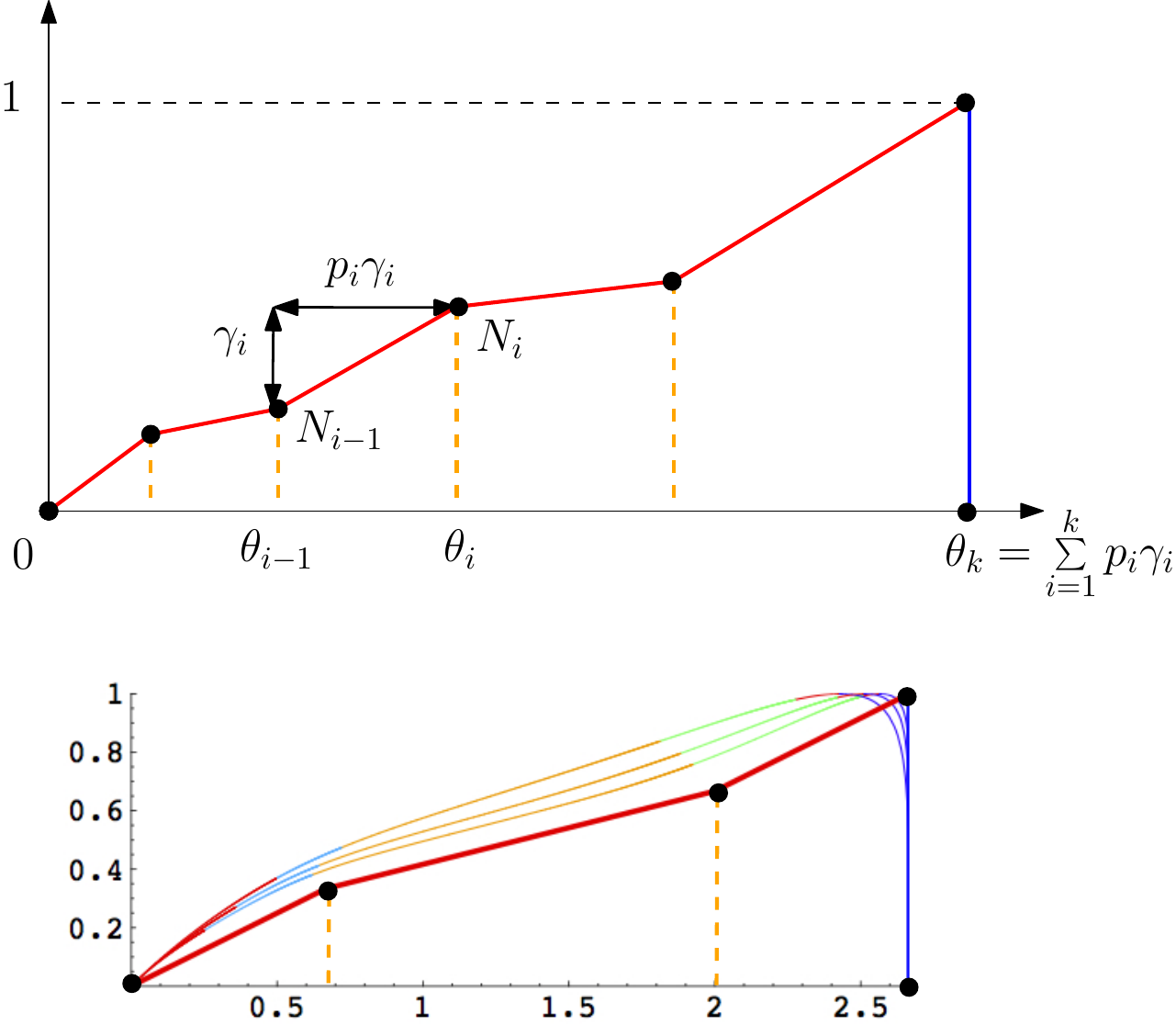}
\end{center}
\caption{\small Top: a schematic picture of the $\qq\to \infty$ limiting shape of the left (red) and right (blue) parts of the arctic curve for a piecewise linear $\al(u)$. The liquid phase below the curve eventually 
crystallizes at $\qq=\infty$ in a configuration as in figure \ref{fig:qinfdiscret}. Bottom: an example of approach of this limit by letting $\qq$ take larger and larger values (here $\qq=20$,
$100$ and $1000$) for
$k=3$, $\gamma_1=\gamma_2=\gamma_3=1/3$, $p_1=p_3=2$ and $p_2=4$.
The different colors of the left part correspond to the contribution of the various intervals of the parameter $t$ in the decomposition \eqref{eq:decompinterbis}.}
\label{fig:qinflimit}
\end{figure}
\medskip

For $\qq\to \infty$, the left part of the arctic curve is now obtained by letting $t$ vary in $]-\infty,1[$ which we may decompose into
\begin{equation}
]-\infty,1[\,=\, ]-\infty,-\qq^{\theta_k}]\cup\left(\prod_{i=1}^k [-\qq^{\theta_{i}},-\qq^{\theta_{i-1}}] \right) \cup [-1,1[\ .
\label{eq:decompinterbis}
\end{equation}
Apart from the external subintervals $]-\infty,-\qq^{\theta_k}]$ and $[-1,1[$ responsible for the extremities $(\theta_k,1)$ and $(0,0)$
of the left part if the arctic curve, the respective portions of arctic curve created by the $k$ intermediate subintervals $[-\qq^{\theta_{i}},-\qq^{\theta_{i-1}}]$, $i=1,\cdots, k$
are now parametrized by $(\al(\tau),\tau)$ for $\tau\in [\sum_{j=1}^{i-1} \gamma_j,\sum_{j=1}^{i} \gamma_j]$. These
are now segments of slope $1/p_i$ joining the points $N_{i-1}$ and $N_{i}$
with coordinates
\begin{equation*}
N_i:=\left(\theta_k-\sum\limits_{j={i+1}}^k p_j \gamma_j ,1-\sum\limits_{j={i+1}}^k \gamma_j\right)=\left(\sum\limits_{j={1}}^i p_j \gamma_j ,\sum\limits_{j=1}^i \gamma_j\right)\ .
\end{equation*}
The concatenation of these segments for $i=1,\cdots, k$ produces the desired piecewise linear curve from $(0,0)$ to $(\theta_k,1)$ displayed in figure \ref{fig:qinflimit},
while the segment joining the point $(\theta_k,1)$ to the point $(\theta_k,0)$ forms the right part of the arctic curve.
Here again, we illustrate in figure \ref{fig:qinflimit} how the arctic curve approaches its limit for increasing values of $\qq$.
As we shall now discuss, the above results still hold in the presence of frozen boundaries with $p_m=1$ or $\infty$ for some $m$, with moreover 
interesting new phenomena.

\subsection{Example of freezing boundary resulting from a fully filled interval}
\label{sec:freez1}
\begin{figure}
\begin{center}
\includegraphics[width=16cm]{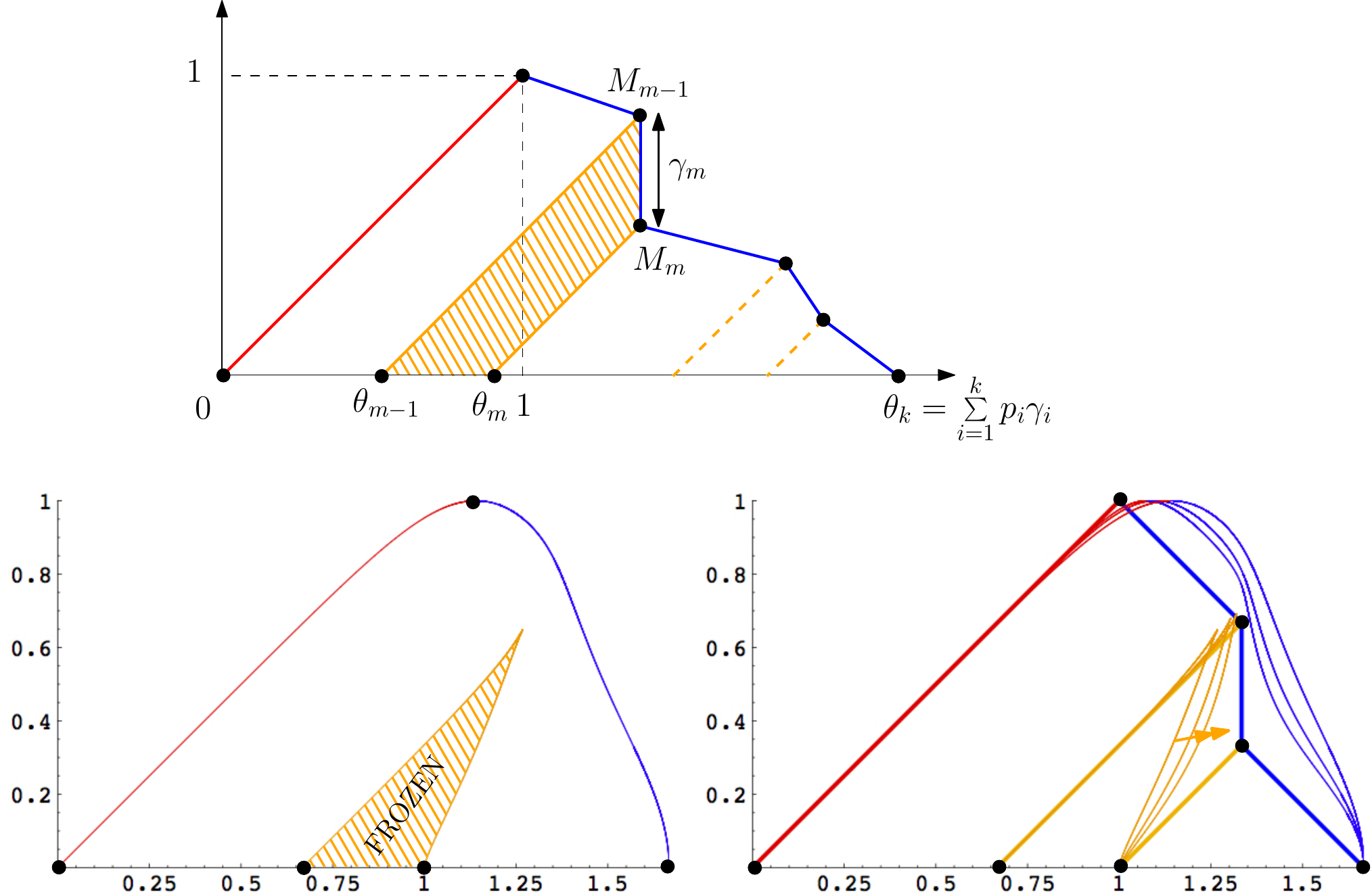}
\end{center}
\caption{\small Top: a schematic picture of the $\qq\to 0$ limiting shape of the left (red) and right (blue) parts of the arctic curve in the presence of a freezing boundary due
to a fully filled interval. The condition $p_m=1$ gives rise to a vertical segment within the right part of the curve. Bottom left: the arctic curve for finite $\qq$ (here  $\qq=10^{-2}$)
also has a new (orange) portion below which the paths are frozen (represented here for $k=3$, $m=2$ with  $\gamma_1=\gamma_2=\gamma_3=1/3$, $p_1=p_3=2$ and $p_2=1$). 
Bottom right: for decreasing $\qq$ (here $\qq=10^{-2}$, $10^{-3}$ and $10^{-5}$), the frozen phase fills the $45^\circ$ strip whose edge is the above vertical segment (dashed region
in the top figure).}
\label{fig:q0freez1}
\end{figure}
\begin{figure}
\begin{center}
\includegraphics[width=16cm]{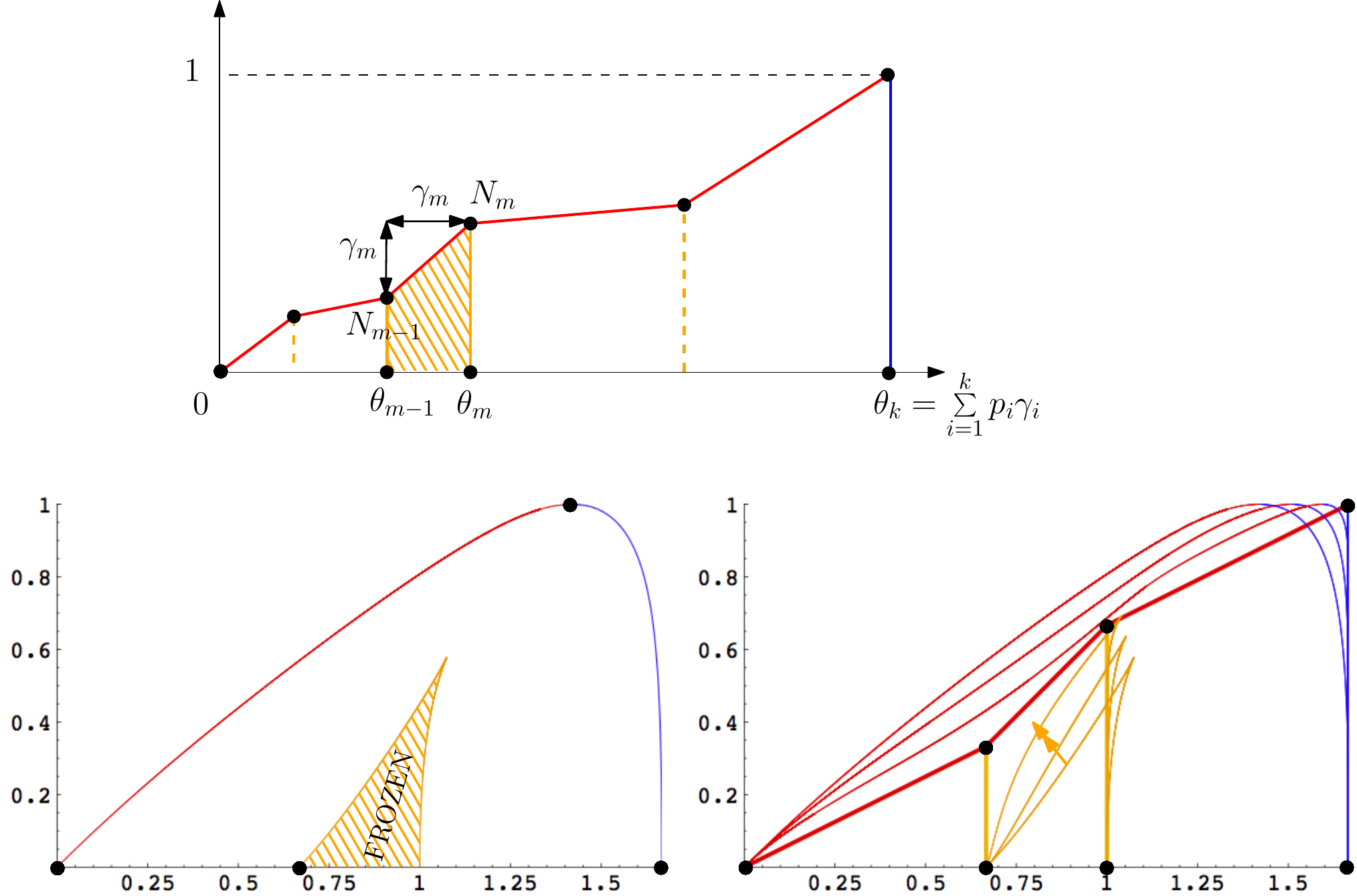}
\end{center}
\caption{\small Top: a schematic picture of the $\qq\to \infty$ limiting shape of the left (red) and right (blue) parts of the arctic curve in the presence of a freezing boundary due
to a fully filled interval. The condition $p_m=1$ gives rise to a $45^\circ$ segment within the left part of the curve. Bottom left: the arctic curve for finite $\qq$ (here  $\qq=5$)
also has a new (orange) portion below which the paths are frozen (represented here for $k=3$, $m=2$ with  $\gamma_1=\gamma_2=\gamma_3=1/3$, $p_1=p_3=2$ and $p_2=1$). 
Bottom right: for increasing $\qq$ (here $\qq=5$, $50$ and $10000$), the frozen phase fills the vertical strip whose upper edge is the above $45^\circ$ segment (dashed region
in the top figure).}
\label{fig:qinffreez1}
\end{figure}

The case of a freezing boundary resulting from a fully filled interval is encountered within the framework of a piecewise liner $\al(u)$ 
in the particular case where $p_m=1$ for some $m$ in $\llbracket1,k\rrbracket$. Here we assume for simplicity that $m\neq1$ and $m\neq k$.
The case $p_1=1$ (respectively $p_k=1$), referred to as "freezing the left (respectively the right) edge" in \cite{DFGUI}, is indeed special and would deserve a more subtle treatment. 
For $p_m=1$, the expression \eqref{eq:xforlinear} is now well defined for $t\in ]\qq^{\theta_{m-1}},\qq^{\theta_m}[$ whenever $\qq>1$
(respectively $t\in ]\qq^{\theta_{m}},\qq^{\theta_{m-1}}[$ whenever $\qq<1$), with expression
\begin{equation}
x(t)=
\left\{\begin{matrix} - \qq^{-1} \prod\limits_{i=1}^{m-1} \left(\frac{t-\qq^{\theta_i}}{t-\qq^{\theta_{i-1}}}\right)^{\frac{1}{p_i}}\times \left(\frac{\qq^{\theta_m}-t}{t-\qq^{\theta_{m-1}}}\right) \times\prod\limits_{i=m+1}^k \left(\frac{\qq^{\theta_i}-t}{\qq^{\theta_{i-1}}-t}\right)^{\frac{1}{p_i}} & \qq>1\\
- \qq^{-1} \prod\limits_{i=1}^{m-1} \left(\frac{\qq^{\theta_i}-t}{\qq^{\theta_{i-1}}-t}\right)^{\frac{1}{p_i}}\times \left(\frac{t-\qq^{\theta_m}}{\qq^{\theta_{m-1}}-t}\right) \times\prod\limits_{i=m+1}^k \left(\frac{t-\qq^{\theta_i}}{t-\qq^{\theta_{i-1}}}\right)^{\frac{1}{p_i}} & \qq<1
\end{matrix}\right.
\label{eq:xforfreez1}
\end{equation}
displaying its negative real value. This in turn creates for finite $\qq$ a new portion of arctic curve emerging above the segment $[\theta_{m-1},\theta_{m}]$ (see for instance the bottom left part of 
figure \ref{fig:q0freez1} or \ref{fig:qinffreez1}) below which the path configuration is frozen. 

Looking at the $\qq\to 0$ limit, the discussion of the previous section still holds\footnote{The actual calculation when $p_m=1$ is slightly more subtle than for for $p_m\neq 1$ since, when estimating 
$\qq^{Y(t)}$ via \eqref{eq:arcticthm}, the dominant part 
$t\, x'(t)-x(t)$ of its numerator cancels exactly at leading order and the calculation must be pushed to the next order (see a similar discussion just below). The corresponding portion of arctic curve is nevertheless not affected by this subtlety.} and now leads for the right part of the arctic curve to a portion with slope $-1/(p_m-1)=-\infty$, 
i.e.\ a vertical segment joining $M_{m-1}$ to $M_m$ (which now have the same $X$-coordinate $1+\sum_{j=0}^{m-1}(p_j-1) \gamma_j$).
More interestingly, the new frozen region below the new portion of arctic curve is deformed so as to fill entirely the $45^\circ$ strip whose edge is the 
above vertical segment (dashed domain in figure \ref{fig:q0freez1}). To understand this property, we start by parametrizing $t\in ]+\qq^{\theta_{m}},+\qq^{\theta_{m-1}}[$ as 
$t=\qq^\tau$ with $\tau\in]\theta_{m-1},\theta_{m}[$ and plug this value
in \eqref{eq:xforfreez1}. This yields
\begin{equation*}
\begin{split}
x(t)&\underset{q\to 0}{\sim}-\qq^{-1+\sum_{i=1}^{m-1}\frac{\theta_i-\theta_{i-1}}{p_i}+\tau-\theta_{m-1}}\left(1+O\left(\qq^{\min(\tau-\theta_{m-1},\theta_m-\tau)}\right)\right)
\\ &=-\qq^{\sum_{i=1}^{m-1}\gamma_i-1+\tau-\theta_{m-1}}\left(1+O\left(\qq^{\min(\tau-\theta_{m-1},\theta_m-\tau)}\right)\right)\to \infty\ .
\end{split}
\end{equation*}
This also implies $t\, x'(t)\sim -\qq^{\sum_{i=1}^{m-1}\gamma_i-1+\tau-\theta_{m-1}}\left(1+O\left(\qq^{\min(\tau-\theta_{m-1},\theta_m-\tau)}\right)\right)$ so that 
$t\, x'(t)-x(t)\sim O\left(\qq^{\sum_{i=1}^{m-1}\gamma_i-1+\tau-\theta_{m-1}+\min(\tau-\theta_{m-1},\theta_m-\tau)}\right)$ which tends to infinity
since the exponent varies between $\sum_{i=1}^{m-1}\gamma_i-1$ and $\sum_{i=1}^{m}\gamma_i-1$ which are both negative.
We deduce
\begin{equation*}
\begin{split}
\qq^{X(t)}&\sim \frac{t^2x'(t)}{-(x(t))^2}\sim \qq^{1-\sum_{i=1}^{m-1}\gamma_i+\theta_{m-1}}\\
\qq^{Y(t)}&\sim \frac{t\, x'(t)-x(t)}{-(x(t))^2}\sim \qq^{1-\sum_{i=1}^{m-1}\gamma_i-\tau +\theta_{m-1}+\min(\tau-\theta_{m-1},\theta_m-\tau)}\ , \\
\end{split}
\end{equation*}
hence 
\begin{equation*}
\begin{split}
X(t)&= 1-\sum_{i=1}^{m-1}\gamma_i+\theta_{m-1} = 1-\sum_{i-1}^{m-1}(p_i-1)\gamma_i \\
Y(t)&=1-\sum_{i=1}^{m-1}\gamma_i+\theta_{m-1}-\tau+\min(\tau-\theta_{m-1},\theta_m-\tau)\\
&=1-\sum_{i=1}^{m-1}\gamma_i+\min(0,\theta_{m-1}+\theta_m-2\tau) \\
\end{split}
\end{equation*}
with $\min(0,\theta_{m-1}+\theta_m-2\tau)$ varying from $0$ to $\theta_{m-1}-\theta_m=-\gamma_m$. This curve is precisely the vertical segment $[M_{m-1},M_m]$ on the right of the dashed domain in
figure \ref{fig:q0freez1}. The new portion of arctic curve therefore sticks to this segment when $\qq\to 0$ but this should still be reconciled with the fact that for 
$t$ \emph{exactly equal to} $\qq^{\theta_m}$ (respectively to 
$\qq^{\theta_{m-1}}$), we have $(X(t),Y(t))= (\theta_m,0)$ (respectively $(\theta_{m-1},0)$), as easily verified from \eqref{eq:xforfreez1} and \eqref{eq:arcticthm}. As we shall now see, the connection
from these points to the segment  $[M_{m-1},M_m]$ is done by the two segments at $45^\circ$ which delimit the dashed domain of figure \ref{fig:q0freez1}. These new segments
arise from values of $t$ in the immediate vicinity of $\qq^{\theta_m}$ (respectively of $\qq^{\theta_{m-1}}$) which are not treated properly by the above estimate. 
For $t\to \qq^{\theta_m}$, a more precise estimate of $x(t)$ is
\begin{equation*}
x(t)\underset{q\to 0}{\sim}-\qq^{-1+\sum_{i=1}^{m-1}\frac{\theta_i-\theta_{i-1}}{p_i}}\left(\frac{t-\qq^{\theta_m}}{\qq^{\theta_{m-1}}}\right) 
\left(\frac{\qq^{\theta_m}}{t-\qq^{\theta_{m}}}\right)^{\frac{1}{p_{m+1}}} =-\qq^{\sum_{i=1}^{m}\gamma_i-1}\left(\frac{t-\qq^{\theta_m}}{\qq^{\theta_{m}}}\right)^{1-\frac{1}{p_{m+1}}} \end{equation*}
which allows to view the contribution of the immediate vicinity of $\qq^{\theta_m}$ by setting $t=\qq^{\theta_m}(1+\qq^\rho)$ for some positive $\rho$.
After some straightforward manipulations, this yields
\begin{equation*}
X(t)=\theta_m+\max\left(1-\sum_{i=1}^{m}\gamma_i -\rho\frac{2p_{m+1}-1}{p_{m+1}},0\right)\ , \qquad Y(t)=\max\left(1-\sum_{i=1}^{m}\gamma_i -\rho\frac{2p_{m+1}-1}{p_{m+1}},0\right)
\end{equation*}
which is the segment from $(\theta_m,0)$ (for $\rho=(1-\sum_{i=1}^{m}\gamma_i )p_{m+1}/(2p_{m+1}-1)$ or larger) to $M_m=(\theta_m+1-\sum_{i=1}^{m}\gamma_i,1-\sum_{i=1}^{m}\gamma_i)$ (for $\rho=0)$. In other words, the immediate vicinity $t=\qq^{\theta_m}$ produces the $45^\circ$ lower segment bordering the frozen dashed region in figure
\ref{fig:q0freez1}. A similar analysis for the immediate vicinity of $\qq^{\theta_{m-1}}$ would now produce the $45^\circ$ upper segment bordering the frozen region and connecting
$(\theta_{m-1},0)$ to $M_{m-1}$.

The fact that the new portion of arctic curve and the right part merge along the vertical segment $[M_{m-1},M_m]$
when $\qq\to 0$ means that the liquid phase narrows and forms a strait around the segment for very small $\qq$ (see figure \ref{fig:q0freez1}, bottom right) before it eventually crystallizes right at $\qq=0$.
\medskip 

The discussion of the $\qq\to \infty$ limit is quite similar and now leads for the left part of the arctic curve to a portion with slope $1/p_m=1$, 
i.e.\ a $45^\circ$ segment joining $N_{m-1}$ to $N_m$. More interestingly, the new frozen region below the new portion of arctic curve is now deformed so as 
to fill entirely the vertical strip below $[N_{m-1},N_m]$ (dashed domain in figure \ref{fig:qinffreez1}). The new portion of arctic curve and the left part therefore merge along the segment $[N_{m-1},N_m]$
when $\qq\to \infty$. In other words, the liquid phase narrows around the segment for very large $\qq$ (see figure \ref{fig:qinffreez1}, bottom right) before it eventually 
crystallizes right at $\qq=\infty$.

\subsection{Example of freezing boundary resulting from a gap}
\label{sec:freez2}
\begin{figure}
\begin{center}
\includegraphics[width=11cm]{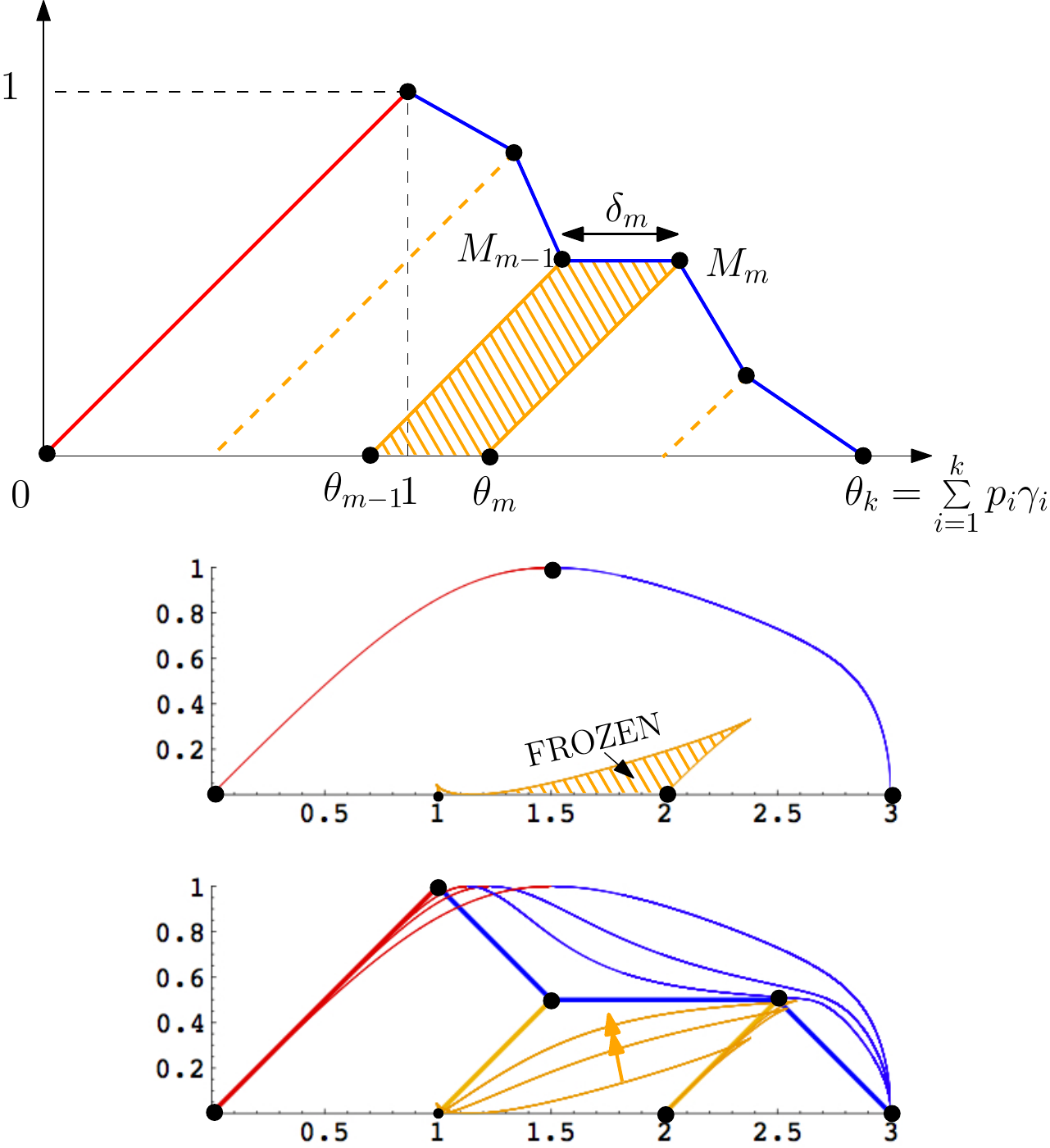}
\end{center}
\caption{\small Top: a schematic picture of the $\qq\to 0$ limiting shape of the left (red) and right (blue) parts of the arctic curve in the presence of a freezing boundary due
to a gap. The condition $p_m=\infty=\delta_m/\gamma_m$ (with $\delta_m$ finite) gives rise to a horizontal segment within the right part of the curve. Middle: the arctic 
curve for finite $\qq$ (here $\qq=.3$) also has a new (orange) portion below which the paths are frozen (represented here for $k=3$, $m=2$ with  $\gamma_1=\gamma_3=1/2$, $\gamma_2\to 0$, 
$p_1=p_3=2$ and $p_2\to \infty$ with $p_2\gamma_2\to \delta_2=1$). 
Bottom: for decreasing $\qq$ (here $\qq=.3$, $.05$ and $.005$), the frozen phase fills the $45^\circ$ strip whose edge is the above horizontal segment (dashed region
in the top figure).}
\label{fig:q0freez2}
\end{figure}
\begin{figure}
\begin{center}
\includegraphics[width=11cm]{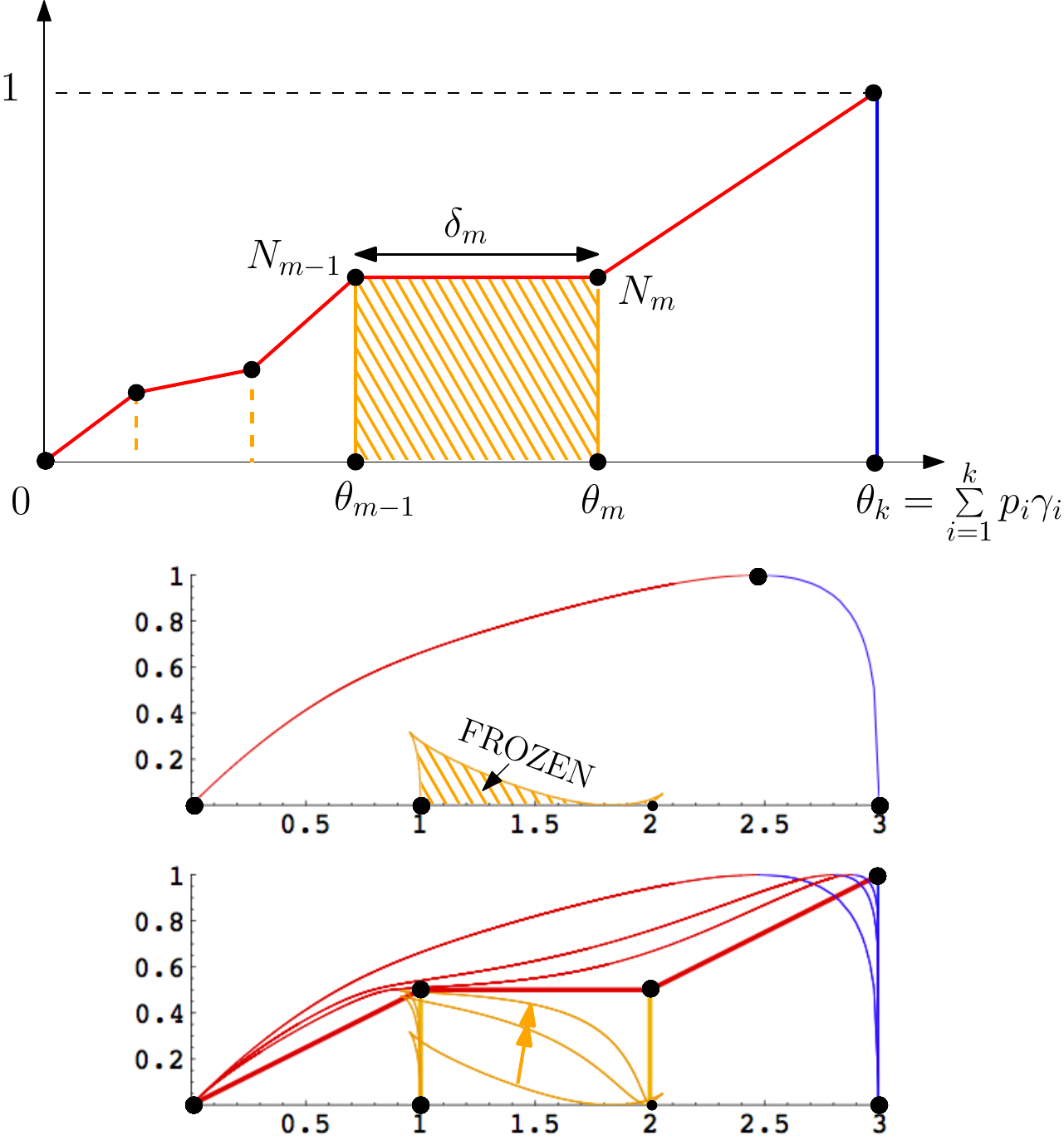}
\end{center}
\caption{\small Top: a schematic picture of the $\qq\to \infty$ limiting shape of the left (red) and right (blue) parts of the arctic curve in the presence of a freezing boundary due
to a gap. The condition $p_m=\infty=\delta_m/\gamma_m$ (with $\delta_m$ finite) gives rise to a horizontal segment within the left part of the curve. Middle: the arctic curve 
for finite $\qq$ (here $\qq=3$) also has a new (orange) portion below which the paths are frozen (represented here for $k=3$, $m=2$ with  $\gamma_1=\gamma_3=1/2$, $\gamma_2\to 0$, 
$p_1=p_3=2$ and $p_2\to \infty$ with $p_2\gamma_2\to \delta_2=1$). 
Bottom: for increasing $\qq$ (here $\qq=3$, $30$ and $300$), the frozen phase fills the vertical strip whose upper edge is the above horizontal segment (dashed region
in the top figure).}
\label{fig:qinffreez2}
\end{figure}

The case of a freezing boundary resulting from a gap is also encountered within the framework of a piecewise linear $\al(u)$, 
now in the case where $p_m\to \infty$, $\gamma_m\to 0$ with $\delta_m=p_m\gamma_m$ finite, for some $m$ 
in $\llbracket2,k-1\rrbracket$ (again we avoid the cases $m\neq1$ and $m\neq k$ which are more subtle). 
For $p_m=\infty$, the expression \eqref{eq:xforlinear} is well defined also for $t\in ]\qq^{\theta_{m-1}},\qq^{\theta_m}[$ whenever $\qq>1$
(respectively $t\in ]\qq^{\theta_{m}},\qq^{\theta_{m-1}}[$ whenever $\qq<1$), with expression
\begin{equation*}
x(t)=
\left\{\begin{matrix}  \qq^{-1} \prod\limits_{i=1}^{m-1} \left(\frac{t-\qq^{\theta_i}}{t-\qq^{\theta_{i-1}}}\right)^{\frac{1}{p_i}}\times\prod\limits_{i=m+1}^k \left(\frac{\qq^{\theta_i}-t}{\qq^{\theta_{i-1}}-t}\right)^{\frac{1}{p_i}} & \qq>1\\
\qq^{-1} \prod\limits_{i=1}^{m-1} \left(\frac{\qq^{\theta_i}-t}{\qq^{\theta_{i-1}}-t}\right)^{\frac{1}{p_i}} \times\prod\limits_{i=m+1}^k \left(\frac{t-\qq^{\theta_i}}{t-\qq^{\theta_{i-1}}}\right)^{\frac{1}{p_i}} & \qq<1
\end{matrix}\right.
\end{equation*}
displaying its positive real value. As before, this creates for finite $\qq$ a new portion of arctic curve emerging above the segment $[\theta_{m-1},\theta_{m}]$ (see for instance the middle part of 
figure \ref{fig:q0freez2} or \ref{fig:qinffreez2}) below which the path configuration is frozen. 

When $\qq\to 0$, our general discussion now leads for the right part of the arctic curve to a portion with slope $-1/(p_m-1)=0$, 
i.e.\ a horizontal segment joining $M_{m-1}$ to $M_m$ (which now have the same $Y$-coordinate $1-\sum_{j=0}^{m-1}\gamma_j$ but $X$-coordinates which
differ by $\delta_m$). As for the new frozen region below the new portion of arctic curve, it is now deformed so as to fill entirely the $45^\circ$ strip whose edge is the 
above horizontal segment (dashed domain in figure \ref{fig:q0freez2}). In particular, the new portion of arctic curve and the right part merge along the horizontal segment $[M_{m-1},M_m]$
when $\qq\to 0$, and the liquid phase narrows around the segment for very small $\qq$ (see figure \ref{fig:q0freez2}, bottom) before it eventually crystallizes right at $\qq=0$.

The $\qq\to \infty$ limit is similar: the left part of the arctic curve now has a portion with slope $1/p_m=0$, 
i.e.\ a horizontal segment joining $N_{m-1}$ to $N_m$. The new frozen region below the new portion of arctic curve is deformed so as 
to fill entirely the vertical strip below $[N_{m-1},N_m]$ (dashed domain in figure \ref{fig:qinffreez2}). In particular, the new portion of arctic curve and the left part merge along the horizontal segment 
$[N_{m-1},N_m]$ when $\qq\to \infty$, meaning once again that the liquid phase narrows around the segment for very large $\qq$ (see figure \ref{fig:qinffreez2}, bottom) before it eventually 
crystallizes right at $\qq=\infty$.

\subsection{$q$-deformation of the ellipse}

\label{sec:ellipse}
\begin{figure}
\begin{center}
\includegraphics[width=11cm]{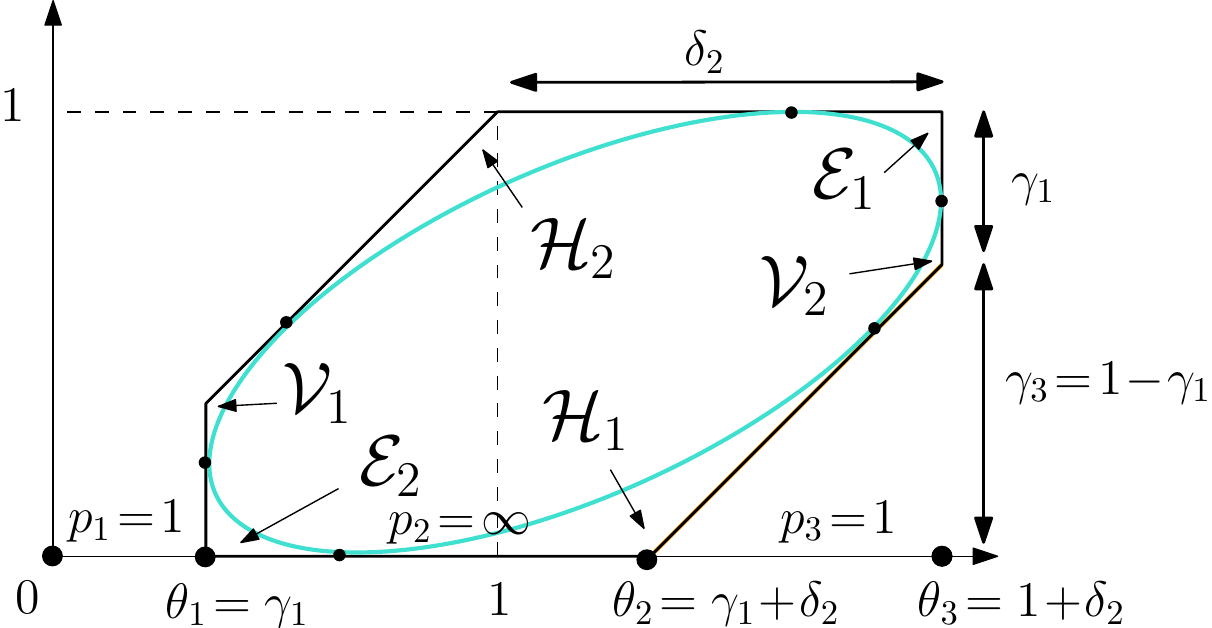}
\end{center}
\caption{\small The phase diagram of NILP configurations for a piecewise linear function $\al(u)$ with $k=3$, $p_1=p_3=1$, $p_2\to \infty $, $\gamma_2\to 0$
with $p_2\gamma_2\to \delta_2$. The paths are frozen \emph{by construction} outside the indicated hexagon with vertical and horizontal sides of respective lengths
$\gamma_1$ and $\delta_2$. At $\qq=1$, the frozen domain extends inside the hexagon and is separated from a central liquid phase by an arctic curve 
whose shape is an ellipse tangent to the six sides of the hexagon. The six regions in-between the hexagon and the ellipse are either empty of all paths
(regions $\mathcal{E}_1$ and $\mathcal{E}_2$), filled with horizontal paths ($\mathcal{H}_1$ and $\mathcal{H}_2$) or filled with vertical paths  ($\mathcal{V}_1$ and $\mathcal{V}_2$).}
\label{fig:ellipseconfig}
\end{figure}

Another interesting and quite studied geometry corres\-onds to paths connecting the opposite sides of a hexagon, which is nothing but 
the path formulation of the classical rhombus tiling problem of a hexagonal domain \cite{CLP}.
This geometry is obtained in our setting by taking an entirely freezing boundary
with a sequence $(a_i)_{1\leq i\leq n}$ made of \emph{two fully filled intervals} of width $\Delta'_1$ and $\Delta'_3=n-\Delta'_1-1$ (so that the total number of paths is $(\Delta'_1+1)+(\Delta'_3+1)=n+1$)\emph{separated by a gap} of width $\Delta_2$.  Using the original path formulation, it is easily seen that 
the paths are in practice frozen outside a hexagon (of total height $n$) with pairwise parallel sides
oriented respectively vertically (with height $\Delta'_1$), horizontally (with width $\Delta_2$) and at $45^\circ$. In other words, the domain $D$ where fluctuations may arise is reduced 
in practice from its
original rectangular shape to a smaller effective domain $D'$ with the above hexagonal geometry. The non-frozen part of the NILP corresponds moreover to a set of $\Delta'_3+1$ paths
whose origins span all the vertices of the rightmost $45^\circ$ side of the hexagon and whose endpoints span all the vertices of the opposite (leftmost $45^\circ$) side.

This situation corresponds after scaling to a piecewise linear function $\al(u)$ as above with $k=3$, $p_1=p_3=1$, $p_2\to \infty $ and $\gamma_2\to 0$
with $p_2\gamma_2=\delta_2$ finite\footnote{Note that this is a situation with a slope $1$ for both the first and the last linear piece. As already mentioned, this 
case steps outside the generic treatment of Sections~\ref{sec:limitslinear}, \ref{sec:freez1} and \ref{sec:freez2}.}. The resulting model therefore depends on two geometrical degrees of freedom $\gamma_1=1-\gamma_3$ and $\delta_2$,
which correspond respectively to 
the length of the vertical and horizontal sides of the hexagon after rescaling (see figure \ref{fig:ellipseconfig}).
At $q=1$ (i.e. $\qq=1$), the frozen domain extends inside the hexagon and surrounds a central liquid phase. The shape of the separating arctic curve is
then an ellipse tangent to the six sides of the hexagon (see for instance \cite{Eynard} for a matrix model derivation or \cite{DFGUI} for a tangent method derivation). The domain lying in-between the hexagon and the ellipse
is split into six parts: two opposite parts $\mathcal{E}_1$ and $\mathcal{E}_2$ correspond to regions empty of all paths, 
two opposite parts $\mathcal{H}_1$ and $\mathcal{H}_2$ correspond to regions filled with horizontal paths and
two opposite parts $\mathcal{V}_1$ and $\mathcal{V}_2$ correspond to regions filled with vertical paths (see figure \ref{fig:ellipseconfig}).
Let us now discuss how these regions evolve whenever $\qq$ decreases to $0$ or increases to $\infty$.

\begin{figure}
\begin{center}
\includegraphics[width=16cm]{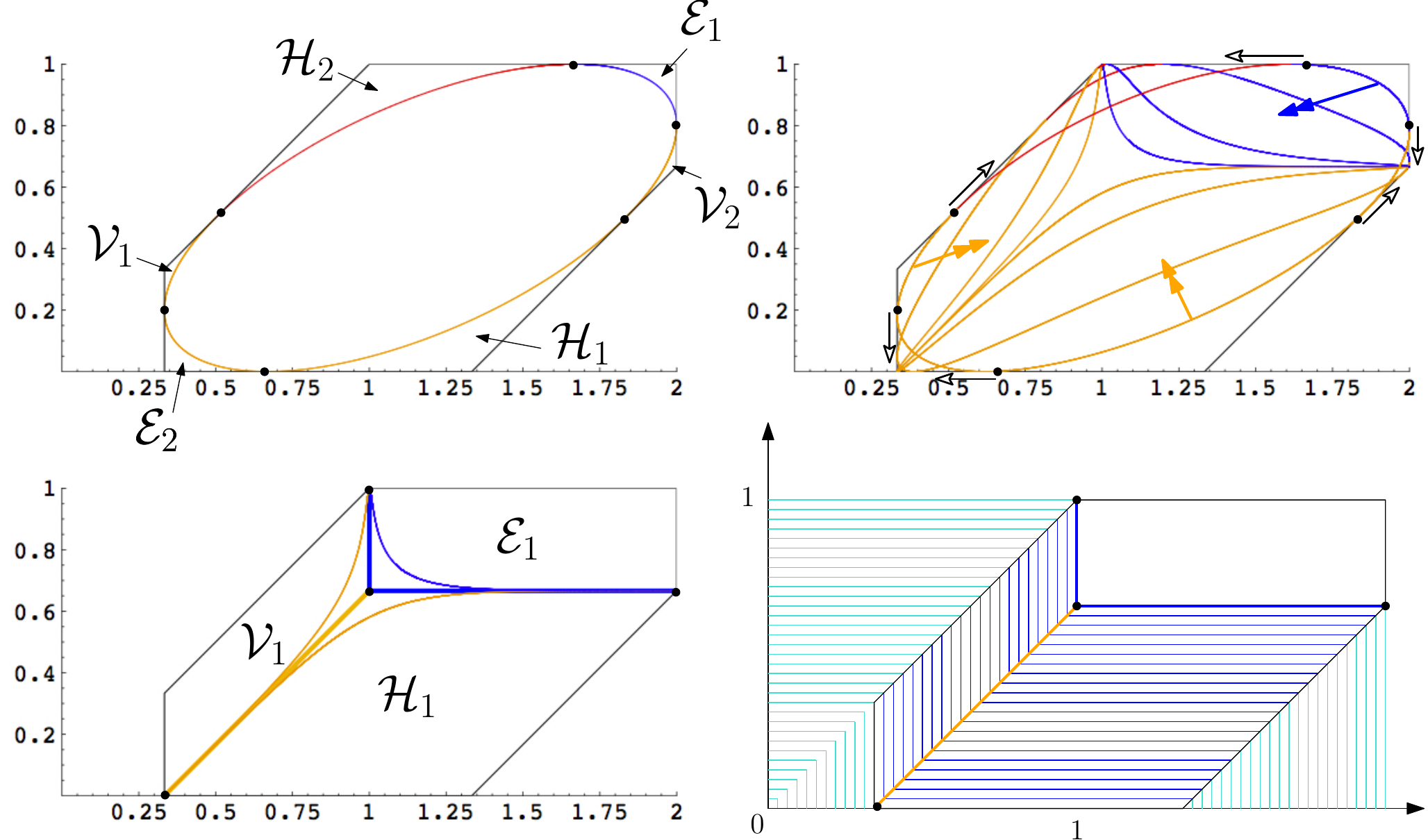}
\end{center}
\caption{\small Deformation of the arctic curve of figure \ref{fig:ellipseconfig} (here for $\gamma_1=1/3$ and $\delta_2=1$) when $\qq\to 0$. Starting from an ellipse at $\qq=1$ (top left) the boundary
of the three domains $\mathcal{E}_2$, $\mathcal{H}_2$ and $\mathcal{V}_2$ are pushed towards the associated hexagon corners while that of
the three domains $\mathcal{E}_1$, $\mathcal{H}_1$ and $\mathcal{V}_1$ are pushed to a central point with coordinates $(1,1-\gamma_1)$ (top right
with $\qq=.8$, $10^{-1}$, $10^{-3}$ and $10^{-7}$)
so that the liquid phase shrinks and reduces to the three indicated segments (bottom left). This splitting of the hexagon into three domains  $\mathcal{E}_1$, $\mathcal{H}_1$ and $\mathcal{V}_1$
is consistent with the $\qq\to0$ most probable configuration (bottom right) where the paths are pushed as much as possible towards the lower left corner. Note 
that the path configuration outside the hexagon (light blue) is frozen by construction for any value of $\qq$. The colors of the arctic curve refer to the domain of variation of the parameter $t$, namely
$]-\infty,\qq^{1+\delta_2}[$ (blue), $]\qq^{1+\delta_2},1[$ (orange) and $]1,+\infty[$ (red).}
\label{fig:q0ellipsedeform}
\end{figure}

The function $x(t)$ describing the situation at hand reads:
\begin{equation*}
x(t)=\qq^{-1}\frac{\left(t-\qq^{\gamma_1}\right)\left(t-\qq^{1+\delta_2}\right)}{\left(t-1\right)\left(t-\qq^{\gamma_1+\delta_2}\right)}
\end{equation*}
and we may easily plot the corresponding arctic curve obtained via \eqref{eq:arcticthm}.

For decreasing $\qq$, the tangency points of the ellipse with the hexagon are found to merge by pairs at three (pairwise non-consecutive) corners of the hexagon 
as indicated in figure \ref{fig:q0ellipsedeform}, so that the three domains $\mathcal{E}_2$, $\mathcal{H}_2$ and $\mathcal{V}_2$
get smaller and eventually disappear when $\qq\to 0$. On the contrary the three domains $\mathcal{E}_1$, $\mathcal{H}_1$ and $\mathcal{V}_1$
inflate so as to invade the liquid phase which reduces when $\qq\to 0$ to the union of three segments $[(\gamma_1,0),(1,1-\gamma_1)]$,
$[(1,1-\gamma_1),(1,1)]$ and $[(1,1-\gamma_1),(1+\delta_2,1-\gamma_1)]$. This splitting of the hexagon in three frozen domains is fully consistent with the path configuration selected
right at $\qq=0$ in which paths are pushed as much as possible towards the lower left corner (see figure  \ref{fig:q0ellipsedeform}).

\begin{figure}
\begin{center}
\includegraphics[width=16cm]{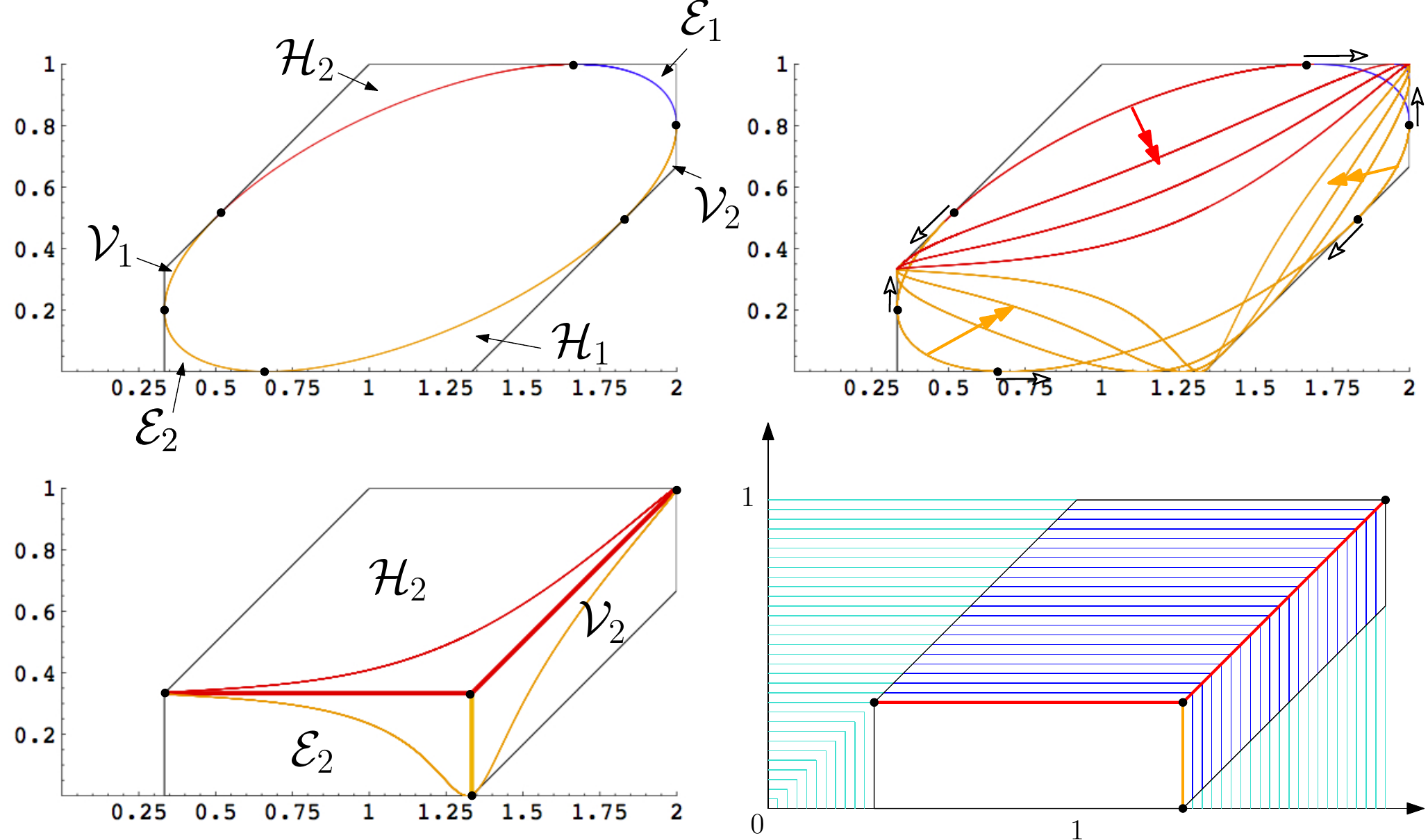}
\end{center}
\caption{\small Deformation of the arctic curve of figure \ref{fig:ellipseconfig} (here for $\gamma_1=1/3$ and $\delta_2=1$) when $\qq\to \infty$. Starting from an ellipse at $\qq=1$ (top left) the boundary
of the three domains $\mathcal{E}_1$, $\mathcal{H}_1$ and $\mathcal{V}_1$ are pushed towards the associated hexagon corners while that of
the three domains $\mathcal{E}_2$, $\mathcal{H}_2$ and $\mathcal{V}_2$ are pushed to a central point with coordinates $(\gamma_1+\delta_2,\gamma_1)$ (top right
with $\qq=1.1$, $10$ , $50$ and $1000$)
so that the liquid phase shrinks and reduces to the three indicated segments (bottom left). This splitting of the hexagon into three domains  $\mathcal{E}_2$, $\mathcal{H}_2$ and $\mathcal{V}_2$
is consistent with the $\qq\to\infty$ most probable configuration (bottom right) where the paths are pushed as much as possible towards the upper right corner. 
The colors of the arctic curve refer to the domain of variation of the parameter $t$, namely
$]\qq^{1+\delta_2},+\infty[$ (blue), $]1,\qq^{1+\delta_2}[$ (orange) and $]-\infty,1[$ (red).}
\label{fig:qinfellipsedeform}
\end{figure}

For increasing $\qq$, the tangency points of the ellipse with the hexagon merge by pairs at the three complementary corners of the hexagon 
as indicated in figure \ref{fig:qinfellipsedeform}, so that these are now the three domains $\mathcal{E}_1$, $\mathcal{H}_1$ and $\mathcal{V}_1$
which get smaller and eventually disappear when $\qq\to \infty$. On the contrary the three domains $\mathcal{E}_2$, $\mathcal{H}_2$ and $\mathcal{V}_2$
inflate, letting the liquid phase reduce when $\qq\to \infty$ to the union of three segments $[(\gamma_1,\gamma_1),(\gamma_1+\delta_2,\gamma_1)]$,
$[(\gamma_1+\delta_2,0),(\gamma_1+\delta_2,\gamma_1))]$ and $[(\gamma_1+\delta_2,\gamma_1),(1+\delta_2,1)]$. This is now fully consistent with the path configuration selected
right at $\qq=\infty$ in which paths are pushed as much as possible towards the upper right corner (see figure  \ref{fig:qinfellipsedeform}).

\section{Conclusion and discussion}
\label{sec:conclusion}

To conclude this paper, let us make a few comments both on the tangent method itself and on its specific results in the present model. 

First, we wish to stress the flexibility of the method, whose implementation for an arbitrary $q$ is not different from what it was at $q=1$.
In particular, the various technical tricks, such as the use of LGV matrices or that of the LU decomposition of \cite{DFLAP} work perfectly.
As a result, the various discrete formulas for the partition function or the one-point function are natural $q$-analogs of
their $q=1$ counterparts computed in \cite{DFGUI} and could have been predicted by some educated guess. Note also that,
after scaling,  the fact that the geodesic
trajectories (whose envelope gives the arctic curve) are not  straight lines is actually not a problem, since the tangency principle underlying 
the method concerns only the splitting point where the perturbed outermost path changes its trajectory. 

In our solution, the way the arctic curve evolves upon varying $q$ is quite interesting, in particular when $q$ becomes either very small or very large.
In a generic case without freezing boundary, the arctic curve is made of only two portions, its right and left parts, which are smoothly deformed
until they reach their limiting curve of figure \ref{fig:qinfdiscretgen} or \ref{fig:q0discretgen}, whose shape directly reflects the distribution $\alpha(u)$
of starting points. In particular, the liquid phase remains of macroscopic size for any finite $q$ and occupies a fairly constant proportion of the allowed domain $D$
for the paths.  In rescaled coordinates, the area of the liquid phase tends indeed to $\int_0^1\al(u)du$ for $q\to 0$ and to the complementary
value $\al(1)-\int_0^1\al(u)du$ for $q\to \infty$. Both values are typically of the order of half of the total area $\al(1)$ of the domain $D$. The situation 
is more interesting in the presence of freezing boundaries with some ``global freezing" phenomenon: the frozen regions induced by freezing boundaries 
start to grow and invade the liquid phase, both for small or for large $q$, therefore creating straits separating macroscopic bodies of this liquid phase. 
The ``global freezing"  becomes even more dramatic 
when the starting point sequence consists of freezing boundaries only (i.e.\ is made of a succession of fully filled intervals separated by gaps).
This occurs for instance in the classical case of Section~\ref{sec:ellipse} where the liquid phase of originally (i.e.\ at $q=1$) elliptic shape gets so squeezed 
that it eventually disappears at $q=0$ or infinity.

As a final question one may wonder if any generalization of the model (e.g. with position-dependent inhomogenous weights) could still be solved using the techniques developed in the present paper, and we keep this as a 
direction of future research.

\bibliographystyle{amsalpha} 

\bibliography{Qarctic}

\end{document}